\documentclass[12pt]{JHEP3} 
\usepackage{amsmath,amsthm,amsfonts,amssymb,
}
\usepackage{graphicx}

\makeatletter
\def\smallunderbrace#1{\mathop{\vtop{\m@th\ialign{##\crcr
   $\hfil\displaystyle{#1}\hfil$\crcr
   \noalign{\kern3\p@\nointerlineskip}%
   \tiny\upbracefill\crcr\noalign{\kern3\p@}}}}\limits}
\makeatother

\def\be{\begin{equation}}\def\ee{\end{equation}}
\def\bea{\begin{eqnarray}}\def\eea{\end{eqnarray}}
\def\p{\partial}
\def\IC{\mathbb{C}}\def\IR{\mathbb{R}}\def\IP{\mathbb{P}}
\def\IZ{\mathbb{Z}}\def\IQ{\mathbb{Q}}\def\IN{\mathbb{N}}
\def\II{\mathbb{I}}\def\It{\tilde{I}}\def\IIB{\mathbb{I}_{\cx B}}

\def\cO{\mathcal{O}}
\def\OpE{\hat E}
\def\kappau{\boldsymbol{\kappa}}
\def\tk{\tx{\kappa}}\def\kh{\hat\kappa}
\def\tq{\theta}
\def\eps{\epsilon}
\def\DL{\mathfrak{L}}
\def\ch{\textrm{ch}}
\def\qcor#1{\langle #1 \rangle}\def\qcord#1{\langle\!\langle #1 \rangle\!\rangle}
\def\cx#1{{\cal#1}}\def\bx#1{{\bar#1}}

\mathchardef\mm="2D
\def\Gr{\operatorname{Gr}}\def\td{{\tilde{d}}}
\def\tr{\operatorname{tr}}

\def\coloneqq{\, :=\,} \def\Mst{\cx M_{st}}

 \def\Td{{\rm Td\,} } \def\An{\bf{A}}\def\tx#1{\tilde #1}\def\DSq{D^2\times_qS^1}\def\SSq{S^2\times_qS^1}
\def\z{Q}\def\taun{\epsilon}\def\hu{u}

\def\qc{{\mathbf q}}
\def\mb{{\bf \ell}}
\def\ml{\mu}
\def\sss#1{\noindent{\bf#1}\\}
\def\chiz2{\chi^{\IZ_2}}
\def\A{v} 
\def\chiztf{\hat \chi}
\def\IID{\II_{\cx D}}
\def\IIL{\mathbb{L}}

\title{BPS Indices, Modularity and Perturbations in Quantum K-theory }
\author{Hans Jockers$^1$, Peter Mayr$^2$, Urmi Ninad$^3$ and Alexander Tabler$^2$\\
$^1\,$PRISMA+ Cluster of Excellence and Institute for Physics (THEP),\\
Johannes Gutenberg Universit\"at, 55099 Mainz, Germany\\
{\tt jockers@uni-mainz.de} \\ \\
$^2\,$Arnold Sommerfeld Center for Theoretical Physics,\\
Ludwig-Maximilians-Universit\"at,
80333 Munich, Germany\\
{\tt mayr@physik.uni-muenchen.de,tabler.alexander@physik.uni-muenchen.de} \\ \\
$^3\,$Bethe Center for Theoretical Physics,\\
Physikalisches Institut, Universit\"at Bonn, 53115 Bonn, Germany\\
{\tt urmi@th.physik.uni-bonn.de} }
\abstract{We study a perturbation family of ${\mathcal N}=2$ 3d gauge theories and its relation to quantum K-theory. A 3d version of the Intriligator--Vafa formula is given for the quantum K-theory ring of Grassmannians. The 3d BPS half-index of the gauge theory is connected to the theory of bilateral hypergeometric $q$-series, and to modular $q$-characters of a class of conformal field theories in a certain massless limit. Turning on 3d Wilson lines at torsion points leads to mock modular behavior. Perturbed correlators in the IR regime are computed by determining the UV--IR map  in the presence of deformations. }
\preprint{BONN--TH--2021--03\\LMU-ASC 14/21\\MITP/21-027}

\begin{document}
\section{Introduction and outline}

\subsection*{Introduction}
If a sufficiently and suitable supersymmetric $d$-dimensional  UV gauge theory is obtained by compactification of a $d+1$-dimensional gauge theory on a circle $S^1$, and the properly localized path integral of the $d$-dimensional  theory computes  intersection theory on a moduli space $\cx M$, the $d+1$-dimensional compactified theory can often be related to K-theory on $\cx M$ \cite{NekPhd}. Such  lifts are studied in the Bethe-gauge correspondence of ref.~\cite{NekSha2}  with fascinating connections to integrable models.  

In this paper we consider the case of a $\cx N=2$ supersymmetric 3d theory compactified on $S^1$. Generically the theory has, amongst others, a Higgs vacuum parametrizing a K\"ahler manifold $X$ (or more generally a stack). The twisted chiral operators of the 3d theory are Wilson and 't Hooft line operators \cite{NekPhd,KW13,DGG14,CV13}, whose Chern characters specify K-theoretic classes in $K(X)$. The product of line operators in the 3d gauge theory defines a quantum deformation of the K-theoretic tensor product $\otimes$ on vector bundles $\cx O_\lambda$ over $X$,
\be\label{opp}
\cx O_\mu * \cx O_\nu =  C_{\mu\nu}^{\ \rho}(Q) \, \cx O_\rho = \cx O_\mu \otimes \cx O_\nu +O(Q)\,.
\ee 
In the weakly coupled UV theory expanded around the Higgs vacuum, the corrections $\sim O(Q)$ arise  from non-perturbative effects, weighted by the exponentials $ Q= e^{-2\pi \zeta}$ of large complexified Fayet--Iliopoulos (FI) parameters $\zeta$. One can view this quantum deformation as a 3d relative of the relation between the twisted chiral ring of the 2d $\mathcal{N}=(2,2)$ supersymmetric  gauged linear sigma model (GLSM) and the topological $A$ model, which  connects a UV gauge theory to a non-linear sigma model in the IR \cite{WitPhases,MP}. The twisted chiral ring \cite{LVW} of the latter theory computes intersections on a compactification of the moduli space of maps from the 2d world-sheet to the vacuum manifold $X$ of the theory. The twisted chiral operators are in 1-1 correspondence with the even cohomology $H^{2*}(X)$ and the chiral ring in the IR describes a quantum deformation of the classical wedge product \cite{Witten:1988xj}. In this sense the K-theoretic product~\eqref{opp} can be viewed as a 3d lift of the 2d quantum product along the lines of refs.~\cite{NekPhd,NekSha2}.

Whereas the weakly coupled GLSM is under control in both 2d and 3d, the IR regime of the 3d theory --- a non-trivial super-conformal fixed point under certain conditions --- is far  less understood in sharp contrast to the 2d case. This is partially compensated in 3d by the existence of a super-conformal index that is  preserved under RG flow.\footnote{The 3d index, and the closely related 3d partition function will be discussed and studied in detail below.} The 3d BPS index allows, in principle, to compute IR quantities from the UV gauge theory, {\it if} one is able to determine the map between the UV gauge theory and the IR degrees of freedom. Apart from very special cases, this map is highly non-trivial, and involves integrating out localized non-perturbative configurations and the generation of contact terms along the RG flow.\footnote{The UV-IR map for the 2d theory is better known as the mirror map. See, e.g., the discussions in refs.~\cite{MP,Losev:1999nt}.} 

A concrete proposal, realizing the general ideas of ref.~\cite{NekPhd}, is the 3d gauge theory/quantum K-theory correspondence of ref.~\cite{JM}, which includes the computation of the UV-IR map. It connects the partition function of the 3d UV gauge theory to the correlators of the permutation symmetric quantum K-theory of ref.~\cite{Giv15all}. The latter is equivariant w.r.t.~permutations of operator insertions, which  reflects the permutation symmetry of particles in the 3d quantum field theory. The correlators are defined on the moduli space $\Mst$ of stable maps, i.e., the moduli space of the non-linear A-twisted sigma model. However, in contrast to the A-model which computes intersection numbers on $\Mst$, the correlators of the 3d gauge theory compute holomorphic Euler numbers, which should be interpreted as the integral degeneracies of 3d BPS state in the IR regime.

The new 3d integral BPS invariants were computed for  the quintic Calabi--Yau and some other toric hypersurfaces in refs.~\cite{JM,JM2,Garoufalidis:2021cha}. Moreover, these BPS degeneracies on the {\it world-volume} were related to the integral {\it target space} invariants of Gopakumar and Vafa \cite{GV98}. The latter count degeneracies of 5d BPS states  in an M-theory compactification on a Calabi--Yau three-fold $X$. The integral 3d invariants reduce to the fractional 2d Gromov--Witten invariants in a small radius limit of $S^1$. The somewhat hidden  integrality properties of the fractional 2d invariants, and the 2d mirror map, emerge in this limit from the integrality of the 3d integral BPS degeneracies. 

In this note we further explore the 3d gauge/quantum K-theory correspondence for the case of Abelian and non-Abelian gauge groups.  For appropriate field content and parameters, these 3d theories have Higgs branches realizing complex Grassmannians. The quantum K-theory of Grassmannians has been studied by now to some detail both from the physics (see, e.g., refs.~\cite{NekSha2,KW13,JM, JMNT,Yos19,Sha19})  and the mathematics \cite{BM08,Wen,RZ18, dong20, Giv20, Yan21} point of view. 

\subsection*{Outline and results}
\begin{itemize}\leftskip-0.7cm
\item In sect.~2 we study the 3d partition function for the non-Abelian theory on $\DSq$ and its large volume limit. The latter defines a cobordism class $\Gamma_q(X)$ that has been proposed to capture the integral structure of the elliptic relatives of D-branes, related to boundary conditions on the loop space $LX$. We compute the $q$-Gamma class of the Grassmannian, which is a 3d lift of the class $\Gamma(X)$ \cite{Libgober,Iritani2,Katzarkov:2008hs}, that describes the integral structure of D-branes on $X$.

\item In the cohomological case, the Intriligator--Vafa formula gives a representation of the quantum cohomology ring of $\Gr(M,N)$ as a polynomial quotient ring in the gauge invariant fields. In sect.~\ref{sec:PQRings} we give a similar closed formula for the quantum K-theory ring of $\Gr(M,N)$, and a natural generalization of the ring from the gauge theory point of view.

\item The half-index $\IIB$ of 2d operators \cite{Gadde:2013wq,Gadde:2013sca,DGP} on the boundary $T^2=\p(\DSq)$ gives an alternative way to study the 3d BPS content, which applies also to non-zero Chern-Simons (CS) levels with Dirichlet boundary conditions (b.c.) on the gauge fields. We compute this index for the same class of 3d gauge theories in sect.~\ref{sec:halfindex} and study the system of difference equations associated with it. We find an  interesting generalization of the UV-IR relation that involves the theory of {\it bilateral} $q$-hypergeometric series. In particular the $q$-series $I$ obtained from the K-theoretic vortex sum and the half-index $\IIB$ with Dirichlet b.c. on the gauge fields solve (almost) the same system of hypergeometric difference equations, with the $I$-function  related to the positive half of the bilateral series $\IIB$.

\item We find an interesting connection of these bilateral series to modular and mock modular functions.\footnote{Relations between K-theoretic $I$-functions and mock modular functions have been first observed in ref.~\cite{RZ18}.} For positive CS levels, the index $\IIB$ converges in the limit of zero masses and FI parameters and gives rise to a large class of modular characters of 2d conformal theories (CFTs) of a particular type studied by Nahm \cite{Nahm}. The CFT classification by ``modular triples'' predicts particular 3d CS levels as candidates for non-trivial 3d superconformal points. The 3d Wilson line algebra becomes isomorphic to the Verlinde algebra of the CFT in this massless limit. Adding Wilson lines at torsion points of the boundary $T^2$, we find a close relation of CFT characters of certain non-unitary minimal models and mock theta functions, which share a common origin from the 3d point of view and arise from the same 
Wilson line algebra.

\item The existing computations for quantum K-theory ring of Grassmannians have been limited so far to the special case of a trivial UV-IR map, and to the {\it unperturbed} theory. For more general CS terms, and for correlators with arbitrary number of insertions of twisted chiral operators and their descendants, one needs to compute the 3d version of the mirror map. Due to the WDVV equations satisfied by quantum K-theory \cite{GivWDVV,Giv15all}, the result can be assembled into a generating function $\cx F(Q,t)$, where $t$ are the perturbations in the IR theory, related to the perturbations in the 3d UV gauge theory by the 3d mirror map.   We compute this map and the perturbed correlators for some Grassmannian examples. 
\end{itemize}

\section{Gauge fields with Neumann boundary conditions\label{sec:PF}}
The 3d gauge theory/quantum K-theory correspondence of ref.~\cite{JM} equates the $I$-function of permutation symmetric quantum K-theory of ref.~\cite{Giv15all} with the vortex sum\footnote{See ref.~\cite{DGH} for the concept of vortex sums in the sense used in this paper.} of a supersymmetric partition function on the 3d space\footnote{The subscript $q$ denotes the twisting of the geometry specified in eq.~\eqref{qdef} below.}  $\DSq$. In the following we derive the K-theoretic $I$-function for the Grassmannian from the 3d gauge theory in two ways. In this section we outline the computation with Neumann b.c.~on the gauge fields to obtain the result announced in ref.~\cite{JMNT}, together with  the so-called $q$-Gamma class for the Grassmannian, which is related to the integral structure of the boundary condition at the boundary elliptic curve $T^2 = \partial(\DSq)$. This 3d field theory computation is only valid for zero effective CS levels, due to the gauge anomaly on the boundary \cite{DGP,YS}. One can make sense of the result for the $I$-function at non-zero CS terms by a factorization of the index for $\SSq$ along the lines of ref.~\cite{BDP}, as was used in ref.~\cite{JM}. A more direct approach to the theory with non-zero effective CS terms will be discussed in sect.~\ref{sec:halfindex}.
 
\subsection{Hypergeometric $q$-series and generalized $I$-functions}
Neumann b.c.~for the gauge fields have been considered in refs.~\cite{Gadde:2013wq,Gadde:2013sca, BDP,YS}.\footnote{Preserving $\mathcal{N}=(0,2)$ supersymmetry at the boundary requires either Neumann or Dirichlet~b.c. for the gauge field, as first pointed out for $\cx N=2$ 3d Abelian vector multiplets in ref.~\cite{Okazaki:2013kaa}.} The $\cx N=2$ supersymmetric geometry is $\DSq$, where 
\be \label{qdef}
q=e^{-\hbar\beta}\,,\ee 
is a weight for the $U(1)$ rotations in $U(1)_E\times U(1)_R$  generated by a loop around $S^1$ \cite{BDP}. Here $U(1)_R$ is the $R$-symmetry, $U(1)_E$ the rotation group on $D^2$ and $\beta$ is the radius of the $S^1$. On the fixed point of localization, the component of the gauge potential $A_\mu$ along the $S^1$ is valued in the Cartan subalgebra of the gauge group $G$.  For  complex Grassmannian $\Gr(M,N)$ we consider the $G=U(M)$ gauge theory with $N$ chiral fields in the fundamental representation. The partition function reduces to a finite dimensional integral 
\begin{equation}\label{LocPathIntegral}
	\begin{aligned}
		Z_{\DSq}=\frac 1 {M!} 
		\int \prod_{a=1}^M\frac{d z_a}{2\pi i z_a}\   e^{-S_{\rm class}}\ Z_{\text{1-loop}} \,,
	\end{aligned}
\end{equation}
where the integration is over $M$ Wilson lines $z_a$, $a=1,\hdots,M$,  in the Cartan subgroup $U(1)^M\subset U(M)$. 
The classical contribution to the path integral $e^{-S_{\rm class}}= Z_{\text{FI}} \cdot Z_{\text{CS}}$ at the  localization locus contains the Fayet--Iliopolous term 
\begin{equation}\label{zfi}
\begin{split}
Z_{\text{FI}} &=  e^{2 \pi  \zeta \tr_R(\sigma)/\ln q}=\tx Q^{- \tfrac{\tr_R (\sigma)}{\log q}}  \  ,\quad  \tx Q=e^{-2\pi \zeta}\ ,\\
\end{split}
\end{equation}
where $\sigma_a := \log z_a$, and the Chern-Simons term\footnote{\label{footnotesum}As in ref.~\cite{JMNT}, for a matrix $\sigma=\textrm{diag}(\sigma_1,\ldots,\sigma_M)$ the trace symbols are defined as $\tr_{U(M)}(\sigma^2)=\sum_a\sigma_a^2$,  $\tr_{U(1)}(\sigma^2)=\frac 1 M (\sum_a\sigma_a)^2$, $\tr_{SU(M)}(\sigma^2)=\tr_{U(M)}(\sigma^2)-\tr_{U(1)}(\sigma^2)$ and $\tr_R(\sigma)=\sum_a\sigma_a$.}
\begin{equation}\label{CSbare}
	\begin{aligned}
		Z_\text{CS}&= e^{
			\tfrac{ \kappa_S}{2\log q} \tr_{SU(M)}(\sigma^2)
			+\tfrac{ \kappa_A}{2\log q} \tr_{U(1)}(\sigma^2)
			- \kappa_R \tr_R(\sigma) }\,.
	\end{aligned}
\end{equation}
The parameters $\kappa_S$ and $\kappa_A$ specify the bare levels for the $SU(M)$ and $U(1)$ subgroups of the gauge group $U(M)$, respectively, while $\kappa_R$ is a level for the mixed gauge/R-symmetry CS term.  The term $Z_{\text{1-loop}}$ contains the one-loop determinants for the chiral and vector multiplets computed in refs.~\cite{Gadde:2013wq,BDP,YS}. The details of the computation are relegated to app.~\ref{app:PFNeum}, where we show that the result can be rewritten as a multi-residue integral
\begin{equation}\label{ZisfI}
Z_{\DSq} = \frac 1 {M!}\, \oint \prod\limits_{a=1}^{M}\, \frac{d \epsilon_a}{2 \pi i } \ f_{\text{Gr}(M,N)} (q,\epsilon) \cdot I^{(\kh_S,\kh_A,\kh_R)}_{\text{Gr}(M,N)} (Q,q,\epsilon) \ .
\end{equation}
Here we have split off the $Q$-independent folding factor $f_{\text{Gr}(M,N)}$ in the integrand, which is the perturbative contribution in the gauge theory. It is related to the boundary condition on $T^2\simeq \partial \DSq$; its geometric meaning will be discussed in sect.~\ref{subsec:qgamma} below. The second factor is the $Q$-dependent vortex sum $I^{(\kh_S,\kh_A,\kh_R)}_{\text{Gr}(M,N)} $ given by the generalized $q$-hypergeometric series
\begin{equation}\label{ISQKGr}
I^{(\kh_S,\kh_A,\kh_R)}_{\text{Gr}(M,N)} (Q,q,\epsilon)= c_0 \sum_{\vec{d} \in \mathbb{Z}^M_{\geq 0}}  (-Q)^{\sum_{a=1}^M \tilde{d}_a }  q^{CS(\tilde{d})}\, \frac{ \prod\limits_{1 \leq a<b \leq M} \! \! q^{\tfrac{1}{2}\tilde{d}_{ab}^2} (q^{\tfrac{1}{2}\tilde{d}_{ab}} - q^{-\tfrac{1}{2}\tilde{d}_{ab}})}{\prod\limits_{a=1}^M\prod\limits_{r=1}^{d_a} (1-q^{r-\epsilon_a})^N}  \ ,
\end{equation}
which is the expression given in ref.~\cite{JMNT}. Here $c_0$ is a $(q,\eps)$- and CS level-dependent normalization factor that will be chosen such that $I_{\Gr(M,N)}|_{\vec{d}=0}=(1-q)(-Q)^{-\sum_a \eps_a}$.\footnote{The factor $(-Q)^{-\sum_a \eps_a}$  in the definition of the $I$-function is included to simplify the difference equations in the following \label{fn:norm}. This is the same factor as in remark 2.11 of \cite{IMT} after using eq.~\eqref{epstoH} below.}  The new parameter $ Q$ has been introduced in \eqref{ISQKGr} such that $ Q = (-1)^{N+M} \tilde Q$ and moreover $\td_a = d_a-\eps_a$ and $\td_{ab}=\td_a-\td_b$. 

The Chern-Simons term $q^{CS(\td)}$ is given in terms of the effective Chern-Simons levels\footnote{The effective CS levels contain the one-loop corrections to the bare CS level, as discussed in ref.~\cite{AHISS}. The one-loop corrections depend in turn on the phase of the 3d gauge theory. If not stated otherwise, we refer to the effective CS levels of the phase with large positive FI parameter (small $|Q|$) and non-negative  real masses in the following.} 
\begin{equation}\label{effCS1}
q^{CS(\td)} = q^{ \frac{1}{2}\hat \kappa_S \tr_{SU(M)} (\tilde{d}^2) + \frac{1}{2}\hat \kappa_A \tr_{U(1)} (\tilde{d}^2)  + \hat \kappa_R \tr_{R}(\tilde{d}) } \ ,
\end{equation}                                                                                             
where the effective levels $\hat \kappa$ are expressed in terms of the bare levels $\kappa $ as 
\begin{equation}\label{effCS2}\textstyle
\hat{\kappa}_S = \kappa_S - M + \frac{N}{2} \quad , \quad \hat{\kappa}_A = \kappa_A + \frac{N}{2} \quad \ \quad \hat{\kappa}_R = \kappa_R + \frac{N}{4} \ . 
\end{equation}

\noindent By the 3d gauge theory/quantum K-theory  correspondence,  the series $I_{\text{Gr}(M,N)} (Q,q,\epsilon)$  computes the $I$-function for the permutation symmetric quantum K-theory  of ref.~\cite{Giv15all} for effective CS levels $\hat\kappa_{i}=0$. For other levels $\hat \kappa_i$, one obtains a three parameter family of generalized $I$-functions. One can check that on the one-parameter slices $(\hat \kappa_S,\hat \kappa_A,\hat \kappa_R)=(\ell_\square,\ell_\square,-\ell_\square/2)$ and $(\hat \kappa_S,\hat \kappa_A,\hat\kappa_R)=(0,M\ell_{\operatorname{det}},-\ell_{\operatorname{det}}/2)$ the above result respectively reproduces the $I$-functions  at level $\ell_\square$ in the fundamental representation of $U(M)$ and at level $\ell_{\operatorname{det}}$ in the determinantal representation of $U(M)$ in quantum K-theory with level structure, as derived in ref.~\cite{Wen} (see also refs.~\cite{Tai,RZ18b,Yos19,Giv20,Yan21}).\\

\subsection{The $q$-Gamma class of the Grassmannian \label{subsec:qgamma}}
The $Q$-independent factor $f_{\text{Gr}(M,N)}$ in the integral eq.~\eqref{ZisfI} has the following  interesting geometric meaning. The residue integral can be viewed as an integral over the Grassmannian $X=\Gr(M,N)$,  using the relation
\be\label{intform}
\frac{1}{M!}\, \oint \prod_{a=1}^{M}\frac{d \tx \epsilon_a}{2 \pi i } \, F(\tx \eps)\, \cx E_X(\tx \eps) = \int_{X}  \tx F(\sigma_\mu(x))\,,\qquad \cx E_X(\tx\eps)= \frac{\prod_{a<b}^M(-\tx \epsilon_{ab}^2)}
		{\prod_{a=1}^M(\tx\epsilon_a)^N}\,,
\ee
introducing the short hands $\tx \eps_a = \log q\,  \eps_a$. 
Here $F(\eps)$ is a function in  $\eps_a$ invariant under the action of the permutation group $S_M$ on the $\eps_a$. It can be expanded as $F(\eps) =F(\sigma_\mu(\eps))=\sum_\mu a_\mu \sigma_\mu(\eps)$, where $\mu$ are representations of $U(M)$ labelled by Young tableaux and $\sigma_\mu$ are the Schur polynomials.\footnote{See app.~\ref{app:SymGrass} for definitions.} The class $\tilde F(\sigma_\mu(x))\in H^*(X,\IQ)$ on the r.h.s.~is obtained by the replacement 
\be\label{epstoH}
\tx\eps_a\to \beta x_a\,,
\ee
with $x_a$ the Chern roots of the dual $S^*$ of the tautological bundle on $X$. The radius $\beta$ can be set to 1, if it is not zero, but it is useful to keep it  as in ref.~\cite{JM} to define the small radius limit $\beta\to 0$, where $q\to 1$. Eq.~\eqref{intform} then turns into a  residue integral representation for the integration formula of ref.~\cite{Martin}, which is derived from the embedding of the symplectic quotient with group $G$ into the symplectic quotient by the maximal torus $T\subset G$. The integral picks out the coefficient  $a_{\mu_\text{top}}$ of the top class in $H^{\dim(X)}(X,\IZ)$ corresponding to the representation  $\mu_\text{top}=(\mu_1,\hdots,\mu_{M})$ with $\mu_i=N-M\ \forall i$.

With this interpretation in mind, the function $f$ can be written as a product of factors involving characteristic classes of $X$ (see app.~\ref{app:CharClasses} for details)\\[-3mm]
\be\label{foldingf}
f_X(q,\eps) =  \Td_q(X)\cdot   \Gamma_q(X) \cdot \cx E_X(\tx \eps) \cdot \frac{q^{\An(\eps)} N(q,\eps) }{\eta(q)^{MN-{\rm{dim}}(G)}}\,,
\ee
where $\eta(q)$ is the Dedekind eta-function.
The first factor represents a $\beta$-dependent version of the Todd class of $X$
\be
		\Td_q(X)=
		\prod_{a=1}^M\bigg(\frac
		{\tx\epsilon_a}
		{(1-q^{-\epsilon_a})}\bigg)^N
		\prod_{a<b}^M\frac
		{(1-q^{\epsilon_{ab}})(1-q^{-\epsilon_{ab}})}
		{-\tx \epsilon_{ab}^2}.
\ee
Except for the normalization factor $\beta$ for the Chern roots, this expression agrees with the Todd class for $\Gr(M,N)$ derived in ref.~\cite{Martin}.
\\
The second factor defines  the $q$-Gamma class of the Grassmannian
	\be
\Gamma_{q}(X)
		=\frac
		{\prod_{a=1}^M\Gamma_q(1+\epsilon_a)^N}
		{\prod_{a<b}^M \Gamma_q(1+\epsilon_{ab})\Gamma_q(1-\epsilon_{ab})}\,,
\ee
where $\Gamma_q(x)$ is the $q$-Gamma function \eqref{defqGamma}. This class has been proposed in ref.~\cite{JM} as a $q$-generalization of the ordinary Gamma class of refs.~\cite{Libgober,Iritani2,Katzarkov:2008hs}, which defines an integral structure in quantum cohomology. The ordinary Gamma class is related to the central charge of D-brane b.c.~on the $S^1$ boundary of $D^2$ in the physics context. The $q$-Gamma class plays an analogous role for the elliptic cousin of a D-brane, an ``E-brane'', associated to the b.c.~on the $T^2$ boundary of $\DSq$. In the small radius limit $\beta\to 0$,  the expression for $\Gamma_{q}(X)$ reduces to the expression obtained in ref.~\cite{GGIritani} for the ordinary Gamma class for $\Gr(M,N)$.

The last factor contains the anomaly term
\be\textstyle
q^{\An(\eps)}=q^{-\tfrac{c_1(\epsilon)}{2}-\ch_2(\epsilon)+ \operatorname{CS}(-\epsilon)}\,,
\ee
where the CS level-dependent term $\operatorname{CS}(-\epsilon)$ is given by restricting to $d_a=0$ in eq.~\eqref{effCS1}, and the Chern characters of $X$ are
\be\label{ChernCharacters12}
\textstyle
\ch_1(\epsilon)=c_1(\eps)=N\sum_{a=1}^M\eps_a\,,\quad  \ch_2(\epsilon)=\frac N 2 \sum_{a=1}^M \epsilon_a^2-\sum_{a<b}^M\epsilon_{ab}^2\, .
\ee
This term represents a familiar global anomaly in the interpretation of the classical limit $Q\to 0$ of the integral as a Dirac--Ramond index on the loop space $LX$ of $X$ \cite{WitLoop}. We refer  to sect.~4 of ref.~\cite{JM} for a discussion and further references. A sign factor, reproducing the one of ref.~\cite{Iritani2} in the 2d limit, and further factors related to normalization and contact terms discussed in \cite{BDP} are collected in $N(q,\eps)$ given in eq.~\eqref{numericalfactor}. 

Note that the perturbative contributions in the folding factor,  including the anomaly factor, depend explicitly on the chosen Neumann b.c.~for the gauge fields. On the other hand, the non-perturbative sector captured by the generalized $I$-function $I^{(\kh_S,\kh_A,\kh_R)}_{\text{Gr}(M,N)} $ depends on the b.c.~for the gauge fields only implicitly via the effective CS levels $\hat\kappa$.\footnote{In addition, both depend explicitly on the b.c.~for the matter fields.} In sect.~\ref{sec:halfindex} we rederive the $I$-function from a different b.c.~for the gauge fields on $T^2$. The result \eqref{indexresult} gives an independent confirmation of the correct relative normalization of the two factors in the split \eqref{ZisfI}, and in particular the normalization of $I^{(\kh_S,\kh_A,\kh_R)}_{\text{Gr}(M,N)} $. 

\subsection{Difference equations}
The vortex sum $I_{\text{Gr}(M,N)}$, or equivalently the partition function $Z$, can be characterized as a solution to a system of difference equations with appropriate asymptotic behavior.\footnote{The relevance and properties of difference equations from the point of the 3d $\cx N=2$ theory have been first studied and elucidated in refs.~\cite{DGG14,Dimofte:2011py,KW13}.}  To this end consider the Abelianized $I$-function\footnote{We suppress the superscript for the CS terms for simplicity in some of the following equations.} $\hat I_{\Gr(M,N)}(Q_a)$ depending on $M$ parameters $Q_a$, obtained from \eqref{ISQKGr} by the replacement 
\be \label{abel} 
  Q^{\sum_a \td_a} \to \prod_{a=1}^M Q_a^{\td_a} \ .
\ee
This amounts to passing to the Abelian quotient with $M$ distinct FI parameters for the gauge group $U(1)^M$. The generalized $q$-hypergeometric series $\hat I(Q_a)$ satisfies the identities
\be\label{Deq1}
\DL_a \hat I _{\Gr(M,N)}(Q_a)=g(\eps)\,, \quad a=1,\hdots, M\,,\qquad g(\eps)\sim (1-q^{-\eps_a})^N\,,
\ee
with the difference operators \cite{JMNT,Giv20}\footnote{Without the extra factor in the normalization mentioned in fn.~\ref{fn:norm}, $p_a$ needs to be replaced by $p_aP_a$ with $P_a=q^{-\eps_a}$.}
\be\label{Deq1b}
 \DL_a =\prod_{b\neq a}(1-qp_{ba})(1-p_a)^N+(-1)^{M-1}Q_a 
q^{\tfrac\kappau2}p_a^{\hat \kappa_S}\prod_{b}p_b^{\tfrac1M (\hat \kappa_A-\hat \kappa_S)}\prod_{b\neq a}(1-qp_{ab})\,.
\ee
Here $p_a$ are the shift operators 
\be 
p_aQ_b = Q_bq^{\delta_{ab}}\, p_a\,,\qquad p_{ab}=p_a/p_b\,,
\ee 
and $\kappau=\hat \kappa_S+\tfrac 1 M (\hat \kappa_A-\hat \kappa_S)+2\hat\kappa_R$. The inhomogeneous term $g(\eps)$ is zero in the cohomology of the Abelianized quotient, due to the relation $(1-q^{-\eps_a})^N=0$ for each $\IP^{N-1}$ factor. This relation is imposed in the gauge theory by the residue integral \eqref{ZisfI}, noting that the folding factor $f_X(q,\eps)$ has $N$-th order pole $\sim (1-q^{-\eps_a})^{-N}$.

The difference operators have been written in ref.~\cite{JMNT} in a somewhat simpler form that will be used in the next section. To this end  one expresses the Abelianized $I$-function as 
\be\label{NAtoA}
\hat I_{\Gr(M,N)}(Q_a)=\Delta \cdot \It_{abel.} (Q_a)\,, \qquad \Delta=\prod_{b\neq a}(p_a-p_b)\,,
\ee
with a generalized Abelian $I$-function for the gauge group $U(1)^M$
\be
\It_{abel.} (Q_b)=c_0\, \sum_{\vec d \in\IZ_{\ge 0}^M} q^{\gamma \sum_{b>a}\td_a\td_b}\prod_a\frac{(-Q_a)^{\td_a}q^{\frac{\alpha}2  \td_a^2+\beta \td_a}}{\prod_{r=1}^{d_a}(1-q^{r-\eps_a})^N}\, .
\ee
The  constants $\alpha,\beta,\gamma$ are related to the effective CS levels $\hat\kappa_i$ in \eqref{ISQKGr} as
\begin{equation} \label{eq:const}
  \alpha = \hat \kappa_S+ \Delta_\kappa+M \ , \quad 
  \beta = \hat \kappa_R-\frac 12 (M-1) \ , \quad
  \gamma =\frac{\hat \kappa_A-\hat\kappa_S}{M}-1=\Delta_\kappa \ .
\end{equation}
This sum satisfies the difference equations $\OpE_a\It_{abel.} (Q_b)=O(\eps_a^N)$ for $a=1,\hdots, M$ with
\bea \label{DiffIt}
\OpE_a&=& (1-p_a)^N\, +Q_a q^{\frac{\alpha}2 + \beta}p_a^\alpha\prod_{b\neq a} p_b^\gamma\, .
\eea
From eq.~\eqref{NAtoA}, the operators $\DL_a$ and $\OpE_a$ are related by a conjugation with the invertible operator  $\Delta(p)$.

\section{Quantum K-theory algebras as polynomial quotient rings} \label{sec:PQRings}
In refs.~\cite{JMNT,Yos19,Sha19} the quantum K-theory algebra of the complex Grassmannian $\Gr(M,N)$ was identified with the Wilson line algebra of a 3d $\mathcal{N}=2$ $U(M)$ gauge theory coupled to $N$ fundamental matter multiplets and with zero effective CS levels. More generally, for non-canonical CS levels one obtains a modified Wilson algebra corresponding to the quantum K-theory with non-zero level of ref.~\cite{RZ18}. The resulting quantum K-theory algebras were represented in ref.~\cite{JMNT} as quotients of polynomial rings in some examples. Below we give a general formula that is similar to the one known for quantum cohomology and valid for all $M,N$ and CS terms.\footnote{Another representation of the ring structure for canonical levels has been given in ref.~\cite{Sha19}. The results obtained in this section should also be closely related to that of ref.~\cite{GKInt}, as pointed out to us by Leonardo Mihalcea. See also ref.~\cite{Gon20} for a relation to the (quantum-)Kirwan map.}

\subsection{Canonical Chern--Simons terms} \label{sec:canCST}
Let us consider the difference operators $\OpE_a$, $a=1,\ldots,M$, annihilating the $I$-functions $\It_{abel.}(Q_a)$ of the Abelianized 3d $\mathcal{N}=2$ $U(1)^M$ gauge theory with $N$ matter fields of charge $+1$ for each $U(1)$ factor. For zero effective CS terms $\hat\kappa_i=0$, the difference operators \eqref{DiffIt} of the associated Abelianized theory can be rewritten as
\begin{equation}\label{deq1}
\OpE_a=\delta_a^N \prod_{b=1 \atop b\neq a}^M (1-\delta_b)  + Q_a (1-\delta_a)^{M-1} \ , \qquad a=1,\ldots, M \ ,
\end{equation}
where $\delta_a = 1-p_a$ is the shifted Wilson lines operator and $Q_a$ is the fugacity of the dual global symmetry $U(1)_\text{top}$ of the $a$-th $U(1)$ factor introduced above. As explained in ref.~\cite{JMNT}, for the purpose of computing the $q$-independent Wilson line algebra, one may take the semi--classical limit $q\to 1$. In this limit the difference operators $\OpE_b$ reduce to $M$ polynomials $E_b(\delta_a,Q_a)$ of $2M$ commuting variables $\delta_a$ and $Q_a$, which realize relations in the fusion products of (shifted) Wilson lines and define the ideal
\begin{equation} \label{eq:Iideal}
  \mathcal{I}^\text{Ab}_{M,N} = \langle\!\langle E_1, \ldots, E_M \rangle\!\rangle \ , \end{equation}
in the polynomial ring $\mathbb{Z}[\delta_a, Q_a]$.

In the non-Abelian theory all fugacities $Q_a$ are set equal to the single fugacity~$Q$ of the single surviving global symmetry $U(1)_\text{top}$ that is dual to the non-Abelian gauge group $U(M)$. Due to the Weyl symmetry $S_M$ of the non-Abelian gauge group $U(M)$, the Wilson lines in the non-Abelian $U(M)$ gauge theory are symmetrized (shifted) Wilson lines of the Abelianized $U(1)^M$ gauge theory. As such they can be expressed in terms of elementary symmetric polynomials $X_1(\delta_a), \ldots, X_M(\delta_a)$ in the variables $\delta_a$. Moreover, the relations of the Abelian (shifted) Wilson lines turn into relations of the non-Abelian (shifted) Wilson lines, as described by the ideal
\begin{equation} \label{eq:defI}
    \mathcal{I}_{M,N} = \left\{ r \in \mathbb{Z}[X_\ell,Q] \, \middle| \,
      \Delta(\delta_a)\, r(X_\ell(\delta_a)) \in \left.\mathcal{I}^\text{Ab}_{M,N}\right|_{Q\equiv Q_1 = \ldots = Q_M} \right\} \ , 
\end{equation}
in the polynomial ring $\IZ[s_a,Q]$. In this definition the additional factor 
\begin{equation} \label{eq:defDelta}
  \Delta(\delta_a) = \prod_{1\le a < b \le M} (\delta_a - \delta_b) \ ,
\end{equation}  
accounts for the charged $W$-boson multiplets upon breaking the non-Abelian gauge group $U(M)$ to its Abelian maximal torus $U(1)^M$ \cite{Kapustin:2009kz,JMNT}. Thus, the algebra of Wilson lines in the discussed non-Abelian gauge theory with canonical CS term is identified with the polynomial quotient ring
\begin{equation}
  \IZ[X_1,\ldots,X_M,Q] / \mathcal{I}_{M,N} \ ,
\end{equation} 
which in turn --- as explicitly confirmed in examples in ref.~\cite{JMNT} --- is isomorphic to the quantum K-theory algebra for the complex Grassmannian $\Gr(M,N)$ at zero level. To arrive at a systematic description of these Wilson line algebras for general $M, N$, we first state:

{\leftskip0.7cm 
\paragraph{Proposal 1:}

The quantum K-theory algebra for the complex Grassmannian $\Gr(M,N)$ at zero level  is isomorphic to the polynomial quotient ring
\begin{equation}\label{proposal1}
  \IZ[X_1,\ldots,X_M,Q] / \mathcal{J}_{M,N} \ , \qquad \mathcal{J}_{M,N} = \langle\!\langle Y_1(Q), \ldots, Y_M(Q) \rangle\!\rangle \ ,
\end{equation}
with the ideal $\mathcal{J}_{M,N}$ generated by
\begin{equation}\label{eq:Yrels}
  Y_\ell(Q) = \frac{1}{\Delta(\delta_a)} 
  \left|\begin{matrix} 
    E_1 & \delta_1^{M-1} & \cdots & \delta_1^{\ell} &  & \delta_1^{\ell-2} &  \cdots &  \delta_1 & 1   \\ 
    E_2 & \delta_2^{M-1} & \cdots & \delta_2^{\ell} &  & \delta_2^{\ell-2} &  \cdots &  \delta_2 & 1   \\ 
    \vdots & \vdots && \vdots && \vdots && \vdots & \vdots \\ 
    E_M & \delta_M^{M-1} & \cdots & \delta_M^{\ell} &  & \delta_M^{\ell-2} &  \cdots &  \delta_M & 1 
   \end{matrix} \right| \ .
\end{equation}
Equivalently, the ideal $\mathcal{J}_{M,N}$ can be expressed compactly as
\begin{equation} \label{eq:JGroth}
   \mathcal{J}_{M,N} = \langle\!\langle \mathcal{O}_{\lambda}-(-1)^M Q \,| \, \lambda \in \Lambda\rangle\!\rangle \ ,
\end{equation}
with the $M$ Grothendieck polynomials $\mathcal{O}_\lambda(\delta_a)$ in the variables $\delta_a$ labelled by the partitions/Young tableaux $\lambda$ in the set 
\be
\Lambda=\{(N-M+1,\smallunderbrace{1,\ldots,1}_{k})\,| k=0,\hdots,M-1\}\ .
\ee
}

\noindent There are a few comments in order about this proposal: First, in both formulations the generators of the ideal $\mathcal{J}_{M,N}$ are symmetric polynomials in the variables~$\delta_a$ and hence are expressible in terms of the elementary symmetric functions~$X_\ell(\delta_a)$. Second, the elementary symmetric polynomials $X_\ell(\delta_a)$ enjoy a geometric interpretation as follows. Setting $\delta_a = 1 - e^{-x_a}$, the variables $-x_a$ are the Chern roots of the universal subbundle $S$ of the complex Grassmannian $\Gr(M,N)$. That is to say the elementary symmetric polynomials $X_\ell(-x_a)$ (as functions of $-x_a$) correspond to the Chern classes~$c_\ell(S)$ of the rank $M$ bundle $S$, which are Poincar\'e dual to the Schubert cycles $\sigma_{1,...,1}$ (partition with $\ell$ boxes) of co-dimension $\ell$ in the complex Grassmannian $\Gr(M,N)$.\footnote{We denote both Schubert cycles and their Poincar\'e dual cohomology classes of the Grassmannian $\Gr(M,N)$ by~$\sigma_\lambda$.} The symmetric polynomials $X_\ell(\delta_a)$ (as functions of $\delta_a$) are the Chern characters of certain coherent sheaves on $\Gr(M,N)$, which are locally free on the subvarieties of their associated Schubert cycles $\sigma_{1,\ldots,1}$.\footnote{The Chern characters of the structure sheaves $\mathcal{O}_{\sigma_\lambda}$ of the Schubert cycles $\sigma_\lambda$ are given by the Grothendieck polynomials $\mathcal{O}_\lambda(\delta_a)$, vindicating their notation. For the elementary symmetric polynomials one finds $X_\ell(\delta_a)=\sum_{k=\ell}^{M} \binom{k-1}{\ell-1} \mathcal{O}_{\smallunderbrace{\scriptstyle1,\ldots,1}_{k}}(\delta_a)$; see also app.~\ref{app:SymGrass}.} Thus, the polynomials $X_\ell(\delta_a)$ are identified with generators of the K-group $K(\Gr(M,N))$. Third, the two given formulations~\eqref{eq:Yrels} and \eqref{eq:JGroth} of the ideal $\mathcal{J}_{M,N}$ are equivalent as shown in app.~\ref{app:QKrels}. After a linear transformation of the generators of the ideal~\eqref{eq:JGroth} one recovers the description~(4.10) and (A.5) of ref.~\cite{JMNT}, which has been deduced from the ideal~\eqref{eq:Iideal}. In app.~\ref{app:QKrels}, we also prove the assertion that the ideals $\mathcal{I}_{M,N}$ and $\mathcal{J}_{M,N}$ are identical. Thus the above formulas generalize the results of ref.~\cite{JMNT} for all values of $N,M$.

\subsection{General Chern--Simons terms} \label{sec:GenCS}
The description in the previous section carries over for more general CS levels as well, which are described by the difference operators  \eqref{DiffIt} of the Abelianized theory.
In the limit $q\to1$ these difference operators give rise to polynomials $E_{\hat\kappa,b}(\delta_a,Q_a)$ in terms of the shifted Wilson lines $\delta_a = 1 - p_a$ according to
\begin{equation} \label{eq:Ekappa}
   E_{\hat\kappa,a} =  
   \begin{cases}
     \delta_a^N + Q_a (1-\delta_a)^\alpha\,\prod_{b\ne a}(1-\delta_b)^\gamma & \text{for $\alpha\ge0, \gamma\ge0$} \ , \\
     \prod_{b\ne a}(1-\delta_b)^{|\gamma|}\,\delta_a^N+ Q_a (1-\delta_a)^\alpha & \text{for $\alpha\ge0, \gamma<0$} \ , \\
     (1-\delta_a)^{|\alpha|}\,\delta_a^N + Q_a \prod_{b\ne a}(1-\delta_b)^\gamma & \text{for $\alpha<0, \gamma\ge0$} \ , \\
     (1-\delta_a)^{|\alpha|}\,\prod_{b\ne a}(1-\delta_b)^{|\gamma|}\,\delta_a^N + Q_a & \text{for $\alpha<0, \gamma<0$} \ .
   \end{cases}
\end{equation}
As in the previous section these polynomials generate both the Abelian ideal $\mathcal{I}^\text{Ab}_{\hat\kappa,M,N}$ in the polynomial ring $\IZ[\delta_a,Q_a]$ of the shifted Wilson lines and the non-Abelian ideal $\mathcal{I}_{\hat\kappa,M,N}$ in the polynomial $\IZ[X_\ell,Q]$ of the elementary symmetric functions~$X_\ell(\delta_a)$, analogously as defined in eqs.~\eqref{eq:Iideal} and \eqref{eq:defI}, respectively. Furthermore, repeating the same arguments presented in app.~\ref{app:QKrels} for canonical CS terms, we can show that for general CS terms the Wilson line algebra can again be represented in terms of polynomial quotient ring as:

{\leftskip0.7cm 
\paragraph{Proposal 2:}

The Wilson line algebra of the 3d $\mathcal{N}=2$ $U(M)$ gauge theory with $N$ fundamental matter multiplets with the effective CS levels $\hat\kappa$ is given by the polynomial quotient ring
\begin{equation}
  \IZ[X_1,\ldots,X_M,Q] / \mathcal{J}_{\hat\kappa,M,N} \ , \qquad \mathcal{J}_{\hat\kappa,M,N} = \langle\!\langle Y_{\hat\kappa,1}(Q), \ldots, Y_{\hat\kappa,M}(Q) \rangle\!\rangle \ ,
\end{equation}
where the ideal $\mathcal{J}_{\hat\kappa,M,N}$ is generated by the symmetric polynomials
\begin{equation}\label{eq:YrelsCS}
  Y_{\hat\kappa,\ell}(Q) = \frac{1}{\Delta(\delta_a)} 
  \left|\begin{matrix} 
    E_{\hat\kappa,1} & \delta_1^{M-1} & \cdots & \delta_1^{\ell} &  & \delta_1^{\ell-2} &  \cdots &  \delta_1 & 1   \\ 
    E_{\hat\kappa,2} & \delta_2^{M-1} & \cdots & \delta_2^{\ell} &  & \delta_2^{\ell-2} &  \cdots &  \delta_2 & 1   \\ 
    \vdots & \vdots && \vdots && \vdots && \vdots & \vdots \\ 
    E_{\hat\kappa,M} & \delta_M^{M-1} & \cdots & \delta_M^{\ell} &  & \delta_M^{\ell-2} &  \cdots &  \delta_M & 1 
   \end{matrix} \right| \ .
\end{equation}
}

\bigskip
\bigskip
\noindent
The structure constants of the Wilson line algebra $\cx A_{(M,N)}^{\hat \kappa}$ depend on the effective CS levels $\hat\kappa_i$, as becomes already apparent in eq.~\eqref{eq:Ekappa}. If the CS~levels $\hat \kappa_i$ lie in a certain range, the algebra $\cx A_{(M,N)}^{\hat \kappa}$ viewed as a $\mathbb{Z}[Q]$-module is freely generated by the same basis of $\binom{N}{M}$ elements as the classical topological K-group $K(\Gr(M,N))$. Then the algebra $\cx A_{(M,N)}^{\hat \kappa}$ becomes a quantum deformation of the classical K-theory of the Grassmannian.\footnote{More generally, the Wilson line algebras can be constructed as $\mathbb{Z}[[Q]]$-modules from the quotients $\mathbb{Z}[[Q]][X_1,\ldots,X_M] / \mathcal{J}_{\hat\kappa,M,N}$. This allows us to study an even broader class of quantum deformations of classical topological K-groups. The generalization to $\mathbb{Z}[[Q]]$-modules becomes for instance important in the context of quantum K-theory of Calabi--Yau manifolds~\cite{JM2}.} For such freely-generated $\mathbb{Z}[Q]$-modules, we refer to the associated range of CS levels as the {\it geometric window},\footnote{This window is different from the one defined in ref.~\cite{JMNT}, which is relevant for the UV-IR map and is further discussed in sect.~\ref{sec:perturbations}.} characterized by
\be\label{window}
 \textrm{geometric window: }\  \dim_{\mathbb{Z}[Q]} \cx A_{(M,N)}^{\hat \kappa} =\binom{N}{M}\ .
\ee
A sufficient condition --- but certainly not a necessary condition --- to be in the geometric window is
\begin{equation} 
   0 \le \alpha < N \ , \quad -1 \le \gamma \le \alpha \ .
\end{equation}
For freely-generated algebras outside the window, one has $\dim_{\mathbb{Z}[Q]} \cx A_{(M,N)}^{\hat \kappa}>\binom{N}{M}$, and the extra generators relate to non-geometric vacua of the underlying 3d gauge theory. An obvious class of such examples arises from CS levels with $\alpha<0$ and, for instance, with $\gamma=0$. Then the Wilson line algebra is freely generated by $\binom{N+|\alpha|}{M}$ elements, which are, for example, given by the Schur polynomials $\sigma_\lambda(\delta_a)$ with partitions $\lambda =( \lambda_1,\ldots,\lambda_M)$ in the range $N+|\alpha| \ge \lambda_1 \ge \ldots \lambda_M \ge 0$. Two explicit examples of this kind are detailed in app.~\ref{app:ExNotInWind}.

\section{Gauge fields with Dirichlet boundary conditions\label{sec:halfindex}}
In this section we consider the $\cx N=2$ 3d gauge theory with Dirichlet b.c.~for the gauge fields and find that the index for Dirichlet b.c.~gives a natural completion of the $q$-series representing the K-theoretic $I$-function to a {\it bilateral} $q$-hypergeometric series. Dirichlet b.c.~were first studied in ref.~\cite{DGP} in the context of the supersymmetric half-index \cite{BDP,Gadde:2013wq,Gadde:2013sca,GPPV,DGP,CDG}, and subsequently in the framework of the closely related 3d partition function in ref.~\cite{Bul20}. There are two new aspects for Dirichlet b.c.~Firstly, the gauge symmetry in the bulk reduces to a global $G$ symmetry on the boundary, and therefore can have a non-zero 2d anomaly, as long as the spectrum satisfies the weaker condition of anomaly freedom in the 3d bulk. A non-zero boundary anomaly allows us to study the case of non-zero effective CS levels without factorization. Secondly, the half-index involves a non-perturbative sum over boundary monopole sectors \cite{DGP}. 

\subsection{Bilaterial hypergeometric $q$-series from the half-index}
The half-index is defined as the supersymmetric index on $\IR_t \times \IR^2$ 
\begin{equation}\label{formalhalfindex}
	\II_{\mathcal{B}}=\tr_{\operatorname{Ops}_{\mathcal{B}}}(-1)^{F} q^{J+\tfrac R 2} y^f\,,
\end{equation}
Here $F$ is the fermion number operator, $J$ is the generator of two-dimensional rotations in spatial plane $\IR^2$, $R$ is the $U(1)_R$-charge, and $y$ and $f$ are general fugacities and charges for the global symmetries respectively. The index depends on the choice of a 2d boundary theory $\cx B$ defined by the b.c.~for the 3d fields on $\IR^2$, and their coupling to 2d fields. It counts the boundary operators in the cohomology of the supercharge operator preserved by the boundary. 

For Neumann b.c.~for the gauge fields, the computation of $\II_{\cal B}$ involves an integration over the Coulomb branch moduli  in the Cartan of the gauge group $G$, to project to gauge-invariant observables, much like in the computation of the partition function. For Dirichlet b.c.~for the gauge fields, the half-index involves a non-perturbative sum over boundary monopole sectors.

As in the previous section, for the Grassmannian $\Gr(M,N)$, we consider the $G=U(M)$ gauge theory with $N$ chiral multiplets with Neumann b.c.~(and no extra 2d fields to begin with). For the theory with generic real masses for the chirals, where the vacua are all massive, the half-index of this theory can be computed as in ref.~\cite{DGP}.\footnote{Half-indices arising form symplectic and orthonormal gauge groups for gauge fields with both Neumann and Dirichlet b.c. have been studied in the context of Seiberg-like dualities in refs.~\cite{Okazaki:2021pnc}.} The geometry of the Grassmannian $\Gr(M,N)$ arises instead for a slice of the parameter space, where the theory has a degenerate vacuum with massless fields. We first consider the mass deformed theory and then discuss the modifications needed to obtain the massless case. 

The half-index of the mass deformed theory with boundary conditions $\cx B=(\cx D,N)$, i.e., Dirichlet for the gauge fields and Neumann for the matter fields, takes the form
\begin{equation}\label{halfindexdirichlet}
	\II_{(\cx D, N)}
	(\z,q,\taun)
	=\frac{1}{(q;q)_\infty^M}
	\sum_{\vec m\in \IZ^M} 
	\frac
	{q^{\tfrac 1 2 \vec m \cdot K\vec m
		+\vec m \cdot K\vec \taun
		-\tk_R \sum_{a=1}^M m_a}
		\z^{-\sum_{a=1}^M m_a}}
	{\prod_{a\neq b}^M (q^{1+\taun_{ab}+m_{ab}};q)_\infty\, \prod_{i=1}^N\prod_{a=1}^M(y_iq^{m_a+\taun_a};q)_\infty}\ ,
\end{equation}  
where $\vec m\in \operatorname{cochar}\big(U(M)\big)=\IZ^M$ counts the monopole sectors, $y_a=q^{\taun_a}$ with $a=1,\ldots, M$ are the fugacities for the Abelian subgroup $U(1)^M$ of the boundary  flavor symmetry $U(M)_\partial$, $\z$ is a fugacity for the topological $U(1)$ symmetry, corresponding to a non-zero the FI term, and $m_{ab}=m_a-m_b$, $\taun_{ab}=\taun_a-\taun_b$. 
The parameters $y_i$ are fugacities for the global $U(N)$ flavor symmetry of the $N$-tuple of chiral multiplets, comprising real masses for these chiral fields. As this symmetry enters the construction of the Grassmannian $\Gr(M,N)\simeq U(N)/(U(M)\times U(N-M))$ as a homogeneous space, we mostly set $y_i=1$ in the following to keep this flavor symmetry unbroken.
The CS levels appearing in eq.~\eqref{halfindexdirichlet} are again {\it effective}  levels \cite{DGP,CDG}. We capture these by the $M\times M$-matrix $K$ with entries
\begin{equation}\label{CSlevelsdirichlet}
   K_{ab}=\tk_S \delta_{ab}+\frac{\tk_A-\tk_S}{M}\ ,
\end{equation}
where $\tk_S$ is the $SU(M)$-level, $\tk_A$ is the Abelian $U(1)$-level and $\tk_R$ the level for the mixed gauge/R-symmetry CS term, similarly as in eq.~\eqref{effCS1}. 
The first factor in the denominator in eq.~\eqref{halfindexdirichlet} accounts for the contribution of the vector multiplet and the second for the contributions from $N$ chiral fields with Neumann b.c.. 

\subsubsection{Difference equations for the half-index}
The half-index \eqref{halfindexdirichlet} with Dirichlet b.c. for the gauge fields contains more information about the 3d gauge theory than the $I$-function \eqref{ISQKGr}. We first characterize the relation between the two in terms of the difference equations, which they satisfy:  Whereas the $I$-function for the target $X$ takes the form of a $q$-hypergeometric power series in $Q$ and satisfies a system of inhomogeneous difference equations, the half-index for Dirichlet b.c.~satisfies the associated homogenous system. Mathematically, the two cases are related to  $q$-hypergeometric series and  {\it bilateral}  $q$-hypergeometric series, respectively.\footnote{See e.g.,~ref.~\cite{Rahman} for background material on $q$-series.} 

To characterize the relation between the two series  in terms of a difference equation, we consider as in eq.~\eqref{abel} an Abelianzed version $\hat \II_{\cx B}(\z_a)$ of the half-index, obtained from the index $\II_{\cx B}$\footnote{To simplify notation, we keep the general subscript $\cx B$ to remind of the dependence of the index on the boundary conditions. Explicit formulas for the index below refer to the case $\cx B=(\cx D,N)$ studied in this section, but the statements about the split of the bilateral series and the difference operators hold more generally.} by sending $\z^{\sum_am_a}\to\prod_a ((-1)^N\z_a)^{m_a-\taun_a}$. For simplicity we set $y_i=1$ in the following. One can  verify, that the  series $\hat \II_{\cx B}$ so defined satisfies the system of difference equations \eqref{Deq1},\eqref{Deq1b} with the following modifications: Firstly, the CS parameters $\hat \kappa_i$ in \eqref{Deq1}  and $\tk_i$ in \eqref{halfindexdirichlet} must be related by
\be\label{kappamatch}
(\hat \kappa_S,\hat \kappa_A,\hat \kappa_R)  = (\tk_S+N,\tk_A+N,\tk_R+\tfrac N2)\simeq-(\tk_S+2M,\tk_A,\tk_R)\,.
\ee
The second equation is an equivalence induced by the reflection $q\to \bar q=q^{-1}$ in the difference operators \eqref{Deq1b}.\footnote{The inversion formula for the relation between the $I$-functions with fugacity $q$ and $\bar q$ is given in eq.~\eqref{inversion}.} The last expression shows, that modulo conventions, the half-index solves analogous difference equations as the Abelianized $I$-function $\hat I (Q_a)$, up to an extra shift of $2M$ in the $SU(M)$ level. This shift is consistent with the different one-loop contributions of the vector multiplets with Neumann and Dirichlet b.c.~as computed in ref.~\cite{DGP}. \def\ps{+}

There is one more essential modification. Namely, the index $\hat \II_{\cx B}(\z_a)$ solves the homogeneous system of difference equations  of eq.~\eqref{Deq1}  {\it without} the inhomogeneous term $g(\eps)$. If we split the sum in eq.~\eqref{halfindexdirichlet} over $\vec m \in \IZ$ into a power series $\hat \II^{\ps}_{\cx B}(\z_a)$ summing non-positive $\vec m\leq 0$,  and the remainder $\hat \II^{\rm rem}_{\cx B}(\z_a^\pm)$, a formal power series in $Q_a$ and $Q_a^{-1}$, i.e., 
\be\label{split}
\hat \II_{\cx B}(\z_a)=\hat \II^{\ps}_{\cx B}(\z_a)+\hat \II^{\rm rem}_{\cx B}(\z_a^\pm)\,,
\ee 
it is straightforward to show that the objects defined above respectively satisfy the difference equations with and without inhomogeneous terms
\be\label{splitdeq}
\DL_a \hat \II_{\cx B}=0\,, \qquad \DL_a \hat \II^{\ps}_{\cx B}(\z_a)=g(\taun_a)\,, \qquad \DL_a \hat \II^{\rm rem}_{\cx B}(\z_a^\pm)=-g(\taun_a)\ , 
\ee
for $a=1,\hdots,M$ and with the CS parameters specified in eq.~\eqref{kappamatch}.  It follows that the half series $\II^{\ps}_{\cx B}(\z_a)$ is equal to the $I$-function, up to a proportionality factor which is invariant under the shifts $Q_a\to qQ_a$ generated by the operators $p_a$. 

The proportionality factor can be obtained by an explicit summation performed in app.~\ref{app:indexcomputation}, where we show that the sum over the monopole sectors can be rewritten as
\begin{equation}\label{indexresult}
	\begin{aligned}
		\II_{\cx B}(\z,q,\taun)=&
		\frac{((-1)^{(N+M+1)}\z)^{\sum_a\taun_a}}{(1-q)}\cdot
                          \frac{1}
                          {(q;q)_\infty^M \prod_{a\neq  b}(q^{1+\taun_{ab}};q)_\infty}\cdot
		\frac{1}{\prod_{a=1}^M(q^{\taun_a};q)_\infty^N}
		\\[1mm]&\cdot\bigg(
		I^{(\tk_S+N,\tk_A+N,\tk_R+\tfrac N 2)}_{\text{Gr}(M,N)} \big((-1)^{M+N}\z,q,\taun)
		+O((1-q^{\taun})^N)\bigg)
		\ .
	\end{aligned}
\end{equation} A similar formula applies for the Abelianized version. Note that the remainder series $\II^{\rm rem}_{\cx B}(\z_a^\pm)$ only contributes to the terms of order $(1-q^{\taun})^N$. 

The $q$-hypergeometric series $I$ given in eq.~\eqref{ISQKGr} converges for small $|Q|$ and has a simple limit $q\to 1$, where it reduces to the generalized hypergeometric series relevant for the  cohomological Gromov--Witten theory \cite{JM}. Instead, the half-index  computes a certain bilateral $q$-hypergeometric series associated to the $q$-hypergeometric series $I$. Due to the dependence on $Q$ and $Q^{-1}$, the bilateral $q$-series converges only for, say, $|q|<1$ and appropriate choice of CS parameters (or $|q|>1$ and inverted CS levels). In particular it does not have a well defined classical limit $Q\to 0$, and also no 2d limit $q\to1$ in general.

As a simple working example,  consider the case $X=\Gr(1,N)=\IP^{N-1}$, i.e., the 3d $U(1)$ gauge theory with $N$ electrons. 
The half-index for general CS terms  is 
\def\x{Q}\be\label{Ipn}
\II_{\mathcal{B}}=\frac{q^{-\tfrac A 2 \eps^2-B\eps}\x^{\eps}}{(q;q)_\infty}\sum_d \frac{\x^{-d-\eps} q^{\tfrac A2 (d+\eps)^2+B(d+\eps)}}{(q^{d+\eps};q)_\infty^N}=:
\frac{q^{-\tfrac A 2 \eps^2-B\eps}\x^{\eps}}{(q;q)_\infty (q^\eps;q)_\infty^{N}} \cdot I (Q,q,\eps)\ .
\ee
The  bilateral $q$-series splits as $I=I^++I^-$ with 
\be
\begin{aligned}
I^-(Q,q,\eps)&= (1-q^\eps)^N\sum_{d>0} \x^{-\hat d}q^{\tfrac A 2 \hat d^2+B\hat d}\prod_{r=1}^{d-1}(1-q^{r+\eps})^N\ , \quad & \hat d &= d+\eps\ ,\\
I^+(Q,q,\eps)&= \sum_{d\geq 0} \frac {\x^{\td}\bar q^{-\tfrac A 2 \td^2+B\td}} {\prod_{r=1}^{d}(1-\bar q^{r-\eps})^N}\ , & \td &= d-\eps\ .
\end{aligned}
\ee
Here $\bar q = q^{-1}$, and $A$ and $B$ parametrize the CS levels. The bilateral $q$-series converges for $A>0$ if $|q|<1$ and for $A<-N$ if $|q|>1$.  The series satisfy the difference equations 
\be \label{deqpn}\DL I = 0\,,\qquad \DL I^\pm  = \pm g(\eps)= \pm \x^{-\eps} q^{\tfrac A2 \eps^2+B\eps}(1-q^\eps)^N\,,
\ee
with
\be
\DL  = (1-\bx p)^N-\bx p^{-A}q^{\tfrac A 2 -B} \,,\qquad \bx pQ = \bar q Q \bx p\,.
\ee
One finds that the $\II_{\cx B}$ also satisfies a difference equation
\be\textstyle
(p^A(1-yp)^N-p_y)\, \II_{\mathcal B} =0\,,\qquad p_yy = q y p_y\,, 
\ee 
which expresses the equivalence of a certain combination of line operators in the topological $U(1)$ and the global $U(1)_\p$.

The bilateral $q$-series $I$ provides infinitely many solutions $I_k$ (linearly independent over $\IC$) to the homogeneous difference equation \eqref{deqpn}, defined by  the expansion 
\be\textstyle\label{abexp}
I=\sum_{k\geq 0}I_k \, \chi_k(y)\,,\qquad \chi_k(y)=(1-q^\eps)^k\,, \quad y=q^\eps\,.
\ee
Since $I^-$ is of order $(1-q^\eps)^N$, the first $N$ solutions $I_0,\hdots,I_{N-1}$ are power series solutions in $Q$. These are the $N=\dim(K(\IP^{N-1}))$ solutions associated with the quantum K-theoretic $I$-function for $\Gr(1,M)$. Imposing the constraint
\be\label{gto0}
g(\eps)=0=(1-q^\eps)^N\,,
\ee
which represents the ideal in the K-theory ring of $\IP^{N-1}$ using eq.~\eqref{epstoH}, these $N$ power series are the only solutions. They have a 2d limit $q\to 1$ where they reduce to the well-known $\dim K(X)$ solutions of the generalized GKZ differential system appearing in the context of  2d mirror symmetry.
The infinitely many extra solutions for $k>N-1$ are bilateral $q$-series and thus have no immediate 2d counterpart, as they do not converge in the limit $q\to 1$.\footnote{The monodromies associated to the solution spaces of the differential equations  in 2d and difference equations 3d are quite different.}

In physics terms, the extra solutions reflect the freedom to couple a given 3d theory to 2d degrees of freedom on the boundary, as indicated by the fact that the factors $\chi_k(y)$ can be interpreted as characters of the boundary symmetry. The physics contents of the bilateral $q$-series will be discussed next.

\subsection{Vacuum condition and the $I$-function}
The homogeneous difference equations represent Ward-identities in the full quantum field theory. Choosing a vacuum puts conditions on the fields at infinity of the spatial plane\footnote{Or on the boundary $T^2$ of the partition function on $\DSq$.} $\IR^2$, that restricts the index to operators compatible with the vacuum.\footnote{See ref.~\cite{Verma} for a detailed discussion of this effect in the $\cx N=4$ supersymmetric theory.} The reduction from the bilateral $q$-series to the $I$-function, which is a normal power series, arises from a choice of vacuum.

The need to impose the condition \eqref{gto0} in the Higgs phase of the physical theory can be understood as follows. The constraint \eqref{gto0} cuts out a particular Higgs vacuum of the theory with target $X=\IP^{N-1}$, and $I^+$ is a semi-classical expansion around the  classical limit $Q\to 0$ of this vacuum.  Indeed $I^+$ is the K-theoretic $I$-function for $\IP^{N-1}$ with non-zero levels, and $Q^n$ is the weight of the sector with vortex number $n\geq 0$ on the Higgs branch. Moreover, the weight $y=q^{\taun}$ represents an equivariant parameter for the global symmetry $U(1)_\p$ on the boundary with Dirichlet b.c.. To be compatible with the Higgs phase, where the chiral fields get vev's and break the symmetry, one needs to consider the limit $q^\taun\to 1$ of the index. Introducing separate fugacities $y_i$, $i=1,\hdots N$, for the $N$ chiral fields as in eq.\eqref{halfindexdirichlet}, the constraint becomes $\prod_{i=1}^N(1-y_i)=0$, which is the ideal in the $T^N$-equivariant cohomology of $\IP^{N-1}$, with the $N$ zeros representing $N$ massive  point-like vacua of the gauge theory. 

For coinciding vacua, there are $N-1$ bosonic zero-modes, and their  fermionic superpartners generate the cohomology of the target $X$. More generally, for $M>1$, the expansion \eqref{abexp} of the  index $\II_{\cx B}$ can be written as\footnote{See app.~\ref{app:SymGrass} for definitions of various symmetric polynomials.}  
\be
\II_{\cx B}=\sum_\mu I_\mu \chi_\mu(y)\,, \qquad \chi_\mu(y)=\sigma_\mu(1-y_a)\,,
\ee
where the sum runs over the Young-tableaux $\mu$ and $\sigma_\mu$ are the Schur polynomials. In a Higgs phase with vacuum $\Gr(M,N)$, the vanishing of the inhomogeneous term implies that 
\be
g(\eps)=0\ \leadsto \chi_\mu(y)\equiv 0 \textrm{ if } \mu\notin \cx B_{(M,N)}\,,
\ee
where $\cx B_{(M,N)}$ is the $M\times (N-M)$ box defined by the largest partition $\mu_\text{top}=(\mu_1,\hdots,\mu_{M})$ with $\mu_i=N-M\ \forall i$. This restricts the solutions of the difference equations \eqref{Deq1} to $N\choose M$ power series solutions, which are the coefficients of the $I$-function for $\Gr(M,N)$. Here we are assuming that the CS terms are in the geometric window \eqref{window}, which is necessary for the theory to have exactly $\Gr(M,N)$ as the vacuum space.


\section{3d (mock) modularity and Wilson lines}
The 3d half-index \eqref{formalhalfindex} counts boundary operators of the 3d $\cx N=2$ supersymmetric theory on $\DSq$ with boundary $\p(\DSq)\simeq T^2$. The modular group $SL(2,\IZ)$ acts naturally on the boundary $T^2$ with complex structure $\tau$. The contribution of a 2d boundary theory on $T^2$ to the half-index is the flavored elliptic genus \cite{Schellekens:1986xh,Witgen,Gadde:2013wq,BeniniEG}. The action of the $SL(2,\IZ)$ on 3d bulk modes is far less obvious. A concept of 3d modularity for the half-index has been proposed in refs.~\cite{CG18,GPPV}, where an interesting class of theories $T[M_3]$ is studied, whose half-indices give  new homological invariants of a three-manifold $M_3$.\footnote{See also refs.~\cite{GPV,MTCi,MTCii} for an intriguing relation between twisted indices of 3d theories and modular tensor categories.}

In this section, we obtain another large class of half-indices with generalized 3d modular behavior from the 3d $\cx N=2$ UV gauge theories considered in this note.\footnote{In the UV, these are generally not related to the class of theories studied in refs.~\cite{CG18,GPPV} in an obvious way, but might be connected by RG flow. In particular some indices obtained below coincide with invariants of particular three manifolds in a certain pattern, suggesting such a relation.} To this end we consider a  limit of the 3d gauge theory, where the fugacities for the global symmetries are of order one and operators of different global symmetry charges contribute at equal weight to the index. More precisely the limit is defined by keeping only the phases of the global fugacities, corresponding to Wilson line backgrounds, except for the fugacity $q$ for the (twisted) spin. This corresponds to setting the dimensionful parameters, such as the real masses $m$ and the FI parameters $\zeta$, to zero. From this limit, to which we refer loosely as the massless limit $\mu\to 0$ in the following, we obtain several classes of half-indices with interesting modular behaviors, related to CFT characters and (mock) modular functions:
\begin{itemize}
\item The massless limit of $U(1)^M$ Chern-Simons matter theories gives $q$-hypergeometric series of the form
\be
\chi_{A,B,C}(q)=\sum_{\mb \in \mathbb{N}^M}\frac{q^{\frac 12 \mb\cdot A  \mb +B \mb+C}}{(q)_{\ell_1}\hdots (q)_{\ell_M}}\,,\qquad  (q)_m=(q;q)_m\,,
\ee
where $A,B,C$ capture non-zero CS levels. This type of sums is known to capture modular characters of rational conformal field theories (CFT) for a judicious choice of the ``modular triple'' $(A,B,C)$ \cite{Nahm,Zagier2007}, with $q=e^{2\pi i \tau}$. 

\item The above type characters includes the non-unitary series of minimal models. The 3d parents can be deformed by adding $\IZ_2$ Wilson lines of global symmetries for the two periods of the boundary $T^2$. The massless limit of the half-indices with Wilson lines are {\it linear combinations of} mock modular thetas, which can be either genuinely mock modular, or modular, by relations between the mock theta functions known as the mock theta conjectures. In particular modular characters of certain $(2,K+2)$ minimal models with discrete symmetry $\IZ_{K+2}$ get related to mock theta functions of order $K+2$ by adding discrete Wilson lines in the 3d theory.

\item More generally, we find that half-indices in $U(1)$ Chern-Simons matter theories with appropriate Wilson line insertions give rise to Appell--Lerch sums 
\be\label{DefAL}\textstyle
A_k(y,z,q)=y^{\frac k2}\sum_{n\in\IZ}\frac{(-)^{kn}q^{\frac k 2 n^2+\frac k 2 n}z^n}{1-yq^n}\, ,
\ee
where the sum runs over the monopole sectors and $z$ and $y$ are 3d Wilson lines in the global symmetry group. These functions are central to the theory of mock modular functions: if $y$ and $z$ are specialized to torsion points $\exp(\IQ+\tau\IQ)$, a normalized version of $A_k$ transforms as a mock modular form and has a non-holomorphic completion to a modular form \cite{Zwegers}. 

\item Deformations of pure $U(1)$ Chern--Simons theory with non-zero level that flow to finite deformations of $U(1)$ WZW model in the IR.

\end{itemize}
The discussed examples share some general features concerning the characters obtained from the massless limit of the half-index:\\[2mm]

\vbox{\leftskip3mm \noindent
- Characters of different representations are related by insertions of Wilson line operators in the 3d path integral or half-index;\\[1mm]
- The 3d Wilson line algebra reduces in the massless limit to the Verlinde algebra;\\[1mm]
- The Wilson line algebra of modular and mock modular characters related by discrete Wilson lines are equivalent as algebras.\\[1mm]
}

\noindent Similar observations have been made earlier in related contexts, as discussed below.\footnote{We would like to thank Sergei Gukov for sharing unpublished work with us \cite{Gukov:unpub}, in which modular properties of the half-index for certain 3d $\mathcal{N}=2$ theories are discussed as well.}

\subsection{Modular triples and $\IZ_2$ Wilson lines \label{sec:triple}}
For the first class of examples we consider the 3d $\cx N=2$ supersymmetric Abelian $U(1)^M$ theory with $N$ fundamentals in each factor. As discussed before, in the massless limit $\eps=0$ the bilateral $q$-series $\IIB$, representing the supersymmetric index for boundary operators, reduces to  
\be\label{DefIt}
\It_{A,B,C}(Q_a,q,\eps_a)=\sum_{\mb \in \mathbb{N}^M}Q^{\tx \mb}\, \frac{q^{\frac 12 \tx \mb\cdot A \tx \mb +B\tx \mb+C}}{((q^{1-\eps_1})_{\ell_1}\hdots (q^{1-\eps_M})_{\ell_M})^N}\,,
\ee
where again $y_a=q^{\eps_a}$ and $Q_a$ stand for the fugacities of the global boundary and topological symmetries, respectively, and $\tx \ell_a=\ell_a-\eps_a$. If the CS levels captured by the $M\times M$ matrix $A$, the $M$-vector $B$ and the constant $C$ are restricted to CS terms of the non-Abelian theory as in eq.~(3.3) of ref.~\cite{JMNT}, the Abelianized $I$-function \eqref{ISQKGr} for the $U(M)$ theory can be written in the form 
\eqref{NAtoA}. In the following, we allow for more general CS terms, corresponding to the generalized $I$-functions of a genuine $U(1)^M$ theory with general CS levels, as considered, e.g., in ref.~\cite{JM}. 

For non-zero CS terms the monopole operators get electrically charged, i.e., they become dyonic. The function $\It_{A,B,C}$, and more generally the bilateral series $\IIB$ associated with it, capture the contribution from the electrically charged fundamental matter and magnetically charged dyon operators. Let $\qc^a=(\qc^a_{e};\qc^a_{m})$, $a=1,\hdots,M$ denote the electric/magnetic charges of a state  in the $a$-th $U(1)$ factor. The function $\It $ in \eqref{DefIt} describes the 3d $U(1)^M$ gauge theory with basic operators of charge
\be
\begin{tabular}{lll}
chiral matter &$\Phi^{(j)}_i$:&$\qc^a_i=(\delta^a_i;0)\,,\qquad  1\leq i \leq M, 1\leq j\leq N\,,$\\
monopoles &$M_k$ :&$\qc^a_k=(A_{ak};\delta_k^a)\,,\quad 1\leq k \leq M\,,$ \\
\end{tabular}
\ee
and bound-states of those.
With an appropriate choice of gauge group $U(1)^M$ and CS levels one can engineer a large set of BPS operators of mutually non-local charges, which contribute to the index in the massless limit with equal weight irrespectively of their global symmetry charges, in particular for the topological $U(1)$ and the global $G_\partial$ symmetry at the boundary associated with the bulk gauge symmetry. The massless limit is abbreviated in the following as $ \ml \to 0$.

Since the half-index $\IIB$ is defined as a formal series in $Q$, the limit $Q\to 1$ can be taken only for positive effective CS terms at $|q|<1$ (or negative for $|q|>1$), such that the series converges due to the factors $q^{\frac 12 \mb A \mb}$ with $\ell$ the flux/monopole number. In this case one may send $Q_a\to 1$ already in the bilateral series $\IIB$. For $N=1$, the massless limit of the sum $\It_{A,B,C}$ takes the form
\be\label{2dsum}
\chi_{A,B,C}(q):= \lim_{\ml \to 0}\It_{A,B,C}(Q_a,q)=\sum_{\mb \in \mathbb{N}^M}\frac{q^{\frac 12 \mb\cdot A  \mb +B \mb+C}}{(q)_{\ell_1}\hdots (q)_{\ell_M}}\,.
\ee
This type of  $q$-series, to which we refer to as a  fermionic sum,\footnote{See ref.~\cite{Feigin} for comments on the combinatorical  interpretation.} has been well-studied in connection with the  characters of 2d rational CFTs,  in particular in refs.~\cite{Nahm,Zagier2007} and refs.~\cite{CNV,CJVY}. For a judicious choice of the data $A,B,C$, the so-called modular triples, $\chi_{A,B,C}(q)$ is a modular character. For these values, the half-index of the 3d gauge theory with CS levels determined by a particular modular triple reduces to the modular character of a rational 2d CFT in the massless limit.

In the CFT context, the matrix $A$ is characteristic for the CFT, while different vectors $B$ at fixed $A$ determine the representation for the character. In the 3d gauge theory, $A$ captures the quadratic CS levels in the gauge group factors and $B$ the mixed gauge/R-symmetry levels. Similarly as in the CFT, the value of $B$ can be changed by inserting  appropriate operators in the 3d partition function.\footnote{This is parallel to the situation found in ref.~\cite{CNV}, where the CFT characters for different representations appear from traces of the so-called monodromy operator $M(q)$ with additional  insertions of line operators.} As reviewed briefly in sect.~\ref{sec:recdiff} below, the  insertion of a Wilson line of $U(1)^M$ charge $(q)_a = -\delta_{ba}$  in the 3d path integral can be represented by the action of the difference operator $p_b$ on the 3d vortex sum $\It$. Inserting more general Wilson lines one obtains in the massless limit the characters for different representations
\be\textstyle
\chi_{A,B_a+n_a,C}(q)={\displaystyle \lim_{\ml\to0}}\left( \prod_a p_a^{n_a}\It_{A,B,C}(Q_a,q)\, \right)\,,\qquad 
n_a\in \IZ^M\,,
\ee
with a shifted value of $B$.
In the framework of the half-index the action of $p_a$  corresponds similarly to the insertion of a  vortex line in the topological $U(1)_a$ \cite{DGP}.\footnote{Since the vortex sum of the non-Abelian theory is obtained by \eqref{NAtoA} to the Abelian sum $\It$, the non-Abelian case is covered by taking linear combinations of Weyl invariant characters of the Abelian theory.}

A concise way to capture the BPS data of the 3d gauge theory are the Ward identities for Wilson line operators of the theory, which descend to the difference operators that annihilate the vortex sum. For the generalized $I$-function  $\It_{A,B,C}$, these take the form \cite{JM}
\be
(1-p_a)^N\It_{A,B,C}=Q_a q^{\frac 12 A_{aa} +B_a}\prod_b p_b^{A_{ba}} \It_{A,B,C} +O(1-q^{-\eps_a})\,.
\ee
As in sect.~\ref{sec:PQRings}, the ideal in the quantum K-theory algebra does not depend on $q$ and can be obtained in the  limit $q\to 1$ of commuting variables $p_a,Q_a$
\be\label{kta}
(1-p_a)^N= Q_a \prod_b p_b^{A_{ba}}\,.
\ee
In a theory with trivial UV-IR map, these are precisely the equations of the thermodynamic Bethe ansatz (TBA), playing the central role in the celebrated Bethe/gauge correspondence of ref.~\cite{NekSha2}. Moreover, taking further the massless limit $|Q_a|\to 1 $, eq.~\eqref{kta} reduces to the TBA equations of the 2d CFT studied in refs.~\cite{Nahm,NahmReck,Zagier2007}. 

As discussed in these papers, the Verlinde algebra of the CFT can be reconstructed from the (CFT limit $|Q|=1$ of the) TBA equations. It follows that the quantum K-theory algebra for the 3d gauge theory with CS levels $(A,B,C)$ becomes isomorphic to the Verlinde algebra of the 2d CFT with data $(A,B,C)$ in the $|Q_a|= 1$  limit.

A large class of modular functions arises from the fermionic sum \eqref{2dsum} with  a matrix $A$ of the form 
$A=C(G)\otimes C^{-1}(G')\,,
$
where $C(G)$ is the Cartan matrix for the group $G$ of ADET type \cite{Nahm,Zagier2007,NahmReck}. The associated $q$-series have been related to superconformal fixed points of 4d $\cx N=2$ quiver theories in refs.~\cite{CNV,CJVY}, and linked to quivers for BPS states of 2d theories classified in ref.~\cite{CV92}. In the present context, we expect the 3d CS matter theories with CS levels determined by the same matrix $A$ and vortex sum \eqref{DefIt} to flow to non-trivial conformal fixed points in the massless limit.

Below we restrict to discuss some interesting examples leading to modular forms, mock modular forms and related bilateral series. A more comprehensive analysis is beyond the scope of this note and we hope to return to these issues in the future. Some of the examples for mock modular functions discussed below have been previously related to K-theoretic $I$-functions in ref.~\cite{RZ18}, and all their examples fall into the general class of the massless limits of 3d gauge theories as discussed in this paper.

\subsubsection{Example: Pure $U(1)$ Chern-Simons theory and WZW models:}
The simplest case is a pure $U(1)$ CS theory with effective level $k$ discussed in ref.~\cite{DGP}. The half-index is\footnote{See app.~\ref{app:sf} for the standard definition of Jacobi theta functions used in this section.} 
\be\label{pureCS}
\IIB = \frac1{(q)_\infty}\sum_{m\in\IZ}\x^my^{km}q^{\frac k2 m^2} = \frac{\theta_3(\x y^k,q^k)}{(q)_\infty}\,.
\ee
The r.h.s.~is the vacuum character of the $U(1)_k$ WZW model, with the numerator coming from the monopole operators $\sim e^{im\phi}$ and the denominator from the $U(1)$ Kac-Moody current $i\p \phi$. The conformal point corresponds to taking the global fugacities to one, i.e., $\x \to 1$ and $y \to 1$. Generalizations to non-Abelian groups have been also studied in ref.~\cite{DGP}.\\

\subsubsection{Example: $U(1)$ CS matter theory with $A=1$}
Consider the level $k=0$ with one 3d chiral of charge one and Neumann b.c. On the one hand, for $|\x|<1$
\be
\II_{(\cx D, N)} =  \frac1{(q)_\infty}\sum_{m\in\IZ}\frac{\x^m}{(yq^m)_\infty} =  \frac1{(q)_\infty (y)_\infty}\sum_{m\in\IZ}\x^m(y)_m 
= \frac{\theta(y\x,q)}{(\x)_\infty\theta(y,q)}\ ,
\ee 
with $\theta(z,q)=(z)_\infty(q/z)_\infty$, using Ramanujan's $_1\psi_1$ summation formula in the last step. This agrees with eq.~(4.18) of ref.~\cite{DGP}. On the other hand
\be\label{ex01}
\II_{(\cx D, N)}  = \frac1{ (q)_\infty (y)_\infty}\sum_{m\in\IZ}\x^{-m}(y)_{-m} = \frac1{(y)_\infty(q)_\infty}\times \sum_{m\in\IZ}\frac{q^{\frac {m^2}2}(-q^{-\frac 12}y\x)^{-m}}{ (y^{-1}q)_m}\ ,
\ee
where the last expression is written in the appropriate factorization for the Higgs vacuum. For $y=1$ the sum over $m$ collapses to
\be\label{freefermion}
\It_{1,0,0}(-q^{\frac 12}\x,q,0)=\sum_{m\geq 0}\frac{(-)^m q^{\frac {m^2-m}2}(q\x^{-1})^m}{(q)_m}=(q\x^{-1} )_\infty\ {\buildrel \x= -q^{\frac12} \over \longrightarrow} \ \frac{q^{\frac 1 {48}}\eta(q)^2}{\eta(q^2)\eta(q^{\frac 12})}\ .
\ee
The r.h.s.~is the case $(k=)A=1$ case of the sum $\chi_{1,0,0}(q)$ in \eqref{2dsum},  the free NS fermion \cite{Nahm,Zagier2007}. This suggests that the same theory can be obtained in a simpler way by starting with Dirichlet boundary conditions for the matter field, which is the b.c.~where the fermionic modes survive at the boundary \cite{DGP}. For $U(1)_\partial$ charge $-1$, R-charge $0$ and effective level $k$, one has
\be\label{DirInd}
\II_{\cx D,D}(y,Q,q) = \frac 1 {(q)_\infty}\sum_{m\in\IZ} q^{\frac k2 m^2}\x^my^{km}(qyq^{m})_\infty =
\frac{(qy)_\infty}{(q)_\infty} \sum_{m\in\IZ} \frac{q^{\frac k 2 m^2}\x^my^{km}}{(qy)_m}\ .
\ee
For $k=1$ and $y=1=\x$ the r.h.s.~is precisely the free fermion $\chi_{1,0,0}(q)$.\\
 
\subsubsection{Example: $(2,5)$ minimal model and order 5 mock theta functions}
Increasing the CS level by one to $k=2$ the index \eqref{DirInd} takes the form 
\be\label{dia}
\II_{\cx D,D}(y,Q,q)   =  N(y,q)\, \sum_{n\in\IZ}\frac {q^{n^2}Q^{n}y^{2n}}{(yq)_n}\,,\qquad N(y,q)=\frac{(qy)_\infty}{(q)_\infty}\,.
\ee
The difference equation for $\IIB$ and the quantum K-theory ring obtained from it are\footnote{To avoid heavy notation we use the mnemonic $\IIB$ for the index and do not explicitly write the CS levels and arguments when the context is clear.}
\be\textstyle 
((1-yp)-Qqp^2y^2)\, \IIB=0\, , \qquad \IZ[p,Q,Q^{-1}]/\langle\!\langle  y^2p^2 - (1-yp)Q^{-1}\rangle\!\rangle\ ,
\ee
Compared to the $U(1)$ gauge theory \eqref{DirInd} with an effective CS~level $A=1$, which geometrically is the quotient $X=\IC/\IC^*$ with the point as its generic orbit, and giving rise to the class $1$ as its single K-theory element, 
the theory with the same spectrum but with non-zero CS level $A=2$ is instead generated by two independent Wilson line operators, say $1,p$. The extra generator reflects the existence of an additional topological vacuum of the 3d gauge theory~\cite{IS}.\\ 

\sss{Trivial Wilson lines}The $q$-series obtained from the  limit $Q=1= y$ of the index with insertions of $1$ and $p^1$ are the Rogers--Ramanjuan functions\footnote{For a review of modular and mock modular functions including many more original references we refer to ref.~\cite{GMrev}.} 
\bea\label{rrf}
\chi_{2,0,0}&=&\lim_{\ml\to0}[\,p^0\IIB\,] = G(q)=\sum_{n\geq0}\frac {q^{n^2}}{(q)_n}\ \,=\frac{1}{\theta(q,q^5)}
\,,\nonumber\\
\chi_{2,1,0}&=&\lim_{\ml\to0}[\,p^1 \IIB\,]= H(q)=\sum_{n\geq 0 }\frac {q^{n^2+n}}{(q)_n}=\frac{1}{\theta(q^2,q^5)}\,.
\eea
which are, up to overall powers of $q$, the characters of the $(2,5)$ minimal model.\footnote{Below we usually suppress these overall factors $q^{h-\frac c {24}}$ with $c$ the central charge and $h$ the conformal dimension of a primary and use characters normalized to a leading term 1.} They transform as a doublet under general $SL(2,\IZ)$ transformations, and are invariant under the subgroup $\Gamma_0(5)$ up to scalar factors (see e.g. \cite{Zagier2007}). The $Q\to 1$ limit $p^2=1-p$ of the quantum K-theory algebra  is isomorphic to the Verlinde algebra $\Phi^2=1+\Phi$ of the minimal model generated by the primary $\Phi$. The same minimal model has been related by a 4d/2d correspondence to the 4d $(A_2,A_1)$ Argyres--Douglas point in refs.~\cite{CNV,Beem1,Cord}. The last equal signs in eqs.~\eqref{rrf} are due to the Rogers--Ramanjuan identities. Note that the product formulas on the r.h.s. have a simple 3d interpretation in terms of $\mathcal{N}=(0,2)$ boundary chirals. Their partition functions computed via localization using zeta function regularization \cite{Tanaka} reproduces the overall factors $q^{-1/60}$ and $q^{11/60}$ of the characters needed for modularity. This follows from the zeta function regularization respects the modularity of the toroidal boundary.
\\

\sss{$\IZ_2$ Wilson lines:  modular/mock modular:}
The discussed gauge theory can be deformed by switching on background Wilson lines in the global $U(1)_t\times U(1)_\p\times U(1)_R$ symmetry with fugacities $Q,y,q$, respectively. The associated Wilson line expectation values take values in their dual tori. Particularly interesting cases arise if we switch on Wilson line expectation values at $\IZ_2$ torsion points, while keeping the real masses associated to these global symmetries at zero. We often refer to such Wilson line expectation values simple as $\IZ_2$~Wilson lines. The key observation is  that 3d Wilson line backgrounds in $U(1)_\partial$ at the $\IZ_2$ torsion points $y=-1$ and $y=q^{\frac 12}$ relate the modular Rogers--Ramanjuan functions to classical mock theta functions of order 5, and these all share the same 3d quantum K-theory algebra.

We start by switching on a half-integral Wilson line in the global $U(1)_\p$ symmetry, corresponding to a fugacity $y=-1$ for the matter field. This gives the fermionic sum 
\be\label{It2}
\IIB(y=-1,Q=1,q)= N(-1,q)\sum_{n\in\IZ}\frac {q^{n^2}Q^{n}}{(-q)_n}\,.
\ee
The homogeneous difference equation and the quantum K-theory ring with two generators are
\be\textstyle
(1+p)\IIB =Qqp^2\IIB \,, \qquad \IZ[p,Q,Q^{-1}]/\langle\!\langle p^2 - (1+p)Q^{-1}\rangle\!\rangle\,.
\ee
At $y=-1$, the bilateral series $\IIB$ has both a positive and a negative part. The $Q\to 1$ limits with insertions are,  omitting the prefactor $N(-1,q)$ 
\be\label{bilmock1}
\chiz2_a:=\lim_{Q\to 1}[p^a\, \IIB] = \sum_{n\in\IZ}\frac {q^{n^2+an}}{(-q;q)_n}:=I_a^++I_a^-\,, \qquad a=0,1\,,
\ee
where $I^+_a$ ($I^-_a$) is the one-sided sum over $n\geq 0$ ($n<0$). The bilateral sums $\chiz2_a$ turn out to be (generalized) theta-functions by Watson's identities \cite{Watson2}:
\bea\label{chibil}
\chiz2_0&=&\A(G(q);-1,2)= \frac12\Big(\theta_4(q^2)G(q)+\theta_3(q^2)G(-q) \Big)+\frac32q^{\frac 34 }\theta_2(q^2)H(q^4)\,,\hskip1cm\\
\chiz2_1&=& \A(H(q);1,2)=-\frac12\Big(\theta_4(q^2)H(q)-\theta_3(q^2)H(-q) \Big)+\frac32q^{-\frac 14 }\theta_2(q^2)G(q^4)\,,\nonumber
\eea
where the two summands in each expression are either even or odd under $q\to-q$ and 
we defined the linear combinations 
\be
\A(f(q);\alpha,\beta)=\alpha\, \theta_4(q^2)f(q)+\beta\, \theta_3(q^2)f(-q)\,. 
\ee
It would be instructive to relate the indices \eqref{chibil} for the theory with $\IZ_2$ Wilson line to a concrete 2d CFT, possibly an orbifold of a non-unitary minimal model. An interesting property of the bilateral indices is that their one-sided parts $I^\pm_a$ in eq.~\eqref{bilmock1} turn out to be classical mock theta functions. It can be easily checked that
\be\textstyle
I_a^+=f_a(q)\,,\qquad I_a^- =  2 \psi_a(q)\,,
\ee
with the order 5 mock thetas \cite{GMrev}
\be
f_a(q)=\sum_{n\geq 0}\frac {q^{n^2+an}}{(-q;q)_n}\,, \qquad \psi_a(q)=\sum_{n\geq 0}q^{\tfrac12(n+1)(n+2(1-a))}(-q;q)_n\,.
\ee
The particular linear combinations in the indices $\chiz2_a=f_a(q)+2\psi_a(q)$ are such that  their shadows  cancel by the mock theta conjectures.\footnote{The shadow of the mock theta functions captures the failure to be a modular theta function; see ref.~\cite{Zagier2} for a nice discussion in the present context. The mock theta conjectures are stated in \cite{GMrev} and proven in ref.~\cite{HickersonProof}.} I.e., the failure of the positive part $I^+_a$ to be modular is related to an obstruction of separating it from its negative partner  $I^-_a$ while keeping modularity. The positive parts $I^+_a$ can also be interpreted as certain K-theoretic $I$-function for the point \cite{RZ18}.\\

\sss{$SL(2,\IZ)$ transforms and K-theory algebras:}
Differently to the case with zero Wilson line, the bilateral indices $\chiz2_a$ do not close under $SL(2,\IZ)$ transformations but transform into another pair obtained from the massless limit of the vortex sums
\be
\chiztf_a(q) := \lim_{Q\to1}\, \left[\sum_{n\in \IZ}Q^{n}\frac{q^{n^2+an}}{(q^{\frac 12 };q)_{n+a}}\right]=:I_a^+(q)+I_a^-(q)\ , \quad a=0,1 \ .
\ee
More precisely, the modular transformations connect the pair $\chiz2_a(q)$, $a=0,1$  in \eqref{bilmock1} with the pair of functions $\chiztf_a(q^2)$ \cite{GMTF}. The bilateral series $\chiztf_a$ obtained from $\chiz2_a$ by a modular transformation enjoy similar properties, namely the bilateral series are the modular theta functions \cite{Watson2}
\be
\chiztf_0( q^4)=\frac 12 \A(G(q);1,1)\,,\qquad 
\chiztf_1( q^4)= \frac {q^{-1}\!}{2} \, \A(H(q);-1,1)\,,
\ee
and the one-sided parts $I^\pm_a$ are related to the order 5 mock theta functions 
\be
F_a(q)=\sum_{n\geq 0} \frac{q^{2n(n+a)}}{(q;q^2)_{n+a}}\,,\qquad 
\phi_a(q)=\sum_{n\geq 0} q^{(n+a)^2}(-q;q^2)\,,
\ee
by
\be
I^+_a(q^2)=F_a(q)\,,\quad I_a^-(q^2)=(a-1)+(-q)^{-a}\phi_a(-q)\,.
\ee
From the point of the 3d gauge theory, the Rogers--Ramanjuan functions and the classical order 5 mock theta functions all arise as specializations of the isomorphic rank 2 Wilson line algebras shown in the table \ref{o5}.
\leftskip-0cm\begin{table}[h]\begin{center}\label{o5}
\hbox{\vbox{\offinterlineskip
\halign{\strut~~$#$~~~\hfil&~~$#$~~~\hfil&~~$#$~~~\hfil&~$#$~~~\hfil\cr
c_n&I^+_{Q=1}& \rm 3d\ difference\ operator &\cx L|_{Q=1=q}\cr
\noalign{\hrule height 1pt }
\z^n\frac{q^{n^2}}{(q)_n}&G(q)&
\cx L = (1-p)-\z p^2q& E_-\cr
\z^n\frac{q^{n^2}}{(-q)_n}&f_0(q)&
\cx L = (1 + p) - \z p^2q& E_+\cr
\z^{-n}q^{\frac 12 (n^2+3n+2)}(-q)_n&\psi_0(q)&
\cx L =(1 + pq^{-1}) - \z p^2q^{-1}& E_+\cr
Q^{2n}\frac{q^{2n^2}}{(q;q^2)_n}&F_0(q)&
\cx L =(1 - pq^{-1}) - \z^2p^2q& E_-\cr
\z^{-2n}q^{n^2}(-q;q^2)_n&\phi_0(q)&
\cx L =(1 + pq^{-1}) - \z^2p^2q^2& E_+\cr
}}}
\end{center}\vskip-1cm
\caption{Level $k=2$ indices and their modular and mock modular limits.}
\vskip0cm\end{table}
\noindent The bilateral sum $\IIB=\sum_{n\in\IZ}c_n$ satisfies $\cx L \II=0$. The last column is the quantum K-theory algebra in the $Q=1$ limit, with $E_\pm=1\pm p -p^2$. I.e. their quadratic polynomials $E_\pm $ are equal, up to a sign, and have eigenvalues $\pm (1 \pm \sqrt{5})/2$. Accordingly, there is a striking similarity between the $S$ and $T$  transformations  of the characters of the parent minimal model and that of the order 5 mock theta functions obtained from $\IZ_2$ Wilson lines. The latter can be combined into vector-valued mock theta functions \cite{Zwegers} transforming with an $6\times 6$ $S$ matrix made from $2\times2$ blocks that contain the $S$ matrix of the $(2,5)$ model and blocks related to it by similarity transformations and scalar factors.\\

\sss{$\IZ_2$ Wilson line for the other half-period:}
The other order 5 mock theta functions  appearing in the $SL(2,\IZ)$ transformation above in fact arise from the $\IZ_2$~Wilson line in the direction $\ln (q)=2\pi i \tau$, $y\to yq^{-\frac 12}$. Namely, the sums $\hat \chi_a(q)$ arise from the massless limits of eq.~\eqref{dia}
\bea
\II^+_{\cx D,D}(y=q^{-\frac 12},Q=1,q)  \leadsto \begin{cases}  N(q^{-\frac 12},q)(1+F_1(q^{\frac 12}))&Q=1\\
N(q^{-\frac 12},q)F_0(q^{\frac 12})&Q=q\end{cases}\,.
\eea
\noindent Intriguingly, these half-integral Wilson lines reduce in the massless limit essentially to the ``half-shift'' transformations studied in ref.~\cite{GM2000}, which relate directly to the original definition of mock theta functions by Ramanujan \cite{GMrev}. This definition involves the asymptotic expansion of a $q$-series $f(q)$ near roots of unity. For instance, for the real root of unity $q=e^{-t}$, $t\to 0^+$, the expansion of $f(q)$  is called closed if 
\be
\ln f(q)=\frac {a_{-1}}{t}+c_0 \ln(t)+\sum_{k=0}^Na_n t ^n+O(t^p)\,,
\ee
for all integers $p>N$ with $N$ finite, i.e., the remaining terms vanish faster than any finite power of $t$. A mock theta function has the property, that its asymptotic expansion at roots of unity is closed and there is no theta function that differs from it by a function that is bounded at all roots of unity. The half-shift of \cite{GM2000} transforms a $q$-series $f(q)$ with a closed expansion to another $q$-series $g(q)$ with (different) closed expansion. From the CFT point of view, this asymptotic expansion determines the effective central charge $c_\text{eff}=\frac 6 {\pi^2}a_{-1}$,\footnote{See e.g. ref.~\cite{Nahm} for a review.} whereas the logarithmic term $c_0 \ln(t)$ indicates the possibly logarithmic nature of the underlying CFT \cite{CG18}.

In the above example the leading terms in the expansions at $t\to 0^+$ of the two functions related by the half-integral Wilson lines are
\begin{small}\bea
\ln H(q)&=&\frac {\pi^2}{15t} + \frac 12 \ln \Big(\frac 2 {5+\sqrt 5}\Big)+\frac {11}{60}t\,,\nonumber\\
\ln F_0(q^{\frac 12})&=&\frac{\pi^2}{15t}+\frac 12 \ln \Big(\frac2 {5+\sqrt5}\Big)-\frac 12 \ln\Big(\frac t \pi\Big) -\frac t{240}\,,\\
\qquad \ln G(q)&=&\ln H(q)-\frac 1 5t + \ln \Big(\frac {1+\sqrt 5}2\Big)\,,\qquad \ln F_1(q^{\frac 12})=\ln F_0(q^{\frac 12})+\frac 3{10}t+ \ln \Big(\frac {1+\sqrt 5}2\Big)\,,\nonumber
\eea\end{small}
\hskip-4pt up to terms of order $O(t^p)$. The asymptotic expansion determines here the effective central charge $c_{\rm eff}=\frac 25$. Note that the very definition of mock theta functions gets closely related to the expansion of the 3d theories around at the roots $q^r=1$, and to the Chern--Simons theory at real coupling constant. This expansion should be compared to the perturbative expansion of the twisted superpotential of the 3d theory associated with the half-index 
\be
\II\sim \exp(\frac 1 \hbar \cx W(\hbar)) = \exp(\frac 1 \hbar W_0+W_1+\hbar W_2+\ldots) \ .
\ee
The closedness condition in the original definition of mock theta functions thus translates to the finiteness of the perturbative expansion of the twisted superpotential. It would be interesting to better understand the significance of the closedness condition from the physics perspective.\\
 
\sss{Order 5 mock theta and homological invariants}
The other pair ($\chi_0(q),\chi_1(q)$) of classical order 5 mock theta functions defined in eq.~(2.2) of ref.~\cite{GMrev} is slightly different. The four functions $\chi_a(\pm q)$ transform under $SL(2,\IZ)$ transformations amongst themselves, up to the typical Mordell integrals for mock objects \cite{GMTF}. In virtue of the mock theta conjectures they can be expressed in terms of the other order 5 mock theta functions obtained above, e.g.,
\be
\chi_0(q)=2F_0(q)-\phi_0(-q)=\sum_{n\geq 0}\frac{q^n}{(q^{n+1})_n}=\sum_{n\geq 0}q^n\frac{(q)_n}{(q)_{2n}}\,.
\ee
The sum has zero effective CS levels and does not extend naturally to a bilateral series. However, it can be obtained from the $I$-function of the $U(1)$ theory computed as in ref.~\cite{JM} and sect.~\ref{sec:PF}, with two matter fields of charges 1 and 2 with Neumann und Dirichlet b.c., respectively:
\be
I_0(Q,q)=\sum_{m\geq0}\frac{Q^n(q,q)_n}{(q,q)_{2n}}\,,
\ee
The series $I_0(Q,q)$  represents a K-theoretic $I$-function and gives as a limit $\chi_0(q)=\lim_{Q\to 1}[pI_0(Q,q)]$. The difference equation satisfied by $I_0(Q,q)$ is
\be\textstyle
\left((1-p^2)(1-p^2q^{-1})-Q(1-pq)\right)\, I_0 =0\,,
\ee
leading to the K-theory algebra $\IZ[p,Q,Q^{-1}]/\langle\!\langle p(1-p)(1-p-p^2)\rangle\!\rangle$.
Except for the trivial roots $p=0,1$, the ideal agrees with the degree 2 polynomial $E_-$ and one has the relation $\chi_1(q)=\lim_{Q\to 1} [(1+p)p(I_0(Q,q)-1)]$.

Interestingly, $\chi_0(q)$ is also essentially the homological block for the Poincar\'e homology sphere $\Sigma(2,3,5)$ with orientation reversal \cite{CG18}, and thus closely related to the  Witten--Reshetikhin--Turaev (WRT) invariants of $\Sigma(2,3,5)$ in the limit $q\to e^{2\pi i /N}$ for $N\in\IZ$  \cite{LawZag}. We find a similar relation for the homology sphere $\Sigma(2,3,7)$ below.

\subsubsection{Example: $(2,7)$ minimal model and order 7 mock theta functions}
More generally, the non-unitary minimal $(p,q)=(2,K+2)$ series with $K$ odd corresponds to the case where $A$ is the rank $n=(K-1)/2$ matrix  $A=C(A_1) \otimes C^{-1}(T_n)$. Here $T_n$ is the Cartan matrix of the tadpole graph, which is equal to the Cartan matrix $C(A_n)$ except that the last diagonal entry is 1. The fermionic sums for the minimal  models and the Verlinde algebra have been studied in the TBA ansatz in \cite{NahmReck}. The (normalized) character of the primary of lowest conformal dimension $h_n = -\frac{K^2-1}{8(K+2)}$ is of the form \eqref{2dsum}
\be\label{sumpq}
\chi_n(q)=\chi_{A,0,0}(q)\,,
\ee  
and the remaining characters are obtained by shifting the value of $B$. By the Andrews-Gordon identities, generalizing the second equal sign in \eqref{rrf}, these sums can be rewritten in the product form 
\be\label{prodpq}
\chi_a(q) =\frac{j(q^{1+a},q^{K+2})}{(q)_\infty}\,,\quad a=0,\hdots,n\,.
\ee
with $j(x,q)=(x;q)_\infty(x^{-1}q;q)_\infty(q;q)_\infty$. The 2d character $\chi_n(q)$ arises in the massless limit of the vortex sum $\eqref{DefIt}$ of the 3d $U(1)^n$ Chern--Simons matter theory, with quadratic CS levels given by the same matrix $A=2C^{-1}(T_n)$. We now check that the relation between minimal models and mock theta functions provided by the massless limits with trivial and $\IZ_2$ Wilson lines extends also to the 3d $U(1)^2$ theory with rank 2, whose massless limit reproduces the characters of the $(2,7)$ minimal model.

To this end we note that the so-called Selberg identities imply that the product \eqref{prodpq} is equal to the first sum in 
\be
\chi_2(q)=N(q)\sum_{n\geq 0} \frac{q^{2n^2}}{(q^2,q^2)_n(-q;q)_{2n}}\quad=\sum_{\vec n \in \IN^2} \frac{q^{\vec n A \vec n}}{(q)_{n_1}(q)_{n_2}}\,,\quad A=\begin{pmatrix}2&2\\2&4\end{pmatrix}\,,
\ee 
where $N(q)=(q;q^2)^{-1}_\infty$. Similar relations hold for $a=0,1$ and will be stated in more general form below. As indicated, this can be viewed as a resummation of the double sum \eqref{sumpq} for the $U(1)^2$ theory  with quadratic CS terms specified by the matrix $A$ to a $U(1)$ theory with different matter content. Indeed the first sum arises from the vortex sum of a $U(1)$ CS matter theory with 3 chirals of $U(1)_\partial$ charge $-1$, R-charges $0$, $0$, $1$ and Dirichlet b.c.
\bea
\It_0(y,Q,q)&=&\sum_{n\geq 0} \frac{q^{n(n+1)}Q^ny^{2n}}{(yq,q)_n(-yq^{\frac 12},q^{\frac 12})_{2n+1}}\,,\nonumber \\
\It_1(y,Q,q)&=&\sum_{n\geq 0} \frac{q^{n(n+1)}Q^n y^{2n}}{(yq,q)_n(-yq^{\frac 12},q^{\frac 12})_{2n}}\,,
\\
\It_2(y,Q,q)&=&\sum_{n\geq 0} \frac{q^{n^2}Q^ny^{2n}}{(yq,q)_n(-yq^{\frac 12},q^{\frac 12})_{2n}}\,.\nonumber
\eea
For trivial Wilson line $y=1$, these have the massless limits $\It_a|_{y=1,Q=1}(q^2)=\chi_a(q)/N(q)$. On the other hand, the restrictions $\It_0(q)|_{\IZ_2}$ to $y=q^{-\frac 12}$ and $Q=1$ give rise to the linear combination 
\bea
2\It_0(q)|_{\IZ_2}=\cx F_0(q) \,,\quad
2\It_1(q)|_{\IZ_2}= \cx F_1(q)+2\,, \quad 
2\It_2(q)|_{\IZ_2}=\cx F_2(q) +2,,
\eea
where
\be
\cx F_0 = \sum_{n\geq 0}\frac {q^{n^2}}{(q^{n+1})_n}\,,\quad
\cx F_1 = \sum_{n\geq 0}\frac {q^{(n+1)^2}}{(q^{n+1})_{n+1}}\,,\quad
\cx F_2 = \sum_{n\geq 0}\frac {q^{n(n+1)}}{(q^{n+1})_{n+1}}\,,
\ee
are the classical order 7 mock theta functions. Again the 3d Wilson lines descend to the  half-shift transformation of refs.~\cite{mci,GM2000} in the limit. In the above we have only considered the positive part of the index sum for simplicity. The bilateral sums associated to the order 7 mock theta function have been studied in ref.~\cite{Choi}.

The difference equation satisfied by $\It_2$ is of degree three,
\be
((1-yp)(1+ypq^{-\frac 12})(1+yp)-Qy^2qp^2)\, \It_2 = (1-y)(1+yq^{-\frac12})(1+y)\,.
\ee
 The quantum K-theory algebra has three elements, say $1,p,p^2$, and these correspond to the three functions $\It_a$, related by the difference operators in the diagram.\\[-9pt]
\begin{center}\includegraphics[height=2.2cm]{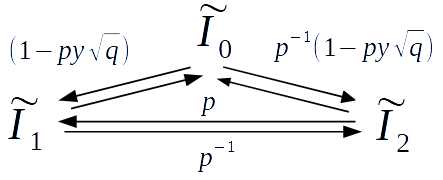}\end{center}\ \\[-22pt]
\noindent All difference operators at the arrows are invertible.   In the massless limit $y=1=Q$ the ideal in the quantum K-theory ring and its solutions are
\be
1+p-2p^2-p^3=0\,,\qquad p_a = -1+2\cos(\frac\pi 7 (2a+1))\,,\ a=0,1,2\,,
\ee
The roots generate the coefficients of the monodromy matrices of the $(2,7)$ minimal model. Indeed the monodromies of the $\cx F_a$ computed in \cite{GMTF} (see also \cite{Zwegers}) are closely related to that of the minimal model, similarly as in the $(2,5)$ example.  

The mock theta function $\cx F_0$ is proposed in ref.~\cite{CG18} as the 
homological block of the three-manifold $M_3$ obtained from the homology sphere $\Sigma(2,3,7)$ via orientation reversal. It would be interesting to understand the relation between the $U(1)$ CS matter systems and the theories $T(M_3)$ considered in ref.~\cite{CG18} systematically.

\subsubsection{Example: $(3,4)$ minimal model and order 3 mock theta functions}
For our final example we come back to the $U(1)$ theory with $k=1$ in \eqref{DirInd}. The index  reduces at the half-torsion points  $y=1$, $Q=q^{\frac a 2}$, $a\in\{-1,0,1\}$ to the characters of the $(3,4)$ minimal model up to an overall power of $q$
\be
\II_{\cx D,D}^{k=1}=\frac{(qy)_\infty}{(q)_\infty}\sum_{n\in\IZ}Q^ny^n\frac{q^{\frac 12 n^2}}{(yq)_n} \ \ 
\longrightarrow \ \
\sum_{n\geq 0}\frac{q^{\frac 12 n(n+a)}}{(q)_n}
\,,\quad a\in\{-1,0,1\}\,,
\ee
with \eqref{freefermion} being the case $a=0$, and $a\pm1$ the other two modular triples in \cite{Zagier2007}. Adding Wilson lines at half torsion points connects the modular characters above to mock theta functions of order 3, which have been already identified in ref.~\cite{RZ18} with K-theoretic $I$-functions of the point. Firstly, the half-torsion point $y=q^{-\frac 12}$, $Q=q^{\frac a 2}$ gives
\be
\II_{\cx D,D}^{k=1,+}=N(q)\sum_{n\geq0}\frac{q^{\frac 12 n(n+a-1)}}{(q^{\frac 12})_n} = N(q)\begin{cases} 
1+q^{-\frac 12 }(1+2\psi(q^{\frac 12}))\,,&a=-1\,,\\
1+\nu(-q^{\frac 12})\,,&a=0\,,\\
1+\psi(q^{\frac 12})\,,&a=1\,,\end{cases}
\ee
with $N(q)=(q^{\frac 12})_\infty/(q)_\infty$ and $\psi(q)$ and $\nu(q)$ defined  in eq.~(2.1) and eq.~(2.4) of ref.~\cite{GMrev} and obtained from the half-shift operation in ref.~\cite{mci}. At the half-torsion point $y=-1$, $Q=1$ we have
\be
\II_{\cx D,D}^{k=1,+}=\frac{(-q)_\infty}{(q)_\infty} \sum_{n\geq0}\phi(q^{\frac 12})\,,
\ee
with $\phi(q)$ the order 3 mock theta functions  in eq.~(2.1) of ref.~\cite{GMrev}. This mock theta function gives the WRT invariant $\tau_N$ of the Seifert manifold $M(2,3,4)$ \cite{Hikami} in the limit $q\to e^{2\pi i/N}$

\subsection{Universal mock theta functions from torsion points}
We now observe a more general relation between Wilson lines in the 3d global symmetry group at torsion points and the half-index of the $U(1)_k$ CS theory at level $k$ with Dirichlet b.c. for the gauge fields in \eqref{pureCS}. 
To this end we consider the dressed  index
\be\label{defLL}
\IIL_k(y,Q,q) :=\frac1{(q)_\infty}\sum_{m\in\IZ}\x^my^{km}q^{\frac k2 m^2}\frac{(yq^{m+1};q)_\infty}{(yq^{m};q)_\infty}=\frac1{(q)_\infty}\sum_{m\in\IZ}\frac{\x^my^{km}q^{\frac k2 m^2}}{1-yq^m}\,.
\ee
Here $Q$ and $y$ are again the fugacities of the topological $U(1)_t$ and boundary $U(1)_\partial$, respectively. The first expression defines $\IIL_k$ as the half-index of the CS matter theory with two matter fields of charges $\pm 1$ under $U(1)_\partial$ and Neumann/Dirichlet b.c., respectively. The contributions in the index cancel up to the single spinless mode in the denominator of the second expression, which can also be viewed as the index of the pure gauge theory with a Wilson line operator $W=1/(1-y)$ inserted.

The r.h.s.~in \eqref{defLL} is the Appell--Lerch sum \eqref{DefAL}, up to a prefactor
\be  \label{IAL} 
\IIL_k(y,Q,q) =\frac{y^{-\frac k2}}{(q)_\infty}A_k(y,z,q)\,,\qquad z=(-q^{-\frac 12 }y)^kQ\,.
\ee
Upon restriction of $y$ and $z$ to torsion points, the sum $A_k$ has mock modular transformation behavior \cite{Zwegers}.\footnote{Appell--Lerch sums have appeared  recently in a related context in ref.~\cite{FerPut}.}  In particular, restricting first the topological fugacity $Q$ to set $z=\pm1$ in \eqref{IAL}, one obtains the one-parameter  functions \cite{GMrev}
\bea
g_3(y,q)&=&\frac 1 {(q)_\infty} \sum_{n\in\IZ}\frac{(-) ^n q^{\frac 32 n(n+1)}}{1-yq^n}=\sum_{n\geq0} \frac{q^{n(n+1)}}{(y;q)_{n+1}(q/y;q)_{n+1}}
\,,\nonumber\\
g_2(y,q)&=&\frac {(-q;q)_\infty} {(q)_\infty} \sum_{n\in\IZ}\frac{(-)^nq^{n(n+1)}}{1-yq^n}=\sum_{n\geq0} \frac{q^{\frac12n(n+1)}(-q,q)_n}{(y;q)_{n+1}(q/y;q)_{n+1}}
\,.
\eea
These are called universal mock theta functions, since the mock modular part of a large class of mock theta functions $f(q)$ of odd (even) order can be obtained by restricting the second fugacity $y$ in $g_3$ $(g_2)$ to torsion points \cite{HickersonProof,GMrev}. Note that they can be written in terms of the dressed $U(1)_k$ index as 
\be
g_3(y,q)=\IIL_3(y,-y^{-3}q^{\frac 32},q)\,,\qquad g_2(y,q)=(-q)_\infty\IIL_2(y,-y^{-2}q,q)\,.
\ee
Moreover, the massless limit of the dressed index with level $k=1$ gives the known Lambert series for an inverse theta function
\be
\IIL_1(y,-y^{-1}q^{\frac 12},q) = \frac 1{(q)_\infty}\sum_{n\in\IZ}\frac{(-)^nq^{\frac 12 n(n+1)}}{1-yq^n}=\frac {(q)_\infty} {\theta(y,q)}\,.
\ee 
The theta function is the elliptic genus for a 2d chiral of charge 1 in the boundary $U(1)$ symmetry \cite{Gadde:2013wq,BeniniEG}.
Further restricting $y$ to half-torsion points, one obtains ordinary modular functions up to overall powers of $q$.

For $k=2$, the restriction of the dressed index to the half-torsion point $y=q^{\frac 12}$
is
\be 
\IIL_2(q^{\frac 12},-1,q)=\frac 1 {(-q;q)_\infty} B(q^{\frac 12})\,, 
\ee
where $B(q)$ the order 2 mock theta function \cite{GMrev}
\be
B(q)= \sum_{n\geq 0}\frac {q^{n^2+n}(-q^2;q^2)_n}{(q;q^2)_{n+1}^2}\,.
\ee
The last identity shows that there are  alternative representations for the same index in terms of either the dressed index $\IIL_k$, or an $U(1)$ CS matter theory for functions involving the universal mock theta functions. In the above example, the sum on the r.h.s.~can be obtained from the index of a 3d gauge theory with two matter fields with Neumann and one field with Dirichlet b.c, giving rise to the $q$-Pochhammers in the denominator and the numerator of the second expression above. 

The same is true for the order 5 examples obtained from the $U(1)$ theory with a matter field with Neumann b.c. and $\IZ_2$ Wilson lines in sect.~\ref{sec:triple}. By the mock theta conjectures these can be alternatively written in terms of $g_3$ restricted to $\IZ_5$ torsion points \cite{HickersonProof}, e.g.,
\bea
f_0(q)&=&-2q^2g_3(q^2,q^{10})+ \theta_4(q^{10})\, G(q)\,,
\nonumber\\
f_1(q)&=&-2q^3g_3(q^4,q^{10})+\theta_4(q^{10})\, H(q)\,,\quad 
\eea
The two descriptions by a either a CS matter system or the index $\IIL_k$ in the above examples appear to indicate a IR duality of the two different 3d gauge theories. 

For the minimal model characters there are the three alternative expressions 
\be\label{GHp}
G(q)=\sum_{n\geq0}\frac {q^{n^2}}{(q)_n}=\frac1{ \theta(q,q^5)}=\frac{\theta_{5,1}(q)}{(q)_\infty}\,,
\ee
and similarly for $H(q)$. The first expression arises from the CS matter system. The theta functions in the second expression are the $\mathcal{N}=(0,2)$ elliptic genus for the $\IZ_5$ Landau Ginzburg orbifold representation of the minimal model. There is also a gauge theory representation for the last expression in \eqref{GHp}.  Consider the massless limit of the index of the pure $U(1)_{k=5}$ CS theory with Wilson line insertions $p^j$, $j=0,\hdots,4$ 
\bea
\IID[U(1)_{k=5},p^{j-1}] &=& \frac {1} {(q)_\infty}\sum_{m\in\IZ}q^{\frac 5 2 m^2}\x ^mq^{m(j-1)}
\ \ {\buildrel \x=-q^{\frac 12 }\over \leadsto} \ \
q^{\frac{(2j-1)^2}{40}}\frac{\theta_{5,j}(q)}{(q)_\infty}\,,\nonumber\\
\theta_{5,j}(z,q)&=&\sum_{n\in\IZ}(-1)^n q^{\frac 5 2 n^2 +\frac {n(2j-1)}2+\frac{(2j-1)^2}{40}}z^n\,,
\eea
with $\theta_{5,j}(q)=\theta_{5,j}(1,q)$. The difference equation satisfied by the index with $j=0$ 
\be
p^5 -\x^{-1}q^{-\frac 52} =0\,,
\ee
is of order 5, but in the special massless limit above, the independent indices with insertions $p^j$,$j=0,\hdots,4$ reduce to two since $\theta_{5,1+i}(q)=-\theta_{5,5-i}(q)$ for $i=1,2$ and  $\theta_{5,3}(q)=0$, thus reproducing the two independent modular characters of the (2,5) minimal model obtained from the order 2 difference equation. It would be interesting to reconstruct the mock modular functions above explicitly as characters of CFTs, which could be non-unitary, non-rational, and logarithmic.\footnote{Interestingly, the summands of eq.~\eqref{DefAL} at fixed $m$ are closely related to the characters of discrete irreducible representations of the $\mathcal{N}=2$ superconformal algebra built on the Ramond vacuum with lowest $U(1)_R$ charge for central charge $\hat c = 1+k$ \cite{Sugawara}
\be
\ch^{\tilde R}_{\rm dis}(m)=\frac{\theta_1(y,q)}{i\eta(q)^3}\frac{q^{\frac k2 m^2}y^{mk}}{1-yq^m}\,.
\ee}

\section{Perturbations and correlators\label{sec:perturbations}}
\subsection{RG flow as a 3d mirror map}
Before we compute the perturbed theory for the Grassmannian theories, we discuss some aspects of the RG flow between the UV gauge theory and the quantum K-theory in the IR related by the correspondence of ref.~\cite{JM}. This concerns the connection between the perturbations of the UV gauge theory and the IR deformations described by the reconstruction theorems of refs.~\cite{GivTon,IMT,GivER,Givental:2015p8} for the $I$ function with non-zero input in quantum K-theory.\footnote{This UV-IR map is the 3d analogue of what is called the mirror map in the physics literature for 2d GLSM \cite{WitPhases,MP}; see refs.~\cite{CK,MB} for background material.}

The UV definition of the gauge theory is in terms of a three-dimensional gauged linear sigma model (GLSM), which is a $\cx N=2$ supersymmetric gauge theory with gauge group $G$ and matter fields in some representations $R$ with canonical kinetic terms. In a Higgs phase, where the massless matter fields parametrize a K\"ahler manifold $X$, the massive gauge fields can be integrated out in the IR limit. The UV theory includes details of the renormalization group (RG) flow to the IR that are not related to the geometry of $X$.\footnote{A well-known example are realizations of the same K\"ahler manifold $X$ embedded in different ambient spaces $W$ as a hypersurface. The different GLSMs depend on the embedding, but their IR limit agrees.} The quantum (twisted) chiral ring associated with the geometry of $X$ emerges in the IR limit.

For the two-dimensional gauge theory with $\mathcal{N}=(2,2)$ supersymmetry, the twisted chiral ring in the IR limit is well understood in terms of the topological $A$ model \cite{Witten:1988xj} and quantum cohomology. The partition function of the 3d $\cx N=2$ gauge theory on $\DSq$ that we consider reproduces correlators of the two-dimensional $A$~model in the small radius considered in ref.~\cite{JM}. The 3d IR phase for finite radius is far less understood, but one can use that the index \eqref{formalhalfindex} is RG invariant. The index of the GLSM is computable in the UV, and can be used to study the IR limit, {\it if} one knows the UV-IR map between the couplings of the UV and IR theories. 

The space of couplings includes perturbations by twisted chiral ring elements $\cx O_i$. Schematically, the perturbed theory computes the correlators 
\be\label{correlators}
\qcord {\cx O_1 \hdots \cx O_N} = \sum_{n=0}^\infty \frac 1 {n!}\qcor {\cx O_1 \hdots \cx O_N t ^n}=\sum_{n=0}^\infty \frac 1 {n!}\sum_{d=0}^\infty Q^d\qcor {\cx O_1 \hdots \cx O_N t ^n}_d\, ,
\ee
in the background with source $t=\sum_\alpha t_\alpha \cx O_\alpha$ and vortex number $d$. The correlators in the IR will be defined by the appropriate cohomology theory on Kontsevich's moduli space $\Mst^d$ of degree $d$ stable maps from a Riemann surface $\Sigma$ to $X$ \cite{Kont}, representing the moduli space of vortices of degree $d$ in the IR theory. In the 2d theory, the correlators are defined as cohomological intersections on $\Mst$ \cite{Witten:1988xj,CK,MB}. By the general arguments of  \cite{NekPhd}, the 3d theory compactified on a circle should compute holomorphic Euler numbers, i.e. a type of quantum K-theory, on $\Mst$, with insertions from twisted chiral operators in $K(X)$ \cite{NekPhd, KW13,CV13}. 

The mathematical framework of ordinary quantum K-theory on $\Mst$ has been developed in refs.~\cite{GivWDVV,Lee:2001mb,GivLee,GivTon,IMT}.\footnote{See refs.~\cite{OkL,Aganagic:2017gsx} for a definition on a different compactification of the moduli space adapted to the UV regime and refs.~\cite{BKK,Pushkar:2016qvw,Koroteev:2017nab} for related works.}
An important generalization, which is needed to compare to the UV gauge theory,  is the permutation equivariant quantum K-theory constructed in ref.~\cite{Giv15all}. It computes Euler numbers of bundles equivariant under the action of the permutation group on the insertions of operators, in parallel with the permutation symmetry of multi-particle states in the three-dimensional gauge theory. If one restricts to symmetric representations of the permutation group, the generating function of 1-point correlators takes the form of a $q$-hypergeometric series of precisely the same type, as it arises in the computation of the partition functions of three-dimensional $\cx N=2$ gauge theories.

\begin{figure}[t]
\begin{center}\includegraphics[height=5.5cm]{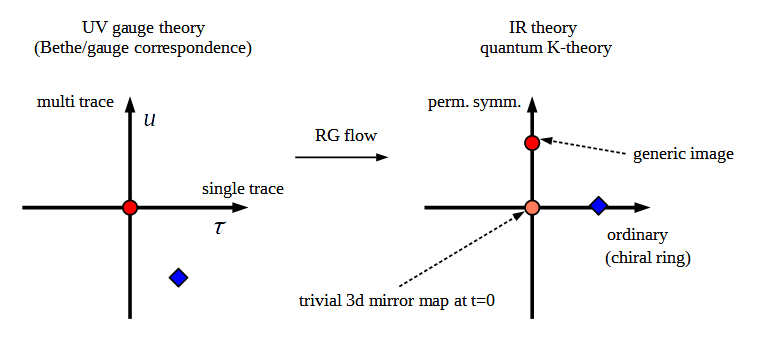}\end{center}
\caption{The figure illustrates the RG flow --- as encoded in the 3d mirror map --- between the UV gauge theory with perturbations by multi- and single-trace operators and the IR quantum K-theory description with the ordinary and permutation symmetric deformations. A generic UV gauge theory without multi- and single-trace perturbations yields a permutation symmetric deformation in quantum K-theory (pair of red dots), whereas an ordinary deformation in quantum K-theory arises generically from a UV theory with both single- and multi-trace perturbations (pair of blue diamonds).  For theories with a trivial 3d mirror map, the unperturbed UV gauge theory (red dot on the left) flows to a quantum K-theory without deformations (orange dot on the right). For such theories, single-trace perturbations in the UV gauge theory are sufficient to yield the ordinary quantum K-theory deformations.} \label{fig:RGFlow}
\end{figure}

The correspondence of ref.~\cite{JM} equates the vortex sum of the unperturbed partition function of the three-dimensional $\cx N=2$ UV gauge theory on $\DSq$ with the perturbed $I$-function of the permutation symmetric quantum K-theory of \cite{Giv15all} (red points in fig.~\ref{fig:RGFlow}).\footnote{In this work, we will restrict to the permutation symmetric case; see ref.~\cite{JM} for the correspondence with the general theory of ref.~\cite{Giv15all}.} The non-zero perturbations in the IR arise from integrating out the high energy modes along the flow. More generally, perturbing the UV gauge theory and applying the RG flow maps out the coupling space of the IR theory. There are two types of perturbations to consider. Firstly, one can integrate in 3d massive matter fields in the UV with twisted mass parameter $\hu$, which generate insertions of multi-trace operators in the partition function. Under the RG flow, these become perturbations $t_{eq}(\hu)$ of the permutation equivariant theory of ref.~\cite{Giv15all}. Secondly, perturbations by single trace operators with parameter $\tau$ in the UV flow to perturbations $t_{ord}(\tau)$ in the ordinary quantum K-theory of refs.~\cite{GivWDVV,Lee:2001mb,GivLee}.  The 3d UV partition function $Z(\hu,\tau)$ perturbed by both types of insertions computes the $I$-functions of the mixed quantum K-theory defined in ref.~\cite{Givental:2015p8}. More precisely, for non-zero effective CS terms, the IR limit of the gauge theory gives a non-zero level in quantum K-theory in the sense ref.~\cite{RZ18}. 

Having obtained the perturbed $I$-function from the vortex sum of the 3d gauge theory, one can determine the UV-IR map between the UV parameters $(\hu,\tau)$ and the IR couplings $t$ as a series expansion in $\hu$ and $\tau$ from a particular term of the $I$-function.\footnote{A (non-trivial) finite representation of the symmetric group distinguishes the equivariant IR perturbation $t_{eq}$ from the ordinary IR perturbation $t_{ord}$ in mixed quantum K-theory, c.f., ref.~\cite{Giv15all}.} This is the {\it input} defined by the decomposition of the $I$-function \cite{GivTon,GivER,Givental:2015p8}
\be\label{definp}
I_X(t)=(1-q)+t +I_{\textrm corr.}(t)\ ,\qquad t \in \cx K_+\ ,\quad I_{\textrm corr.}\in \cx K_-\ ,
\ee
with $X=\Gr(M,N)$  and\footnote{Note that $\mathbb{C}[t]$ denotes the polynomial ring in the variable $t$ with coefficients in $\mathbb{C}$, $\mathbb{C}[[t]]$ is the ring of formal power series in $t$ with coefficients in $\mathbb{C}$, and $\mathbb{C}(t)$ denotes the field of rational functions in the variable $t$ over $\mathbb{C}$.}
\be
\begin{aligned}
  \mathcal{K}_+&=K(X)\otimes \mathbb{C}[q,q^{-1}]\otimes\mathbb{C}[[Q]]  \ ,  \\
  \mathcal{K}_- &=K(X) \otimes  \left\{\,r(q) \in \mathbb{C}(q) \,\middle|\, \text{$r(0)\ne\infty$ and $r(\infty)=0$} \right\} \otimes\mathbb{C}[[Q]] \ .
\end{aligned}  
\ee
The formal series $t(\hu,\tau,Q,q)$ captures the UV-IR map between the UV parameters $(\hu,\tau)$ and the perturbations $t$ in the IR. From the physics point of view, the split \eqref{definp} is a reorganization of the 3d multiplets in terms of twisted spin $J_{tw}=J+\frac R 2$. The coefficients in $K(X)$ represent gauge charges, the factor $\IC[q,q^{-1}]$ takes into account different $R$-charges, and the dependence on the classical exponentiated FI parameter $Q$ captures non-perturbative corrections to the UV-IR map arising from integrating out point-like vortices.\footnote{This was discussed first in ref.~\cite{MP} in the 2d case and for other dimensions, e.g.,~in ref.~\cite{Losev:1999nt}.} The coefficients of the series $t(u,\tau,Q,q)$ are itself degeneracies of BPS states in the 3d UV gauge theory  and thus integral \cite{JM}. In the small radius limit, the 3d UV-IR map resums to the 2d mirror map, thus linking the integrality properties of the 2d mirror map to integral 3d BPS degeneracies.

For a special class of theories, the UV-IR map is trivial at the origin of the couplings, i.e., the $I$-function computed from the unperturbed 3d gauge theory is already at $t=0$ (orange dot in fig.~\ref{fig:RGFlow}). For Grassmannian targets, this requires the CS levels to lie in another window, the {\it input window}, which requires the condition $|\kappa_S+\Delta_\kappa|\leq \frac N2$ given in ref.~\cite{JMNT}, and another condition on $\hat\kappa_R$, which is satisfied for $\hat\kappa_R=0$. Only in this special situation, the basic observables of the UV and IR theories match on the nose, and the twisted chiral ring in the IR can be computed directly in the UV theory. In particular, one can then use the Bethe-gauge correspondence of ref.~\cite{NekSha2} to compute the twisted chiral ring for zero perturbation. 

In the general case one has to study the details of the non-trivial UV-IR flow. Even in the theories in the input window with trivial UV-IR map at zero deformation, the 3d mirror map becomes non-trivial for non-zero deformations. For a judicious choice of UV perturbations, the input in the IR is purely in the ordinary K-theory (blue diamonds in fig.~\ref{fig:RGFlow}). The twisted chiral ring of the ordinary quantum K-theory carries the structure of a Frobenius algebra \cite{GivWDVV}. The aim of the following section is to compute this twisted chiral ring and the related IR correlators for UV gauge theories with Higgs phase a Grassmannian $X=\Gr(M,N)$. 

\subsection{Non-Abelian reconstruction of the deformed theory}
One way to perturb the UV gauge theory, is to integrate in new massive charged test particles. The contribution of a chiral field to the single particle index in the finite representation $R$ of the gauge group $G$ and $f$ of the global symmetry group is~\cite{Kapustin:2011jm,DGP}
\be\label{SingleChiralPert}
I_\text{chiral}(z,y,q)=\pm \sum_{\rho \in w(R)} \left(z^\rho y^fq^{(\Delta-1)/2}\right)^{\pm 1}\frac{q^{\frac 12 }}{1-q}\,,
\ee
where we sum over the weights $w(R)$ of the representation $R$, the plus (minus) sign holds for Neumann (Dirichlet) b.c., and $\Delta$ is the $U(1)_R$ charge. As before, $z_a$ are the fugacities for the Cartan torus of the gauge group $G$, and we define $z^\rho=\prod_a z_a^{\rho(a)}$ for any weight $\rho$ of the representation $R$. Furthermore, $y^f$ are the fugacities for the global symmetry group. The single trace deformation is defined in ref.~\cite{JM} as an insertion of the exponential \def\cR{{\cal R}}
\be\label{stop}
  \cR_{s.t.}(z,\tau,q)=e^{I_\text{chiral}(z,\tau,q)}\ ,
\ee
where the notation $\tau$ is reserved for mass parameter of the chiral field to be integrated in. In general it is a combination of fugacities $y$ of the global symmetry group {\it before} perturbation and a new $U(1)$ global flavor symmetry that comes with the new field. Here and in the following, we often spell out only the novel fugacities $\tau$ (and $\hu$ below) that enjoy in the UV the interpretation as twisted masses of the new chiral fields.

The contribution from multi-particle states is obtained from the single particle index by taking the plethystic exponential (see, e.g.,~ref.~\cite{Aha}). For the plus sign in eq.~\eqref{SingleChiralPert} one obtains
\bea\label{mtop}
\cR_{m.t.}(z,\hu,q)&=&\exp\left(\sum_r \frac{I_\text{chiral}(z^r,\hu^r,q^r)}r\right)\\&=&
\exp\left(\sum_{\rho,r}\frac{(z^\rho \hu \,q^{\Delta/2})^r}{r(1-q^r)}\right)
=
\frac1 {\prod_{\rho}(z^\rho \hu \,q^{\Delta/2};q)_\infty}\ .\nonumber
\eea
The last expression is, up to a regularization factor, equal to the one-loop determinant \eqref{oneloopdets} of the chiral  field with twisted mass $\hu$ with Neumann b.c.~computed in refs.~\cite{DGP,YS}. A similar comment applies to the case of a chiral field with Dirichlet b.c.~\cite{Bul20}.

\subsubsection{Reconstruction by difference operators \label{sec:recdiff}}
The above deformations can be related to the reconstruction of the perturbed $I$-function by difference operators by rewriting the insertion of the operators \eqref{stop},\eqref{mtop} in the localized path integral \eqref{LocPathIntegral} as a difference operator acting on the vortex sum~$I$ \cite{JM}. Taking the sum \eqref{ZisfI} over the residues at the poles \eqref{poles}, the fugacities $z_a$ are replaced by $z_a=q^{-n_a+\eps_a}=q^{-n_a}P_a^{-1}$. Here $P_a=e^{-\beta x_a}$ is the Chern character of the bundle $O(-1)$ by \eqref{epstoH}. In the Abelian case, one can write $\rho=-q_a$ for negative $U(1)$ charges $q_a<0$ and represent the insertion of $\cR_{m.t.}$ as the difference operator\footnote{The extra factors $P_a$ arise from a different normalization $ I_{\Gr(M,N)}|_{Q=0}=(1-q)$ of the $I$-functions in ref.~\cite{Givental:2015p8} and in this section; see remark in fn. \ref{fn:norm}.}
\be\label{RecDeq}
\exp\left(\sum_{r\geq 1} \frac{\psi_r(\hu\,q^{\Delta/2})\prod_a (P_aq^{Q_a\p Q_a})^{q_ar}}{r(1-q^r)}\right)\,,
\ee
with the Adams operator $\psi_r(u\,q^{\Delta/2})=u^r q^{r\Delta/2}$. The r.h.s.~is the reconstruction operator for the permutation symmetric theory in  ref.~\cite{Givental:2015p8}. Similarly, the single trace deformation obtained by restricting to the term $r=1$ in  the sum in eq.~\eqref{RecDeq} gives the reconstruction operator in ordinary quantum K-theory of refs.~\cite{GivTon,IMT,GivER}.

To obtain the perturbed $I$-function for non-Abelian gauge group $U(M)$ one can perturb  the Abelianized $U(1)^M$ $I$-function $\hat I$ in \eqref{abel}, use the action of the $M$ $U(1)$ difference operators $\delta_a$ for reconstruction, and pass to the locus $Q_a=Q$ at the end of the computation, if possible. This was used ref.~\cite{JMNT} to compute the quantum K-theory algebra \eqref{proposal1} for the Grassmannian in some examples, and studied in detail in refs.~\cite{Giv20,Yan21} in the meantime. 

Below we follow a different path by perturbing directly in the non-Abelian gauge theory, avoiding  Abelianization altogether. This potentially generalizes the reconstruction theorems \cite{GivTon,GivER,Givental:2015p8} to cases, which do not allow for Abelianization, e.g.,~cases related to strongly coupled phases of non-Abelian gauge theories.

\subsubsection{Reconstruction in the non-Abelian theory}
From the 3d gauge theory point of view, the case of a non-Abelian group is not any different, and one can use eqs.~\eqref{stop} and \eqref{mtop} to  integrate in matter in any representation $R$ of the non-Abelian gauge  group $G$. The non-Abelian formalism also directly applies to the deformation of the bilateral $q$-series associated with the sum over monopoles  for Dirichlet b.c.

For the gauge group $U(M)$ the insertion of the contribution~\eqref{mtop} for matter fields in the finite irreducible representations $\mu$ yields with the twisted masses $\hu_\mu$ and $\tau_\mu$ the respective multi- and single trace factors
\be\label{Rop}
\begin{aligned}
\cR_{m.t.}(z,u,\vec m) &=  \exp\left(\sum_{r=1}^\infty
\left(\frac{\sum_\mu \psi_r\left(\hu_\mu\sigma_\mu(z\,q^m)\right)}{r(1-q^r)}\right)\right)\ , \\
\cR_{s.t.}(z,\tau,\vec m) &= \exp\left(\frac{\sum_\mu \tau_\mu\sigma_\mu(z\,q^m)}{1-q}\right) \ ,
\end{aligned}
\ee
in the sector with monopole number $\vec m\in\IZ^M$.\footnote{More generally, for a gauge group $G$ the monopole number $\vec m$ takes values in the cocharacter lattice of the Lie group $G$.} Here $\sigma_\mu$ is the Schur polynomial  in the fugacities $z$ of the boundary gauge symmetry,\footnote{Here, the fugacity $z$ refers to the gauge symmetry $G$ at the boundary arising from Neumann b.c.~of the gauge boson. For Dirichlet b.c.~of the gauge boson the boundary symmetry $G$ is global.}
and we have absorbed the factor $q^{\frac\Delta 2}$ in the definition of the parameters $\hu_\mu$ and $\tau_\mu$. Moreover, the Adams operation~$\psi_r$ acts as $\psi_r(x)=x^r$ on all fugacities $x$.

To compute the perturbed twisted chiral ring in the ordinary quantum K-theory, we deform the $I$-functions with zero input by single traces with deformation parameter
\be
 \tau_\mu(t)\in \IC[q,q^{-1}]\otimes \IC[[Q]]\quad \text{with} \quad \left.\tau_\mu(t)\right|_{Q=0} = t_\mu \in\IC\ .
\ee
As a working example we consider the case $X=\Gr(2,4)$ with canonical CS terms. Our starting point is the $I$-function \eqref{ISQKGr}
\begin{equation}\label{J24}
  J_{\Gr(2,4)}(t=0)=(-Q)^{\eps_1+\eps_2}I^{(0,0,0)}_{\text{Gr}(2,4)} (Q,q,\epsilon)= c_0 \sum_{d_1,d_2 \geq 0}  (-Q)^{d_1+d_2} \, \frac{ q^{\tfrac{1}{2}\tilde{d}_{12}^2} (q^{\tfrac{1}{2}\tilde{d}_{12}} - q^{-\tfrac{1}{2}\tilde{d}_{21}})}{\prod\limits_{a=1}^2\prod\limits_{r=1}^{d_a} (1-q^{r-\epsilon_a})^4}.
\end{equation}
where $z=q^\eps$ is the fugacity for the boundary and we recall the definitions $\td_a = d_a-\eps_a$ and $\td_{ab}=\td_a-\td_b$. Since $I_{\Gr(2,4)}(0)|_{\cx K_+}=(1-q)$, the unperturbed UV theory  computes the $I$-function at zero input $t=0$, i.e., the UV-IR map is trivial for zero deformation. 
\def\tord{t} Our goal is to find the $J$-function\footnote{The reason for the change of notation from $I$ to $J$ is the following. Geometrically, the concepts of the $I$-function and the $J$-function differ in the their definition as generating functions of correlators on the moduli spaces of quasi-maps and stable maps, respectively. Technically, the two are connected by the 3d mirror map, which involves a change of frame and a change of coordinates to the flat parameters $t$ (in the sense of the 3d $tt^*$ equations of ref.~\cite{CV13}). When the parameterization by flat coordinates or the stable map compactification of the moduli space is important, the proper concept is that of the $J$-function.}   with ordinary perturbations $ J_{\Gr(2,4)}(\tord)$. It is the function that satisfies
\begin{equation}\label{formalreconstructed}\textstyle
	 J_{\Gr(2,4)}(\tord)=(1-q)+\tord+J_{corr.}(\tord)
\end{equation}
where $ J_{corr.}(\tord)\in \cx K_-$ is the generating function for the correlators \eqref{correlators} 
\begin{equation}
	 J_{corr.}(\tord)= \sum_{d=0,n=0\atop (d,n)\neq(0,0),(1,0)}^\infty\frac{Q^d}{n!}\Big\langle \frac{\cx O_\mu}{1-q L},\tord^n\Big\rangle_{0,n+1,d}g^{\mu\nu}\cx O_\nu \ ,
\end{equation}
where $g^{\mu\nu}$ is the inverse of the classical bilinear form defined in eq.~\eqref{defgtwisted} below.
We parametrize the input $\tord\in \cx K_+$ as  
\begin{equation}\label{etord}
	\tord=\!\!\sum_{\mu\in \cx B_{(M,N)}}\!\! t_\mu\cx O_\mu
	=t_1\cx O_1+t_2\cx O_2+t_{1,1}\cx O_{1,1}+t_{2,1}\cx O_{2,1}+t_{2,2}\cx O_{2,2}\ ,\quad t_\mu\in \IC \ ,
\end{equation}
where $\cx O_\mu=\cx O_\mu(1-q^{-\eps})$ are Grothendieck polynomials labeled by partitions $\mu\in \cx B_{(M,N)}$ fitting into the $M\times (N-M)$ box. This ansatz is related by a linear basis transformation to the expression \eqref{Rop} in terms of the Schur polynomials; the Grothendieck polynomials represent the Chern characters of the structure sheaves of the Schubert cycles and provide the natural basis from the geometric point of view (see app.~\ref{app:KtheoryGr} for some details).

The $J$-function resulting from this deformation is of the form
\begin{equation}
	\begin{aligned}\label{reconstructedGr24}
		J_{\Gr(2,4)}\big(\tord(\tau)\big)
&= c_0 \sum_{d_1,d_2 \geq 0} \cR_{s.t.}(\tau,\eps,\vec d) (-Q)^{d_1+d_2} \, \frac{ q^{\tfrac{1}{2}\tilde{d}_{12}^2} (q^{\tfrac{1}{2}\tilde{d}_{12}} - q^{-\tfrac{1}{2}\tilde{d}_{21}})}{\prod\limits_{a=1}^2\prod\limits_{r=1}^{d_a} (1-q^{r-\epsilon_a})^4},
	\end{aligned}
\end{equation}
where 
\begin{equation}
	\cR_{s.t.}(\tau,\eps,\vec d)=\exp \bigg(\frac{1}{1-q}\!\sum_{\mu\in \cx B_{(M,N)}}\!\!\tau_\mu \cx O_\mu(1-q^{d-\eps})\bigg).
\end{equation}
The choice of the UV deformation $\tau(t)$ that flows to the IR input \eqref{etord} in eq.~\eqref{formalreconstructed} can be computed recursively 
as follows:
\begin{enumerate}
	\item We start with $\tau_0^{(0)}=0$ and $\tau_\mu^{(0)}=\mathfrak{t}\cdot t_\mu$ for non-trivial Young tableaux $\mu$,\footnote{The $t_0$-dependence is determined by the string equation, see ref.~\cite{Lee:2001mb}, and it is therefore dropped for simplicity.} 
	where $\mathfrak t$ is a convenient parameter to keep track of homogeneous powers of $t_\mu$.
	\item In each step $i>0$, we set 
	\begin{equation}
		\tau_\mu^{(i+1)}(Q,q)=\tau_\mu^{(i)}(Q,q)-c_\mu^{(i)}(Q,q),
	\end{equation}
	where the correction $c_\mu^{(i)}$ is given by
	\begin{equation}
		c_\mu^{(i)}(Q,q)=J_{\Gr(2,4)}(\tau^{(i)})|_{\cx K_+}-(1-q)-\mathfrak t\cdot t_\mu\,.
	\end{equation}
	\item We repeat until $c^{(m)}(Q,q)=\cx O(Q^{c_Q},\mathfrak t^{c_\mathfrak{t}})$ for some $m$ and some chosen cutoffs $c_Q,c_\mathfrak{t}$.\footnote{The procedure terminates at each finite order in $Q$, since for $c_\mu^{(i)}=\cx O(Q^\ell)$ we find $c_\mu^{(i+1)}=\cx O(Q^k)$, where $k\geq \ell$ with equality only for finitely many $i$'s. In other words, if a correction term $c^{(m)}$ is of order $Q^\ell$, then all subsequent correction terms leave the lower $Q$-order terms in $\tau(Q,q,t)$ unchanged.}
\end{enumerate}

\noindent In the  example we find, up to orders $Q, \mathfrak{t}^2$
\begin{equation} \label{mimaex}
	\begin{aligned}
		\tau( Q,t)=&
		Q \mathfrak{t}^2  (-t_{2} t_{2,1}-\tfrac{1}{2} t_{2}^2)\cx{O}_{0}
		+\big(\mathfrak{t}t_1+Q \mathfrak{t}^2  (t_{2} (t_{2,1}-t_{2,2})-\tfrac{1}{2} t_{2,1}^2+\tfrac{t_{2}^2}{2})\big)\cx{O}_{1}
		\\&
		+\big( \mathfrak{t}t_2+Q \mathfrak{t}^2  (\tfrac{1}{2} t_{2,1}^2-t_{2,1} t_{2,2})\big)\cx{O}_{2}
		+\big( \mathfrak{t}t_{1,1}+Q \mathfrak{t}^2 (\tfrac{1}{2} t_{2,1}^2+t_{2} t_{2,2}) \big)\cx{O}_{1,1}
		\\&
		+\big( \mathfrak{t}t_{2,1}+Q \mathfrak{t}^2 (t_{2,1} t_{2,2}-\tfrac{1}{2} t_{2,1}^2)\big) \cx{O}_{2,1}
		+ \mathfrak{t}t_{2,2} \cx{O}_{2,2}\,,
	\end{aligned}
\end{equation}
illustrating that the 3d mirror map is highly non-trivial even for canonical CS levels. The correlator terms of the deformed $I$-function has the form
\begin{equation}\label{Jcorr}
	J_{corr.}(\tord)=\sum_{\mu\in \cx B_{(M,N)}} J_{corr.}^{\mu}(\tord) \;\cx O_\mu(1-q^{-\eps}).
\end{equation}
Suppressing some notation, the coefficients are up to print-friendly orders $Q, \mathfrak{t}^2$:
{\allowdisplaybreaks
\scriptsize
	\begin{align*}	
		J^{0}\!=&
		Q \Big(
		\tfrac{-q-1}{(q-1)^3}
		+
		\mathfrak{t} \big(\tfrac{t_{1,1}}{(q-1)^2}+\tfrac{t_{2,1}}{1-q}+\tfrac{(-q-1) t_{1}}{(q-1)^3}+\tfrac{t_{2}}{(q-1)^2}\big)
		+
		\mathfrak{t}^2 \big(t_{1} (\tfrac{t_{1,1}}{(q-1)^2}+\tfrac{t_{2,1}}{1-q}+\tfrac{t_{2}}{(q-1)^2})+\tfrac{t_{2} t_{1,1}}{1-q}+\tfrac{(-q-1) t_{1}^2}{2 (q-1)^3}\big)\Big),
		\\
		J^{1}\!=&
		Q \Big(
		-
		\tfrac{2 (q^2+2 q)}{(q-1)^4}
		+
		\mathfrak{t} \big(\tfrac{q t_{1,1}}{(q-1)^3}+\tfrac{t_{2,2}}{1-q}+\tfrac{(-q^2-3 q) t_{1}}{(q-1)^4}+\tfrac{q t_{2}}{(q-1)^3}\big)
		\\&
		+
		\mathfrak{t}^2 \big(t_{1} (\tfrac{q t_{2,1}}{(q-1)^2}+\tfrac{t_{2,2}}{1-q})+\tfrac{t_{1,1}^2}{2 (q-1)^2}+\tfrac{t_{2} t_{2,1}}{1-q}+\tfrac{t_{1,1} t_{2,1}}{1-q}-\tfrac{q t_{1}^2}{(q-1)^4}+\tfrac{t_{2}^2}{2 (q-1)^2}\big)\Big),
		\\
		J^{2}\!=&
		\tfrac{\mathfrak{t}^2 t_{1}^2}{2 (1-q)}
		+
		Q \Big(
		+\tfrac{-3 q^3-8 q^2+q}{(q-1)^5}
		+
		\mathfrak{t} \big(\tfrac{q t_{1,1}}{(q-1)^3}+\tfrac{q t_{2,1}}{(q-1)^3}-\tfrac{q t_{2,2}}{(q-1)^2}+\tfrac{(-q^3-4 q^2+q) t_{1}}{(q-1)^5}+\tfrac{q t_{2}}{(q-1)^3}\big)
		\\&
		+
		\mathfrak{t}^2 \big(t_{1} (-\tfrac{q t_{1,1}}{(q-1)^4}+\tfrac{q t_{2,1}}{(q-1)^3}-\tfrac{q t_{2}}{(q-1)^4})+\tfrac{q t_{1,1}^2}{2 (q-1)^3}-\tfrac{t_{2,1}^2}{2 (q-1)}+\tfrac{t_{1,1} t_{2,2}}{1-q}-\tfrac{q t_{1}^2}{2 (q-1)^4}+\tfrac{q t_{2}^2}{2 (q-1)^3}\big)\Big),
		\\
		J^{1,1}\!=&
		\tfrac{\mathfrak{t}^2 t_{1}^2}{2 (1-q)}
		+
		Q \Big(
		+\tfrac{-3 q^3-8 q^2+q}{(q-1)^5}
		+
		\mathfrak{t} \big(\tfrac{q t_{1,1}}{(q-1)^3}+\tfrac{q t_{2,1}}{(q-1)^3}-\tfrac{q t_{2,2}}{(q-1)^2}+\tfrac{(-q^3-4 q^2+q) t_{1}}{(q-1)^5}+\tfrac{q t_{2}}{(q-1)^3}\big)
		\\&
		+
		\mathfrak{t}^2 \big(t_{1} (-\tfrac{q t_{1,1}}{(q-1)^4}+\tfrac{q t_{2,1}}{(q-1)^3}-\tfrac{q t_{2}}{(q-1)^4})+\tfrac{q t_{1,1}^2}{2 (q-1)^3}-\tfrac{t_{2,1}^2}{2 (q-1)}+\tfrac{t_{2} t_{2,2}}{1-q}-\tfrac{q t_{1}^2}{2 (q-1)^4}+\tfrac{q t_{2}^2}{2 (q-1)^3}\big)\Big),
		\\
		J^{2,1}\!=&
		\mathfrak{t}^2 (t_{1} (\tfrac{t_{1,1}}{1-q}+\tfrac{t_{2}}{1-q})+\tfrac{t_{1}^2}{2 (q-1)})
		+
		Q \Big(\tfrac{-5 q^4-19 q^3+3 q^2+q}{(q-1)^6}
		\\&
		+
		\mathfrak{t} \big(\tfrac{(q^3-4 q^2-q) t_{1,1}}{(q-1)^5}+\tfrac{(3 q^2+q) t_{2,1}}{(q-1)^4}-\tfrac{q (q+1) t_{2,2}}{(q-1)^3}+\tfrac{(-q^3-6 q^2-q) t_{1}}{(q-1)^5}+\tfrac{(q^3-4 q^2-q) t_{2}}{(q-1)^5}\big)
		\\&
		+
		\mathfrak{t}^2 \big(t_{1} (\tfrac{(-3 q^2-q) t_{1,1}}{(q-1)^5}+\tfrac{q (q+1) t_{2,1}}{(q-1)^4}+\tfrac{(-3 q^2-q) t_{2}}{(q-1)^5})+\tfrac{q (q+1) t_{1,1}^2}{2 (q-1)^4}+\tfrac{t_{2,1}^2}{2 (q-1)}+\tfrac{t_{2,1} t_{2,2}}{1-q}+\tfrac{(q^3+6 q^2+q) t_{1}^2}{2 (q-1)^6}+\tfrac{q (q+1) t_{2}^2}{2 (q-1)^4}\big)\Big),
		\\
		J^{2,2}\!=&
		\mathfrak{t}^2 \big(\tfrac{t_{1,1}^2}{2 (1-q)}+\tfrac{t_{1} t_{2,1}}{1-q}+\tfrac{t_{2}^2}{2 (1-q)}\big)
		+
		Q \Big(
		-
		\tfrac{2 (3 q^4+14 q^3+3 q^2)}{(q-1)^6}
		\\&
		+
		\mathfrak{t} (\tfrac{(q^4-8 q^3-3 q^2) t_{1,1}}{(q-1)^6}+\tfrac{2 (2 q^3+q^2) t_{2,1}}{(q-1)^5}+\tfrac{(-q^3-q^2) t_{2,2}}{(q-1)^4}+\tfrac{(-q^5-3 q^4+19 q^3+5 q^2) t_{1}}{(q-1)^7}+\tfrac{(q^4-8 q^3-3 q^2) t_{2}}{(q-1)^6})
		\\&
		+
		\mathfrak{t}^2 (t_{1} (-\tfrac{2 (2 q^3+q^2) t_{1,1}}{(q-1)^6}+\tfrac{(q^3+q^2) t_{2,1}}{(q-1)^5}-\tfrac{2 (2 q^3+q^2) t_{2}}{(q-1)^6})+\tfrac{(q^3+q^2) t_{1,1}^2}{2 (q-1)^5}+\tfrac{(q^4+7 q^3+2 q^2) t_{1}^2}{(q-1)^7}+\tfrac{(q^3+q^2) t_{2}^2}{2 (q-1)^5})\Big).
	\end{align*}
}
The next non-trivial example of $\Gr(2,5)\simeq \Gr(3,5)$ is presented in app.~\ref{app:Gr25Pert}.

\subsection{Topological correlators on the sphere with deformations}
The product of line operators in the 3d gauge theory defines a quantum deformation of the tensor product $\otimes$ on vector bundles $E,F$ over $X$,
\be
\cx O_\mu * \cx O_\nu = \cx O_\mu \otimes \cx O_\nu +O(Q)\,.
\ee 
In the IR limit, the quantum product of the theory deformed only by single trace operators is the product of ordinary quantum K-theory, which carries the structure of a commutative Frobenius algebra
\be
G(\cx O_\mu * \cx O_\nu,\cx O_\rho)=G(\cx O_\mu, \cx O_\nu*\cx O_\rho)\,.
\ee
Here $G$ is the non-constant, $Q$-dependent pairing
\be\label{gpairingbyGWpotential}
G(\cx O_\mu ,\cx O_\nu)=\p_{t_0}\p_{t_\mu}\p_{t_\nu} \cx F\,,\qquad
\cx F(Q,t) = \sum_{n=0}^\infty\sum_{d=0}^\infty \Big\langle \frac {t^n}{n!}\Big\rangle_{g=0,n,d}\,,
\ee
where $t=t_\mu\cx O_\mu$ is the deformation, and the genus zero $K$-theoretic correlators on the r.h.s.~are defined  in refs.~\cite{GivWDVV,GivLee,RZ18} as holomorphic Euler characteristics on Kontsevich's moduli space $\Mst^d$ of degree $d$ stable maps \cite{Kont}. 

From the point of the 3d gauge theory, the pairing $G(\cx O_\mu ,\cx O_\nu)=G_{\mu\nu}$ is computed by the two point function on $S^2\times S^1$ of the topologically twisted theory, which can be obtained by a localization computation in the UV theory \cite{BZ,ClK}. The 3d $tt^*$ structure of \cite{CV13} implies, that the topological sphere correlator $G_{\mu\nu}$ can be obtained by gluing two $D^2\times_qS^1$ partition functions along their $T^2$ boundaries.\footnote{This is the $tt^*$-fusion, i.e., $G_{\mu\nu}(Q,t)$ is the 3d version of the topological metric $\eta_{\mu\nu}$ in the notation of \cite{CV13}, which is known to be constant and equal to the classical pairing in the 2d limit. See also sect.~7 of ref.~\cite{JM} for a more detailed discussion. In the context of localization computations, this factorization has been first observed in ref.~\cite{Pas}.} This factorization property descends in the IR limit and the holomorphic limit of ref.~\cite{CV13} to a certain identity relating the the K-theoretic $J$-function and the topological metric $G$. This relation is a strong consistency check and can be used to compute the topological correlators of the deformed theory on $S^2\times S^1$ in the IR coordinates, which is not possible so far by localization methods without the knowledge of the mirror map~\eqref{mimaex}.

The IR image of the factorization identity that we will use is the relation 
\be\label{GTid}
G(\cx O_\mu,\cx O_\nu)(Q,t)=g(T(Q,\bar q,t) \circ \cx O_\mu,T(Q,q,t) \circ \cx O_\nu)\,,\qquad \bar q = q^{-1}\,,
\ee
obtained in ref.~\cite{IMT} from the WDVV equations of quantum K-theory \cite{GivWDVV}. The pairing appearing on the r.h.s.~is the classical holomorphic  Euler characteristic
\be\label{defgtwisted}
 g_{\mu\nu}=g(\cx O_\mu,\cx O_\nu)=\chi(X,\cx O_\mu\otimes \cx O_\nu \otimes D)=\lim_{Q\to0} G(\cx O_\mu,\cx O_\nu)\,,
\ee
which, as indicated, coincides with the classical limit of $G_{\mu\nu}=G(\cx O_\mu,\cx O_\nu)$. 
Here we have included a twist by a line bundle $D$ compared to \cite{IMT}, which is needed to cover the case of non-zero effective CS terms, corresponding to non-zero level theory of ref.~\cite{RZ18}. The operator $T$ is a map\footnote{$T$ is the 3d analogue of what is called the period matrix in the context of 2d mirror symmetry.} in  ${\rm End}(K(X))\otimes\IQ(q)[[Q,t]]$ related to the derivatives of the $J$-function by
$$\textstyle
T(Q,q,t)\circ \cx O_\mu=\frac{\p}{\p t_\mu}J(Q,q,t)\,.
$$
Note that the r.h.s.~of \eqref{GTid} apparently depends on $q$, but the l.h.s.~does not if the deformation $t(q)$ is $q$-independent; the $q$-independence of the r.h.s.~gives a strong consistency check on the result for the deformed $J$-function.

For a $q$-independent deformation $t$, one can get exact expressions in $Q$ by taking the $q \to 0$ limit of \eqref{GTid}, where
\be
G_{\mu\nu} = g(\cx O_\mu,T(Q,0,t)\cx O_\nu)\ .
\ee
 In the first entry we have used the relation $\lim_{\bar q \to \infty}T={\rm id}$, which follows from \eqref{definp}. In the second entry, only the terms with degrees $|d_{ab}|<1$ in the $I$-function \eqref{ISQKGr} survive in the $q\to0$ limit for positive CS levels in the input window. As an example we consider the case $X=\Gr(2,4)$ of the previous section at zero effective CS terms, corresponding to the level zero theory with $D=\cx O(X)$. Computing the $q\to0$ limit of the $I$-function and its first order variations, the exact result for the $n=0$ terms in the ordered basis \def\O{\mathcal{O}}$ \{\O_0,\O_1,\O_2,\O_{1,1},\O_{2,1},\O_{2,2}\}$ can be computed to\footnote{The result can also be obtained from the results of ref.~\cite{BCLM}, which gives  a nice geometric interpretation in terms of rational curves of $X$. Moreover it agrees with that of ref.~\cite{Yos19} obtained from the unperturbed $S^2\times S^1$ partition function.}
\begin{equation}\label{Gr24GPairing}
G_{\mu\nu}(t=0)=\frac1 {1-Q}\, \left(\begin{matrix}
		1 & 1 & 1 & 1 & 1 & 1 \\
		1 & 1 & 1 & 1 & 1 & Q \\
		1 & 1 & 1 & Q & Q & Q \\
		1 & 1 & Q & 1 & Q & Q \\
		1 & 1 & Q & Q & Q & Q \\
		1 & Q & Q & Q & Q & Q^2
	\end{matrix}\right)\,.
\end{equation}
The pole at $Q=1$ signals the failure of the perturbative expansion for non-positive CS  level. 
To display the perturbed expression, we expand  the sphere metric as 
\be\label{gpairingformalexpansion}
 G_{\mu\nu}(t)=\frac1{1-Q}\sum_{d,n\geq 0}G_{\mu\nu}^{(d)}Q^d\,,
\ee
with the zeroth-order in perturbations $t$ captured by \eqref{Gr24GPairing}. At first order in $t$ one finds
\bea
	G_{\mu\nu}^{(0)}|_{t^1}&=&
	\left(
	\begin{smallmatrix}
		U(t) & U(t)-t_{2,2} & 
		t_{1}+t_{2} & t_{1,1}+t_{1} & t_{1} & 0 \\
		U(t)-t_{2,2} & t_{1,1}+t_{1}+t_{2} & t_{1} & t_{1} & 
		0 & 0 \\
		t_{1}+t_{2} & t_{1} & 0 & 0 & 0 & 0 \\
		t_{1,1}+t_{1} & t_{1} & 0 & 0 & 0 & 0 \\
		t_{1} & 0 & 0 & 0 & 0 & 0 \\
		0 & 0 & 0 & 0 & 0 & 0
	\end{smallmatrix}
	\right),\nonumber\\
G_{\mu\nu}^{(1)}|_{t^1}&=&\left(
	\begin{smallmatrix}
		0 & t_{2,2} & t_{1,1}+t_{2,1}+t_{2,2} & t_{2}+t_{2,1}+t_{2,2} & U(t)-t_{1} &  U(t)-t_{2,2}\\
		t_{2,2} & t_{2,1}+t_{2,2} & U(t)-t_{1} & U(t)-t_{1} &U(t) & U(t)-t_{2,2}\\
		t_{1,1}+t_{2,1}+t_{2,2} & U(t)-t_{1} &  U(t)-t_{2,2} &  U(t) &  U(t)-t_{2,2} & t_{1}+t_{1,1} \\
		t_{2}+t_{2,1}+t_{2,2} &  U(t)-t_{1} &  U(t) &  U(t)-t_{2,2} &  U(t)-t_{2,2} & 1+t_{1}+t_{2} \\
		 U(t)-t_{1} &  U(t) &  U(t)-t_{2,2} &  U(t)-t_{2,2} & t_{1}+t_{2}+t_{1,1} & t_{1} \\
		 U(t)-t_{2,2} &  U(t)-t_{2,2} & t_{1}+t_{1,1} & t_{1}+t_{2} & t_{1} & 0
	\end{smallmatrix}
	\right) \ ,\\
G_{\mu\nu}^{(2)}|_{t^1}&=&\left(
	\begin{smallmatrix}
		0 & 0 & 0 & 0 & 0 & t_{2,2} \\
		0 & 0 & 0 & 0 & 0 & t_{2,2} \\
		0 & 0 & t_{2,2} & 0 & t_{2,2} & t_{2,1}+t_{2,2}+t_{2} \\
		0 & 0 & 0 & t_{2,2} & t_{2,2} & t_{1,1}+t_{2,1}+t_{2,2} \\
		0 & 0 & t_{2,2} & t_{2,2} & t_{2,1}+t_{2,2} & 
		U(t)-t_{1} \\
		t_{2,2} & t_{2,2} & t_{2,1}+t_{2,2}+t_{2} & t_{1,1}+t_{2,1}+t_{2,2} 
		&U(t)-t_{1} & U(t)
	\end{smallmatrix}
	\right),\quad  G_{\mu\nu}^{(3)}|_{t^1}=G_{\mu\nu}^{(4)}|_{t^1}=0\ ,\nonumber
\eea
with $U(t)=t_1+t_2+t_{1,1}+t_{2,1}+t_{2,2}$.\footnote{Note that the perturbed metric satisfies $G_{\mu\nu}(t)=\p_{t_\mu}\p_{t_\nu}G_{00}(t)$ by eq.~\eqref{gpairingbyGWpotential} and the string equation of ref.~\cite{Lee:2001mb}. Hence we can succinctly capture higher-order dependence in $(Q,t)$ by expanding the single component $G_{00}(Q,t)=\frac{1}{1-Q}\sum_{d\geq 0}G_{00}^{(d)}(t)Q^d$. We print out higher-order $G_{00}^{(d)}(t)$'s for $\Gr(2,4)$ in app. \ref{app:pertpairingGr24}
}

For non-zero effective CS terms, we have to specify the bundle $D$ in \eqref{defgtwisted} in terms of the effective CS levels $\hat \kappa_i$ in eqs.~\eqref{CSbare} and \eqref{effCS2}. $D$ is the restriction to zero degree $d=0$ of the twisting bundle of the level $\ell$ K-theory called $\cx D^{R,\ell}$ in ref.~\cite{RZ18}; in the  notation there $D=(\det \cx R)^{-\ell}$. A short computation shows that the bundle $D$ obtained from the CS terms of the 3d gauge theory depends only on the CS  level $\hat \kappa_R$ for the mixed gauge symmetry/$R$-symmetry CS term, and is given in terms of the tautological bundle as 
\be
D = (\det S)^{2\hat \kappa_R}\,.
\ee
We have verified the $q$-independence of the r.h.s.~of \eqref{GTid} for non-zero CS terms in the input window for the example.

\bigskip
\noindent{\bf Acknowledgments:} 
We would like to thank Tudor Dimofte, Gerald Dunne, Sergei Gukov, Marcos Mari\~n{}o, Leonardo Mihalcea, Eric Sharpe, and Danu Thung for discussions, correspondences and comments.
The work of H.J. is supported by the Cluster of Excellence ``Precision Physics, Fundamental Interactions and Structure of Matter'' (PRISMA+ --- EXC 2118/1) within the German Excellence Strategy (project ID 39083149).
The work of P.M. is supported by the German Excellence Cluster Origins. 
U.N. is supported by the graduate school BCGS, 
and A.T. is supported by the DFG.\\[-1cm]

\newpage
\appendix
\section{Appendix}
\subsection{Symmetric polynomials and K-theory of Grassmannian }

\subsubsection{Bases of symmetric polynomials \label{app:SymGrass}}
We recall some useful bases of the space of symmetric functions in the formal variables $z_1,\ldots, z_M$, to be identified with ($K$-theoretic) Chern roots ($\delta_a$) $x_a$ later.
The elementary symmetric polynomials $X_\ell(z_a)$ generate the ring of symmetric polynomials $\IZ[z_a]^{S_M}$, such that we have the ring isomorphism $\IZ[X_\ell] \xrightarrow{\sim} \IZ[z_a]^{S_M}$, $X_\ell\mapsto X_\ell(z_a)$. 

\paragraph{Schur polynomials}

The Schur polynomials $\sigma_\lambda(z)=\sigma_\lambda(z_1,\ldots,z_M)$, where $\lambda$ is a partition/Young diagram $\lambda=\lambda_1\geq\lambda_2\geq\ldots \geq \lambda_M$, are defined by the formula
\begin{equation}\label{defSchurPolys}
	\sigma_\lambda(z)=\frac{\det (z_a^{\lambda_b+M-b})_{ab}}{\Delta(z)},
\end{equation}
with the Vandermonde determinant $\Delta(z)$ as in eq.~\eqref{eq:defDelta}. They satisfy
\begin{equation}\label{SchurMultStr}
\sigma_\lambda(z)\sigma_\rho(z)= \sum_{\mu}c_{\lambda\rho}^\mu \sigma_\mu (z),
\end{equation}
where  $c_{\lambda\rho}^\mu$ are the Littlewood-Richardson coefficients labeled by partitions, computable by Pierri and Giambelli formulas. The Schur polynomials labeled by ``vertical'' partitions $(1,\ldots,1)$ ($\ell$-times) are precisely the elementary symmetric polynomials $X_\ell(z_a) = \sigma_{1,\ldots,1}(z_a)$, and the Schur polynomials form an integral basis for the  free $\IZ$-module $\IZ[z_a]^{S_M}$.

\paragraph{Grothendieck polynomials}
The symmetric $\beta$-Grothendieck polynomials $G_\lambda(z_a;\beta)$ are labeled by a partition/Young diagram $\lambda = (\lambda_1\ge\lambda_2\ge\ldots \ge 0)$ and a parameter $\beta$, and they can be defined in terms of a generalized Weyl formula as 
\begin{equation}
	G_\lambda(z_a; \beta) =
	\frac{1}{\Delta(z_a)} 
	\left|\begin{matrix} 
		z_1^{\lambda_1 + M-1} && z_1^{\lambda_2 + M-2}(1+\beta z_1) && \cdots && z_1^{\lambda_M}(1+\beta z_1)^{M-1}  \\ 
		z_2^{\lambda_1 + M-1} && z_2^{\lambda_2 + M-2}(1+\beta z_2) && \cdots && z_2^{\lambda_M}(1+\beta z_2)^{M-1}  \\
		\vdots && \vdots &&&& \vdots  \\ 
		z_M^{\lambda_1 + M-1} && z_M^{\lambda_2 + M-2}(1+\beta z_M) && \cdots && z_M^{\lambda_M}(1+\beta z_M)^{M-1} 
	\end{matrix} \right| \ .
\end{equation}
with $\Delta(z)$ as defined in eq.~\eqref{eq:defDelta}. For $\beta=0$ and $\beta=-1$ the $\beta$-Grothendieck polynomials~$G_\lambda(z_a; \beta)$ become the Schur polynomials $\sigma_\lambda(z_a)$ \eqref{defSchurPolys} and the Grothendieck polynomials $\mathcal{O}_\lambda(z_a)$, respectively, i.e.,
\begin{equation} \label{eq:SchurGroth}
	\sigma_\lambda(z_a) = G_\lambda(z_a; 0) \ , \qquad
	\mathcal{O}_\lambda(z_a) = G_\lambda(z_a; -1) \ .
\end{equation}
Grothendieck polynomials $\cx O_\lambda(z_a)$ are integral linear combinations of Schur polynomials $\cx O_\lambda(z_a)=\sum_{\rho}K_{\lambda\rho}\;\sigma_\rho(z_a)$ hence  form an integral basis of $\IZ[z_a]^{S_M}$ as well. They satisfy
\begin{equation}\label{GrotMultStr}
	\cx O_\lambda (z)\cx O_\rho (z)= \sum_{\mu}C_{\lambda\rho}^\mu \cx O_\mu(z),
\end{equation}
where the  structure constants $C_{\lambda\rho}^\mu$ are related to the $c_{\lambda\rho}^\mu$ by the linear transformation $K_{\rho\lambda}$, and describe the classical algebra of Wilson loops.

\subsubsection{Cohomology and characteristic classes}
\label{app:CharClasses}
The cohomology ring~$H^*(\Gr(M,N),\IZ)$ of the complex Grassmannian $\Gr(M,N)$ is generated by the  the Chern classes $c_\ell(S)$, $\ell=1,\ldots,M$, of the universal subbundle $S$ of $\Gr(M,N)$ of rank $M$. These are identified by definition with the elementary symmetric polynomials $X_\ell(-x_a)$, where $-x_a$ are the Chern roots of $S$. The relations are easily expressed in terms of Schur polynomials: assigning to the elementary symmetric polynomial $X_\ell(-x_a)$ the corresponding cohomology class defines a surjective ring homomorphism $\IZ[-x_a]^{S_M} \to H^*(\Gr(M,N),\IZ)$.  This ring homomorphism sends a Schur polynomial $\sigma_\lambda(-x_a)$, to the (Poincar\'e dual of the) Schubert cycle $\sigma_\lambda$ if the partition $\lambda$ is contained in the $M\times(N-M)$ rectangle $\cx B_{(M,N)}$ (i.e., if $\lambda_1\leq N-M$) and otherwise to~$0$. The $\dim H^*\big(\Gr(M,N)\big)=\binom{N}{M}$ Schur polynomials $\sigma_\lambda(x)$ labeled by partitions $\lambda \in \cx B_{M,N}$ form a  basis for the cohomology ring with multiplicative structure given by \eqref{SchurMultStr}.

The characteristic classes of (the tangent bundle of) $\Gr(M,N)$ can be computed in terms of the Chern roots $-x_a$ of $S$ as follows.
The tangent bundle satisfies $T\text{Gr}(M,N)\cong \operatorname{Hom}(S,Q)$, where $Q$ is the quotient bundle of rank $(N-M)$ defined by the Euler sequence $0\to S\to \underline{\mathbb C}^{N}\!\to Q\to 0$, where $\underline{\mathbb C}^{N}$ denotes the trivial bundle over $\Gr(M,N)$. The sequence implies the $K$-theory relation $
[T\text{Gr}(M,N)]
=[\operatorname{Hom}(S,\underline{\mathbb C})^{\oplus N}]-[\operatorname{Hom}(S,S)]
=N[S^*]-[S^*\otimes S]
$ (see, e.g., \cite{Galkin:2014laa}\footnote{The relation can also be derived from the symplectic quotient construction $\Gr(M,N)\cong\operatorname{Hom}(\IC^M,\IC^N)/\!/U(M)$ \cite{Martin}.}). For characteristic classes $g$ defined by multiplicative sequences $g(x)$ this implies that
\begin{equation}\label{mulcl}
	g(X)=\frac{\prod_{a=1}^M g( x_a)^N}{\prod_{a<b}^M g(x_{ab})g(x_{ba})}, \quad x_{ab}=x_a-x_b\ ,
\end{equation}
where again $x_a$ are the Chern roots of $S^*$. The right-hand side can be decomposed $g(X)=\sum_{\mu\in \cx B_{M,N}} g_\mu \sigma_\mu(x)$ for some coefficients $g_\mu$ and for Schur polynomials $\sigma_\mu$  representing cohomology classes.

\noindent For example, the total Chern class, generated multiplicatively by $g(x)=1+x$ is
\begin{equation}\label{ChernClassFormal}
	\begin{aligned}
		c(X)
		&=\frac{\prod_{a=1}^M (1+x_a)^N}{\prod_{a<b}^M (1-x_{ab}^2)}
		\\&=1+N\sigma_{1}(x)
		+({\scriptstyle\binom{N}{2}}+M-1)\sigma_{2}(x)
		+({\scriptstyle\binom{N+1}{2}}-M-1)\sigma_{1,1}(x)
		+\ldots
	\end{aligned}
\end{equation}
The map from Chern classes of a vector bundle $V$ to its Chern characters is
\begin{equation}
	\ch(V)=\operatorname{rank}(V) +c_1(V)+\tfrac 1 2 \big(c_1^2-2c_2\big)(V)
	+\tfrac 1 6 \big(c_1^3-3c_1c_2+3c_3\big)(V)
	+\ldots
\end{equation}
and we find from \eqref{ChernClassFormal} that the Chern character of (the tangent bundle of) $X=\Gr(M,N)$ is
\begin{equation}\label{ChernCharacter}
	\ch(X)=\operatorname{dim}X 
	+N\sigma_1(x)
	+(1-M+\tfrac N 2)\sigma_2 (x)
	+(1+M-\tfrac N 2)\sigma_{1,1} (x)
	+\ldots \ ,
\end{equation}
from which eqs.~\eqref{ChernCharacters12} follow using the notation of fn.~\ref{footnotesum}.

Two other important multiplicative classes appearing in the result \eqref{foldingf} are the Todd class $\Td(X)$ and the $q$-Gamma class $\Gamma_q(X)$ defined by the series of the functions
\be\label{ToddGammaDef}
\Td(X): \ g(x)=\tfrac{x}{1-e^{-x}}\,,\qquad \Gamma_q(X):\ g(x)=\Gamma_q(1+x)\,,
\ee
where $\Gamma_q(x)$ is defined in \eqref{defqGamma}.

\subsubsection{K-theoretic classes} \label{app:KtheoryGr}
As in the case of $H^*(X)$, $X=\Gr(M,N)$ we may use the K-theoretic Chern roots $ [S]=P_1+\ldots+P_M$ to describe $K(X)$. 
Defining $\delta_a = 1 - P_a$, the symmetric polynomials in the variables $\delta_a$ are elements in the K-theory ring $K(X)$  and --- due to the Chern character homomorphism which sends $\ch(P_a)=e^{-x_a}$ --- Chern characters of coherent sheaves on $\Gr(M,N)$. As in the cohomological case, the symmetric polynomials in $\delta_a$ in fact generate $K(X)$ and the relations are described as follows: the (surjective) ring homomorphism $\IZ[\delta_a]^{S_M} \to K(\Gr(M,N))$ sends the Grothendieck polynomial $\mathcal{O}_\lambda(\delta_a)$  to the structure sheaf $\mathcal{O}_{\sigma_\lambda}$ of the Schubert cycle $\sigma_\lambda$ if the partition $\lambda$ is contained in the rectangle $M\times(N-M)$ and to $0$ otherwise \cite{MR686357}.\footnote{The Schur polynomials $\sigma_\lambda(\delta_a)$ with $\lambda$ in the $M\times(N-M)$ rectangle $\cx B_{M,N}$ are mapped to coherent sheaves $\mathcal{E}_{\sigma_\lambda}$, which are locally free on the Schubert cycle $\sigma_\lambda$. The Schur polynomials $\sigma_\lambda(\delta_a)$ form an integral basis for $K(\Gr(M,N))$ as well.} The $\dim K\big(\Gr(M,N)\big)=\binom{N}{M}$ Grothendieck polynomials $\cx O_\lambda(\delta)$, $\lambda\in \cx B_{M,N}$ form a basis of $K(X)$ with classical multiplicative structure given by eq.~\eqref{GrotMultStr}.

\subsection{Relations among quantum K-theory generators} \label{app:QKrels}
Let us now show the claims made in Proposal~1 in Sect.~\ref{sec:canCST}. First, we show that the definition of the ideal $\mathcal{J}_{M,N}$ in terms of the generators~\eqref{eq:Yrels} is equivalent to the ideal~\eqref{eq:JGroth} in terms of the Grothendieck polynomials. By explicitly inserting into the generators $Y_\ell(Q)$ the expression $E_a$ of eq.~\eqref{deq1}, one finds (using standard properties of the determinant)
\begin{equation} \label{eq:Yrel2}
\begin{aligned}
  Y_\ell(Q) =& \frac{\prod_{b=1}^M (1-\delta_b)}{\Delta(\delta_a)} 
  \left|\begin{matrix} 
    \delta_1^N(1-\delta_1)^{-1} &  \delta_1^{M-1} & \cdots & \delta_1^{\ell} & \delta_1^{\ell-2} & \cdots &  \delta_1 & 1 \\ 
    \vdots & \vdots && \vdots & \vdots && \vdots & \vdots \\ 
    \delta_M^N(1-\delta_M)^{-1} &  \delta_M^{M-1} & \cdots & \delta_M^{\ell} & \delta_M^{\ell-2} & \cdots &  \delta_M & 1 
   \end{matrix} \right| \\
   &\qquad\qquad +\frac{Q}{\Delta(\delta_a)}
     \left|\begin{matrix} 
    (1-\delta_1)^{M-1} &  \delta_1^{M-1} & \cdots & \delta_1^{\ell} & \delta_1^{\ell-2} & \cdots &  \delta_1 & 1 \\ 
    \vdots & \vdots && \vdots & \vdots && \vdots & \vdots \\ 
   (1-\delta_M)^{M-1} &  \delta_M^{M-1} & \cdots & \delta_M^{\ell} & \delta_M^{\ell-2} & \cdots &  \delta_M & 1 
   \end{matrix} \right|\\[2ex]
   =&  \frac{1}{\Delta(\delta_a)}
   \left|\begin{matrix} 
    \delta_1^N & (1-\delta_1)\delta_1^{M-1} &  \cdots & (1-\delta_1)\delta_1^{\ell} & (1-\delta_1)\delta_1^{\ell-2} & \cdots & (1-\delta_1)\\ 
    \vdots & \vdots && \vdots & \vdots && \vdots\\ 
    \delta_M^N & (1-\delta_M)\delta_M^{M-1}  & \cdots & (1-\delta_M)\delta_M^{\ell} & (1-\delta_M)\delta_M^{\ell-2} & \cdots & (1-\delta_M)
   \end{matrix} \right| \\
   &\qquad\qquad -(-1)^{M} \binom{M-1}{\ell-1} Q \ .
\end{aligned}
\end{equation}
In the last step the factor $(1-\delta_a)^{M-1}$ in the first column of the determinant of the $Q$-dependent term is expanded out, and it becomes proportional to the Vandermonde determinant, which cancels the prefactor $\Delta(\delta_a)^{-1}$. We now concentrate on the $Q$-independent part of $Y_\ell(Q)$. For the specific polynomial $Y_1$ we find 
\begin{equation} \label{eq:Yrel3}
\begin{aligned}
  Y_1(0)=&\frac{1}{\Delta(\delta_a)} \left| 
    \begin{matrix} 
       \delta_1^N & (1-\delta_1) \delta_1^{M-1} & (1-\delta_1) \delta_1^{M-2}& \cdots & (1-\delta_1) \delta_1 \\
       \vdots & \vdots &\vdots && \vdots  \\ 
       \delta_M^N & (1-\delta_M) \delta_M^{M-1} & (1-\delta_M) \delta_M^{M-2} & \cdots & (1-\delta_M) \delta_M
    \end{matrix} \right| \\[1ex]
   = & \frac{1}{\Delta(\delta_a)} \left| 
   \begin{matrix} 
       \delta_1^N & (1-\delta_1) \delta_1^{M-1} & (1-\delta_1)^2 \delta_1^{M-2}& \cdots & (1-\delta_1)^{M-1} \delta_1 \\
       \vdots & \vdots &\vdots && \vdots  \\ 
       \delta_M^N & (1-\delta_M) \delta_M^{M-1} & (1-\delta_M)^2 \delta_M^{M-2} & \cdots & (1-\delta_M)^{M-1} \delta_M
    \end{matrix} \right| \\[1ex]
    = & \mathcal{O}_{N-M+1,\smallunderbrace{\scriptstyle1,\ldots,1}_{M-1}}(\delta_a) \ , 
\end{aligned}
\end{equation} 
where the second line is obtained by successively subtracting columns from each other, such that in the last step the expression can be identified with the stated Grothendieck polynomial according to definition~\eqref{eq:SchurGroth}. For the general polynomials~$Y_\ell(Q)$ we arrive via induction from $Y_{\ell-1}(0)$ to $Y_{\ell}(0)$ at the identity
\begin{equation}
  Y_\ell(0) = \mathcal{O}_{N-M+1,\smallunderbrace{\scriptstyle1,\ldots,1}_{M-\ell}}(\delta_a)
  + \sum_{k=1}^{\ell-1} \binom{M-\ell+k-1}{k}  \mathcal{O}_{N-M+1,\smallunderbrace{\scriptstyle1,\ldots,1}_{M-\ell+k}}(\delta_a) \ .
\end{equation}   
Inserting this result into eq.~\eqref{eq:Yrel2} and observing  $\sum_{k=0}^{\ell-1} \binom{M-\ell+k-1}{k} = \binom{M-1}{\ell-1}$, one finds that the generators the Grothendieck generators~\eqref{eq:JGroth} are readily obtained from the generators $Y_\ell(Q)$ by a linear transformation of determinant one. This shows that both formulations of the ideal $\mathcal{J}_{M,N}$ stated in Proposal~1 are indeed equivalent.

\bigskip

Second, we want to show that the ideals $\mathcal{I}_{M,N}$ and $\mathcal{J}_{M,N}$ --- respectively given in eq.~\eqref{eq:defI} and in Proposal 1 --- are actually the same. To this end we consider any element $r \in \mathcal{J}_{M,N}$ that enjoys the expansion $r = \sum_\ell \alpha_\ell(\delta_a) Y_\ell(Q)$ with $\alpha_\ell \in \IZ[\delta_a,Q]^{S_M}$. Then with eq.~\eqref{eq:Yrels} we find that
\begin{equation}
  \Delta(\delta_a) r = \sum_{\ell} \alpha_\ell(\delta_a) 
    \left|\begin{matrix} 
    E_1 & \delta_1^{M-1} & \cdots & \delta_1^{\ell} &  & \delta_1^{\ell-2} &  \cdots &  \delta_1 & 1   \\ 
    E_2 & \delta_2^{M-1} & \cdots & \delta_2^{\ell} &  & \delta_2^{\ell-2} &  \cdots &  \delta_2 & 1   \\ 
    \vdots & \vdots && \vdots && \vdots && \vdots & \vdots \\ 
    E_M & \delta_M^{M-1} & \cdots & \delta_M^{\ell} &  & \delta_M^{\ell-2} &  \cdots &  \delta_M & 1 
   \end{matrix} \right| \ .
\end{equation}
Expanding  the determinants in the first column yields the expansion $\Delta(\delta_a) r = \sum_b \beta_b(\delta_a) E_b(\delta_b)$, which is in the ideal  $\left.\mathcal{I}^\text{Ab}_{M,N}\right|_{Q\equiv Q_1 = \ldots = Q_M}$. Therefore, we find that $r \in \mathcal{I}_{M,N}$ according to eq.~\eqref{eq:defI}, which implies $\mathcal{J}_{M,N} \subset \mathcal{I}_{M,N}$.

Next, we consider any element $s \in \mathcal{I}_{M,N}$. Then we have the expansion $\Delta(\delta_a) s = \sum_b \beta_b(\delta_a) E_b(\delta_a)$. Note that $\Delta(\delta_a)s$ is anti-symmetric in the variables $\delta_1,\ldots,\delta_M$, whereas the coefficients $\beta_b(\delta_a)$ are anti-symmetric polynomials in $\delta_1,\ldots,\widehat{\delta_b},\ldots,\delta_M$. These symmetry properties ensure that the polynomial $\Delta(\delta_a) s$ admits the expansion
\begin{equation}
  \Delta(\delta_a) s =\!\!\!\!\!\!  \sum_{\substack{0\le i_1 < i_2 < \ldots < i_{M-1}\\ \text{finite}}}\!\!\!\!\!\! \gamma_{i_1,\ldots,i_{M-1}}(\delta_a) 
   \sum_{b=1}^M (-1)^b 
   \left|\begin{matrix} 
     \delta_1^{i_{M-1}} & \delta_1^{i_{M-2}} & \cdots & \delta_1^{i_2} &  \delta_1^{i_1}   \\ 
     \vdots & \vdots && \vdots & \vdots \\
     \delta_{b-1}^{i_{M-1}} & \delta_{b-1}^{i_{M-2}} & \cdots & \delta_{b-1}^{i_2} &  \delta_{b-1}^{i_1} \\
     \delta_{b+1}^{i_{M-1}} & \delta_{b+1}^{i_{M-2}} & \cdots & \delta_{b+1}^{i_2} &  \delta_{b+1}^{i_1} \\
     \vdots & \vdots && \vdots & \vdots \\
    \delta_M^{i_{M-1}} & \delta_M^{i_{M-2}} & \cdots & \delta_M^{i_2} &  \delta_M^{i_1} 
   \end{matrix} \right| \, E_b(\delta_a) \ ,
\end{equation} 
where the (finitely many) coefficient polynomials $\gamma_{i_1,\ldots,i_{M_1}}(\delta_a)$ are symmetric in $\delta_1,\ldots,\delta_M$. Therefore, we can express the polynomial $s$ as
\begin{equation} \label{eq:sExp1}
  s = \!\!\!\!\!\!  \sum_{\substack{0\le i_1 < i_2 < \ldots < i_{M-1}\\ \text{finite}}}\!\!\!\!\!\! \frac{\gamma_{i_1,\ldots,i_{M-1}}(\delta_a)}{\Delta(\delta_a)} 
  \left| \begin{matrix}
    E_1 & \delta_1^{i_{M-1}} & \delta_1^{i_{M-2}} & \cdots & \delta_1^{i_2} &  \delta_1^{i_1}   \\
    \vdots &\vdots & \vdots && \vdots & \vdots \\
    E_M & \delta_M^{i_{M-1}} & \delta_M^{i_{M-2}} & \cdots & \delta_M^{i_2} &  \delta_M^{i_1} 
    \end{matrix} \right| \ .
\end{equation}
Now we recall that the elementary symmetric polynomials $X_\ell(\delta_a)$ are generated by the product
\begin{equation}
  \prod_{c=1}^M (t - \delta_c) = t^M - X_1(\delta_a) t^{M-1} + X_2(\delta_a) t^{M-2} - \ldots + (-1)^M X_M(\delta_a) \ .
\end{equation}
Since the left-hand side vanishes for $t=\delta_b$, we have the identity
\begin{equation}
  \delta_b^M = X_1(\delta_a)  \delta_b^{M-1} - \ldots + (-1)^{M-1} X_M(\delta_a) \ , \quad b=1,\ldots,M \ .
\end{equation}
Thus, we can rewrite any monomial $\delta_b^K$ with $K \ge M$ as a linear combination of the monomials $1, \delta_b, \delta_b^2, \ldots, \delta_b^{M-1}$ with coefficients in the ring of symmetric polynomials in $\delta_1,\ldots,\delta_M$. Using the multi-linearity of the determinant, we can reduce all monomials~$\delta_b^{i_\ell}$ appearing in the columns of the determinant in eq.~\eqref{eq:sExp1} to degrees smaller than $M$. Therefore, the polynomial $s$ has an expansion of the form
\begin{equation} \label{eq:sExp2}
\begin{aligned}
  s &= \!\!\!\!\!\!  \sum_{0\le i_1 < i_2 < \ldots < i_{M-1}<M}\!\!\!\!\!\! \frac{{\tilde\gamma}_{i_1,\ldots,i_{M-1}}(\delta_a)}{\Delta(\delta_a)} 
  \left| \begin{matrix}
    E_1 & \delta_1^{i_{M-1}} & \delta_1^{i_{M-2}} & \cdots & \delta_1^{i_2} &  \delta_1^{i_1}   \\
    \vdots &\vdots & \vdots && \vdots & \vdots \\
    E_M & \delta_M^{i_{M-1}} & \delta_M^{i_{M-2}} & \cdots & \delta_M^{i_2} &  \delta_M^{i_1} 
    \end{matrix} \right| \ .\\[2ex]
    &= \sum_{\ell=1}^M  \tilde\gamma_{0,1,\ldots,\widehat{\ell-1},\ldots,M-1}(\delta_a)\,Y_\ell(\delta_a) \ ,
\end{aligned}    
\end{equation}
with symmetric polynomials $\tilde\gamma_{0,1,\ldots,\widehat{\ell-1},\ldots,M-1}(\delta_a)$. Therefore, we find that $s \in \mathcal{I}_{M,N}$ and hence that $\mathcal{I}_{M,N} \subset \mathcal{J}_{M,N}$. Altogether, this asserts that the ideals $\mathcal{I}_{M,N}$ and $\mathcal{J}_{M,N}$ are identical.

\subsection{Examples of Wilson line algebras outside the geometric window} \label{app:ExNotInWind}
To illustrate the richness of the structure of the Wilson line algebras arising from distinct choices of CS levels, we detail here two examples of Wilson line algebras based on the 3d gauge theory associated to the Grassmannian $\Gr(2,4)$ that have CS level outside the geometric window:

\bigskip
\bigskip
As our first example, we consider the CS levels for the parameters $\alpha=-1$, $\beta=\frac 12$, $\gamma=0$. The resulting Wilson line algebra $\cx A_{(2,4)}^{\hat \kappa}$ is freely generated as $\mathbb{Z}[Q]$-module of dimension~$\binom52$, and an explicit basis of Wilson line generators is given in terms of the Schur polynomials $\sigma_{\lambda_1,\lambda_2}$ with partitions $3\ge \lambda_1 \ge \lambda_2 \ge 0$. Compared to CS level in the geometric window (c.f., the end of sect.~\ref{sec:GenCS}), there are four additional generators, namely $\sigma_3$, $\sigma_{3,1}$, $\sigma_{3,2}$, and $\sigma_{3,3}$. The products in this Wilson line algebra are given by 
\begingroup%
\allowdisplaybreaks
{\footnotesize
\begin{align*}
 \sigma _1^2&=\sigma _2+\sigma _{1,1}\ , \qquad
 \sigma _1 \sigma _2=\sigma _3+\sigma _{2,1}\ ,\qquad
 \sigma _1 \sigma _{1,1}=\sigma _{2,1}\ ,\qquad
 \sigma _1 \sigma _3=\sigma _3+\sigma _{3,1}+Q\ ,\\ 
 \sigma _1 \sigma _{2,1}&=\sigma _{2,2}+\sigma _{3,1}\ ,\qquad
 \sigma _1 \sigma _{3,1}=\sigma _{3,1}+\sigma _{3,2}+Q \left(\sigma_1-1\right) \ , \qquad
 \sigma _1 \sigma _{2,2}=\sigma _{3,2}\ ,\\
 \sigma _1 \sigma _{3,2}&=\sigma _{3,2}+\sigma _{3,3}+Q \left(\sigma _2-\sigma _1\right)\ ,\qquad
 \sigma _1 \sigma _{3,3}=\sigma _{3,3}+Q \left(\sigma _3-\sigma _2\right)\ ,\\
 \sigma _2^2&=\sigma_3+\sigma _{2,2}+\sigma _{3,1}+Q\ ,\qquad
 \sigma _2 \sigma _{1,1}=\sigma _{3,1} \ ,\qquad
 \sigma _2 \sigma _3=\sigma _3+\sigma _{3,1}+\sigma _{3,2} + Q \left(\sigma _1+1\right)\ ,\\
 \sigma _2 \sigma _{2,1}&=\sigma _{3,1}+\sigma _{3,2}+Q \left(\sigma _1-1\right)\ ,\qquad
 \sigma _2 \sigma _{3,1}=\sigma _{3,1}+\sigma _{3,2}+\sigma _{3,3}+Q\left(\sigma _2+\sigma _{1,1}-1\right)\ ,\\
 \sigma _2 \sigma _{2,2}&=\sigma _{3,2}+Q \left(\sigma _2-\sigma _1\right)\ ,\qquad
 \sigma _2 \sigma _{3,2}=\sigma_{3,3}+\sigma _{3,2}-Q \left(\sigma _1-\sigma _3+\sigma _{1,1}-\sigma _{2,1}\right)\ ,\\
 \sigma _2 \sigma _{3,3}&=\sigma _{3,3}-Q \left(\sigma _2-\sigma _3+\sigma _{2,1}-\sigma _{3,1}\right)\ ,\qquad
 \sigma _{1,1}^2=\sigma _{2,2}\ ,\qquad
 \sigma _3 \sigma _{1,1}=\sigma _{3,1}+Q \left(\sigma _1-1\right)\ ,\\
 \sigma _{1,1} \sigma _{2,1}&=\sigma _{3,2}\ , \qquad
 \sigma _{1,1}\sigma _{3,1}=\sigma _{3,2}+Q \left(\sigma _2-\sigma _1\right)\ ,\qquad
 \sigma _{1,1} \sigma _{2,2}=\sigma _{3,3}\ ,\\
 \sigma _{1,1} \sigma _{3,2}&=\sigma _{3,3}+Q \left(\sigma _3-\sigma _2\right)\ ,\qquad
 \sigma _{1,1} \sigma _{3,3}=Q^2\ ,\\
 \sigma _3^2&=\sigma _3+\sigma _{3,1}+\sigma _{3,2}+\sigma _{3,3}+Q \left(\sigma _1+\sigma_2+1\right)\ ,\qquad
 \sigma _3 \sigma _{2,1}=\sigma _{3,1}+\sigma _{3,2}+Q \left(\sigma _2+\sigma _{1,1}-1\right)\ ,\\
 \sigma _3 \sigma _{3,1}&=\sigma _{3,1}+\sigma_{3,2}+\sigma _{3,3}+Q \left(\sigma _3+\sigma _{1,1}+\sigma _{2,1}-1\right)\ ,\qquad
 \sigma _3 \sigma _{2,2}=\sigma _{3,2}-Q \left(\sigma _1-\sigma _2+\sigma _{1,1}-\sigma _{2,1}\right) \ ,\\
 \sigma _3 \sigma _{3,2}&=\sigma _{3,2}+\sigma _{3,3}-Q \left(\sigma _1-\sigma _3+\sigma _{1,1}-\sigma _{2,2}-\sigma _{3,1}\right)\ ,\\
 \sigma _3\sigma _{3,3}&=\sigma _{3,3}-Q \left(\sigma _2-\sigma _3+\sigma _{2,1}+\sigma _{2,2}-\sigma _{3,1}-\sigma _{3,2}\right)\ ,\qquad
 \sigma _{2,1}^2=\sigma _{3,2}+\sigma _{3,3}+Q \left(\sigma _2-\sigma _1\right)\ ,\\
 \sigma _{2,1} \sigma _{3,1}&=\sigma _{3,2}+\sigma _{3,3}-Q \left(\sigma _1-\sigma _3+\sigma_{1,1}-\sigma _{2,1}\right)\ ,\qquad
 \sigma _{2,1} \sigma _{2,2}=\sigma _{3,3} + Q \left(\sigma _3-\sigma _2\right)\ ,\\
 \sigma _{2,1} \sigma _{3,2}&=\sigma _{3,3}+Q\left(-\sigma _2+\sigma _3-\sigma _{2,1}+\sigma _{3,1}\right)+Q^2\ ,\qquad
 \sigma _{2,1} \sigma_{3,3}=Q^2 \sigma _1\ ,\\
 \sigma _{3,1}^2&=\sigma _{3,2}+\sigma _{3,3}+Q\left(-\sigma _1+\sigma _3-\sigma _{1,1}+\sigma _{2,2}+\sigma _{3,1}\right)+Q^2\ ,\\
 \sigma _{2,2} \sigma _{3,1}&=\sigma _{3,3}-Q \left(\sigma _2-\sigma _3+\sigma _{2,1}-\sigma _{3,1}\right)\ ,\\
 \sigma_{3,1} \sigma _{3,2}&=\sigma _{3,3}-Q\left(\sigma _2-\sigma _3+\sigma _{2,1}+\sigma _{2,2}-\sigma _{3,1}-\sigma _{3,2}\right)+Q^2\sigma _1\ ,\\
 \sigma _{3,1} \sigma _{3,3}&=Q^2 \sigma _2 \ ,\qquad
 \sigma _{2,2}^2=Q^2\ ,\qquad
 \sigma _{2,2} \sigma _{3,2}=Q^2 \sigma _1\ ,\qquad
 \sigma _{2,2}\sigma _{3,3}=Q^2 \sigma _{1,1}\ ,\\
 \sigma _{3,2}^2&=Q^2 \left(\sigma _2+\sigma _{1,1}\right)\ ,\qquad
 \sigma _{3,2} \sigma _{3,3}=Q^2 \sigma _{2,1}\ ,\qquad
 \sigma _{3,3}^2=Q^2 \sigma _{2,2} \ .
\end{align*}}%
\endgroup

\bigskip
\bigskip
As our second example, we consider the Wilson line algebra $\cx A_{(2,4)}^{\hat \kappa}$ for CS levels arising from $\alpha=0$,  $\beta=0$, $\gamma=-2$. The resulting Wilson line algebra is freely generated in terms of a basis of dimension $\binom42+1$. In addition to the generators $\sigma_\lambda(\delta_a)$ with $\lambda=(\lambda_1,\lambda_2)$ in the range $2\ge \lambda_1 \ge \lambda_2 \ge 0$ (which are familiar from the CS level in the geometric window), there is one additional generator $\rho$ that we choose to be
\begin{equation}
  \rho = \frac14 \sigma_3 + \frac14 Q \sigma_1 +\frac12 Q \ .
\end{equation}
In terms of these generators the multiplication becomes:
\begin{equation*}
{\footnotesize
\begin{aligned}
 \sigma _1^2&=\sigma _{1,1}+\sigma _2 \ ,  \qquad
 \sigma _1 \sigma _2=\sigma _{2,1}+4 \rho -Q \left(\sigma _1+2\right)\ ,\qquad
 \sigma _1 \sigma _{1,1}=\sigma _{2,1}\ ,\\
 \sigma _1 \sigma _{2,1}&=\sigma _{2,2}+3 \rho -Q \left(\sigma _1+2\right)\ ,\qquad
 \sigma _1 \sigma _{2,2}=2 \rho -Q\left(\sigma _1+2\right)\ ,\\
 \sigma _1\rho&=2 \rho+Q \left(\sigma _{1,1}+\sigma _{2,1}-\sigma _1-\sigma _2-2\right) \ ,\qquad
 \sigma _2^2=\sigma _{2,2}+8 \rho + Q \left(3 \sigma _{1,1}+4 \sigma _{2,1}-6 \sigma _1-5 \sigma _2-8\right) \ ,\\
 \sigma _2\sigma _{1,1}&=3 \rho -Q\left(\sigma _1+2\right)\ ,\qquad
 \sigma _2\sigma _{2,1}=6 \rho+Q \left(2 \sigma _{1,1}+3 \sigma _{2,1}-5 \sigma _1-4 \sigma _2-6\right) \ ,\\
 \sigma _2\sigma _{2,2}&=3 \rho+ Q \left(\sigma _{1,1}+2 \sigma _{2,1}-3 \sigma _1-3 \sigma _2-3\right) \ ,\qquad
 \sigma _2\rho=3 \rho+Q \left(\sigma_{2,1}+\sigma _{2,2}-\rho -3 \sigma _1-2 \sigma _2-3\right) \ ,\\
 \sigma _{1,1}^2&=\sigma _{2,2} \ ,\qquad
 \sigma _{1,1}\sigma _{2,1}=2 \rho -Q \left(\sigma _1+2\right)\ ,\qquad
 \sigma _{1,1}\sigma _{2,2}=\rho -Q \left(\sigma _1+1\right)\ ,\\
 \sigma _{1,1}\rho&=\rho+Q \left(\sigma_{1,1}+\sigma _{2,1}-\sigma _1-\sigma _2-1\right)\ ,\qquad
 \sigma _{2,1}^2=4\rho +Q \left(\sigma _{1,1}+2 \sigma _{2,1}-4 \sigma _1-3 \sigma _2-4\right) \ ,\\
 \sigma _{2,1}\sigma _{2,2}&=2 \rho + Q \left(\sigma _{2,1}-2 \sigma _1-2 \sigma _2-2\right) \ ,\qquad
 \sigma _{2,1}\rho=2 \rho+Q \left(\sigma _{2,1}+\sigma _{2,2}-\rho -2 \sigma _1-2 \sigma _2-2\right) \ ,\\
 \sigma _{2,2}^2&=\rho -Q \left(\sigma _1+\sigma _2+1\right)\ ,\qquad
 \sigma _{2,2}\rho=\rho+Q \left(\sigma _{2,2}-\rho -\sigma _1-\sigma _2-1\right) \ ,\\
 \rho^2&=\rho -Q \left(\sigma _1+\sigma_2+1\right) \ .
\end{aligned}}
\end{equation*}

\subsection{Partition functions, indices and $q$-series}
\subsubsection{Special functions \label{app:sf}}
In this appendix subsection we recall some basic functions. The $q$-Pochhammer symbol is defined by
\begin{equation}\label{qpochdefinition}
	(z;q)_\infty\coloneqq 
	\begin{cases}
		\prod_{r=0}^\infty (1-q^r z)&,\quad |q|<1, \;z\in \IC,
		\\
		\prod_{r=1}^\infty \frac{1}{1-q^{-r}z}&,\quad |q|>1,\; z\in\IC\setminus\{q^k\}_{k\in\IZ_{> 0}}.
	\end{cases}
\end{equation}
The inversion formula 
\begin{equation}\label{qpochinversionidentity}
(z;q)_\infty=\frac 1 {(q^{-1}z;q^{-1})_\infty}, \quad|q|\gtrless 1 
\end{equation}
follows directly from the definition. The $q$-Pochhammer symbol for general index is defined as
\begin{equation}\label{qpochgeneralindex}
(z;q)_\alpha \coloneqq \frac{(z;q)_\infty}{(q^\alpha z;q)_\infty},\quad |q|\gtrless 1,\; z,\alpha\in \IC \text{ (generic)}\ ,
\end{equation}
which implies the usual definitions
\begin{equation}
	(z;q)_n=
	\begin{cases}
		\prod_{r=0}^{n-1} (1-q^r z) &, \quad n\geq 0
		\ ,
		\\
		\prod_{r=1}^{-n}\frac{1}{1-q^{-r} z} &, \quad n<0
		\ .
	\end{cases}
\end{equation}
We also write $(z)_n=(z;q)_n$ if the second argument of the Pochhammer symbol is $q$.

We define the $q$-Gamma function by 
\begin{equation}\label{defqGamma}
	\begin{aligned}
		\Gamma_q(x)=
		\begin{cases}
			(1-q)^{1-x}\frac{(q;q)_\infty}{(q^x;q)_\infty},\, \text{ for } |q|<1 \, ,
			\\ q^{\binom{x-1}{2}}\Gamma_{q^{-1}}(x), \, \text{ for }|q|>1 \; .
		\end{cases} 
	\end{aligned}
\end{equation}
The Jacobi theta functions are
\bea
\theta_1(z,q)&=&\sum_{n\in\IZ}(-1)^{n+\frac 12}q^{\frac12 (n+\frac 12)^2}z^{n+\frac12} =-iq^{\frac 18}z^{\frac 12}(q)_\infty(qz)_\infty(z^{-1})_\infty\,,\nonumber\\
\theta_2(z,q)&=&\sum_{n\in\IZ}q^{\frac12 (n+\frac 12)^2}z^{(n+\tfrac 12)}=
 q^{\frac 18}z^{\frac 12}(q)_\infty(-qz)_\infty(-z^{-1})_\infty\,, \nonumber\\
\theta_3(z,q)&=&\sum_{n\in\IZ}q^{\frac {n^2}2}z^n =(q)_\infty(-q^{1/2}z)_\infty(-q^{1/2}z^{-1})_\infty\,, \nonumber\\
\theta_4(z,q)&=&\sum_{n\in\IZ}(-1)^nq^{\frac {n^2}2}z^n =(q)_\infty(q^{1/2}z)_\infty(q^{1/2}z^{-1})_\infty\,.
\eea
For argument $z=1$ we also write $\theta_i(1,q)=\theta_i(q)$. We also use the notation
\begin{equation}\label{qpochtheta}
  \theta(z,q)= (z;q)_\infty(q z^{-1};q)_\infty\,.
\end{equation}
Finally, we collect some identities that are useful in the computations below:
\begin{subequations}
	\begin{align} 
	\label{qpochidentity:qandindexinversion}
	(z;q)_{\alpha}&=\frac{1}{(q^{-1}z;q^{-1})_{-\alpha}}\ ,
	\\
	\label{qpochidentity:indexinversion}
	\frac{\tq(z,q)}{\tq(q^{n}z,q)}&=(z;q)_n(qz^{-1};q)_{-n}=(-1)^nq^{\tfrac 12 n(n-1)}z^n\ ,
	\\
	\label{qpochidentity:qzinversion}
	(z;q^{-1})_n&=(z^{-1};q)_n (-1)^nq^{-\tfrac 1 2 n(n-1)}z^{n}\ .
	\end{align}
\end{subequations}
Here $z, \alpha\in\IC$ are generic and $n\in\IZ$.

\subsubsection{Partition function for Neumann boundary conditions \label{app:PFNeum}}
\footnote{The computations in this and the next section have also been reported in the Ph.D.~theses of U.N. and A.T. \cite{phdU,phdA}.} In addition to the classical contributions \eqref{zfi} and \eqref{CSbare}, we have the 1-loop determinants from the vector multiplet with gauge group $G=U(M)$ and the $N$ chiral multiplets, both types of fields with Neumann b.c.
\begin{equation}
	Z_\text{1-loop}=Z_\text{vect}\cdot Z_\text{chir, N}^N\ ,
\end{equation}
in the partition function \eqref{LocPathIntegral}. The contributions have been computed via supersymmetric localization \cite{Gadde:2013wq,BDP,YS}. In terms of  Wilson lines $z_a=e^{\sigma_a}$, $a=1,\ldots, M$, these are\footnote{We include  the factor $(q;q)_\infty^M$ in the vector multiplet contribution~$Z_\text{vect}$ arising from the maximal torus of the gauge group $U(M)$ as explained in refs.~\cite{DGP,CDG, Bul20}; this factor is omitted in ref.~\cite{YS} and some versions of ref.~\cite{Yos19}.}
\begin{equation}\label{oneloopdets}
	\begin{aligned}
		Z_\text{vect}&=(q;q)_\infty^{M}\prod_{a\neq b}^M e^{\tfrac{\sigma_{ab}^2}{4\log q}} (e^{\sigma_{ab}};q)_\infty\ ,
		\\
		Z_\text{chir, N}&=\prod_{a=1}^M e^{-\tfrac{\log q}{24}+\tfrac{\sigma_a}{4}-\tfrac{\sigma_a^2}{4\log q}} 
		\frac{1}{(e^{\sigma_a};q)_\infty}\ .
	\end{aligned}
\end{equation}
where $\sigma_{ab}=\sigma_a-\sigma_b$. 
For  large positive Re$(\zeta)$, i.e. small $|Q|$ and $|q|<1$, we pick the poles of the $q$-Pochhammer symbol in the chiral contribution. 
Setting
\begin{equation}\label{poles}
	\sigma_a=-\log q (d_a-\eps_a)=-\log q \;\td_a\,, 
\end{equation}
there is a pole at $\eps_a=0$ for each $a=1,\ldots, M$ and each $d_a\in \mathbb{Z}_{\geq 0}$. The partition function integral \eqref{LocPathIntegral} becomes a sum of integrals around the poles
\begin{equation}\label{PFfirstsubstitution}
	\begin{aligned}
		Z_{\DSq}=\frac{(\log q)^M}{M!} q^{-\tfrac{MN}{24}}\, (q;q)_\infty^{M}
		\oint \prod_{a=1}^M\frac{d \eps_a}{2\pi i }\  
		\sum_{\vec d \in \mathbb{Z}_{\geq 0}} 
		\tx Q^{\sum_{a=1}^M \td_a}
		q^{\overline{CS}(\td)}
		\frac{\prod_{a\neq b}^M (q^{\td_{ab}};q)_\infty}{\prod_{a=1}^M (q^{-\td_a};q)_\infty^N}
		\,,
	\end{aligned}
\end{equation}
where  $ \tilde Q=e^{-2\pi \zeta}$ and where the explicit $q$-exponent
\begin{equation}
	\overline{CS}(\td)=\frac{\kappa_S+M-\tfrac N 2}{2}\tr_{SU(M)}(\td^2)+\frac{\kappa_A-\tfrac N 2 }{2}\tr_{U(1)}(\td^2)+(\kappa_R-\tfrac N 4)\tr_R(\td)\,.
\end{equation}
contains bare Chern-Simons terms, as well as shifts from the 1-loop determinants \eqref{oneloopdets}.\footnote{We have used the identity
\begin{equation*}
	\begin{aligned}
		\sum_{a\neq b}^M \td_{ab}^2&=2 M\tr_{SU(M)}(\td^2)\,.
	\end{aligned}
\end{equation*}}
The infinite $q$-Pochhammer in the denominator may be rewritten using eqs.~\eqref{qpochgeneralindex}, \eqref{qpochidentity:indexinversion} and $d_a\geq 0$:
\begin{equation}\label{chiralIdentity}
	\begin{aligned}
		\prod_{a=1}^M\frac{1}{(q^{-\td_a};q)_\infty^N}
		&=
		\frac{q^{
				-\tfrac{N}{2}\tr_{SU(M)}(\epsilon^2) 
				-\tfrac{N}{2}\tr_{U(1)}(\epsilon^2)
				+\tfrac{N}{2} \tr_{R}(\epsilon)	
		}}{\prod_{a=1}^M(q^{\epsilon_a};q)_\infty^N}\times
		\\&\hspace{.4cm}
		\times
		\frac{(-1)^{N\sum_{a=1}^M d_a} q^{
				\tfrac{N}{2}\tr_{SU(M)}(\td^2) 
				+\tfrac{N}{2}\tr_{U(1)}(\td^2)
				+\tfrac{N}{2} \tr_{R}(\td)
		}}{\prod_{a=1}^M\prod_{r=1}^{d_a}(1-q^{r-\epsilon_a})^N}
	\end{aligned}
\end{equation}
and similarly using \eqref{qpochidentity:indexinversion} the infinite $q$-Pochhammer in the numerator can be expressed as
\begin{equation}\label{nonAbelianIdentity}
	\begin{aligned}
		\prod_{a<b}^M
		(q^{\td_{ab}};q)_\infty (q^{-\td_{ab}};q)_\infty
		&=
		q^{M\tr_{SU(M)}(\eps^2)}\prod_{a<b}^M
		\frac{(q^{\eps_{ab}};q)_\infty (q^{-\eps_{ab}};q)_\infty}{q^{\tfrac 1 2 \eps_{ab}^2}(q^{-\tfrac 12 \eps_{ab}}-q^{\tfrac 1 2 \eps_{ab}})}
		\times 
		\\&\hspace{.4cm}
		\times 
		q^{-M\tr_{SU(M)}(\td^2)}
		(-1)^{(M+1)\sum_{a=1}^M d_a}
		\prod_{a<b}^M 
		q^{\tfrac 1 2 \td_{ab}^2}(q^{\tfrac{1}{2}\td_{ab}}-q^{-\tfrac{1}{2}\td_{ab}})\ .
	\end{aligned}
\end{equation}
Substituting these expressions back in \eqref{PFfirstsubstitution} and defining 
\begin{equation}
	Q=(-1)^{N+M}\tx Q\ ,
\end{equation}
we obtain the result
\begin{equation}
Z_{\DSq} = \frac 1 {M!}\, \oint \prod\limits_{a=1}^{M}\, \frac{d \epsilon_a}{2 \pi i } \ f_{\text{Gr}(M,N)} (q,\epsilon) \cdot I^{(\kh_S,\kh_A,\kh_R)}_{\text{Gr}(M,N)} (Q,q,\epsilon) \ .
\end{equation}
Here, we have the normalized $I$-function
\begin{equation} \label{IfctAppendix}
I^{(\kh_S,\kh_A,\kh_R)}_{\text{Gr}(M,N)} (Q,q,\epsilon)= c_0 \sum_{\vec{d} \in \mathbb{Z}^M_{\geq 0}}  (-Q)^{\sum_{a=1}^M \tilde{d}_a }  q^{CS(\tilde{d})}\, \frac{ \prod_{a<b}^{M}   q^{\tfrac{1}{2}\tilde{d}_{ab}^2} (q^{\tfrac{1}{2}\tilde{d}_{ab}} - q^{-\tfrac{1}{2}\tilde{d}_{ab}})}{\prod\limits_{a=1}^M\prod\limits_{r=1}^{d_a} (1-q^{r-\epsilon_a})^N}  \ ,
\end{equation}
with effective Chern-Simons levels 
\begin{equation}
	\begin{aligned}
		&CS(\td) = \frac{1}{2}\hat \kappa_S \tr_{SU(M)} (\tilde{d}^2) + \frac{1}{2}\hat \kappa_A \tr_{U(1)} (\tilde{d}^2)  + \hat \kappa_R \tr_{R}(\tilde{d}) \ ,
		\\
		&\hat{\kappa}_S = \kappa_S - M + \frac{N}{2} \quad , \quad \hat{\kappa}_A = \kappa_A + \frac{N}{2} \quad \ \quad \hat{\kappa}_R = \kappa_R + \frac{N}{4} \ . 
	\end{aligned}
\end{equation}
The normalization factor is
\begin{equation}
	c_0(q,\epsilon,\hat \kappa_S,\hat\kappa_A,\hat\kappa_R)=\frac{1-q}{q^{CS(-\epsilon)}\prod_{a<b}^M q^{\tfrac 1 2 \eps_{ab}^2}(q^{-\tfrac 1 2 \eps_{ab}}-q^{\tfrac 1 2 \eps_{ab}})}\ ,
\end{equation}
and the remaining $Q$- and $d_a$-independent terms have been collected in the \emph{folding factor}\footnote{Note that the folding factor $f_{\text{Gr}(M,N)} (q,\epsilon)$ depends on the CS levels as well, which is not explicitly indicated for ease of notation.} 
\begin{equation}
	\begin{aligned}
		f_{\text{Gr}(M,N)} (q,\epsilon)&=
		\big(\log q\, (q;q)_\infty\big)^M  q^{-\tfrac{MN}{24}}
		q^{	\big(M-\tfrac{N}{2}\big)\tr_{SU(M)}(\epsilon^2) 
			-\tfrac{N}{2}\tr_{U(1)}(\epsilon^2)
			+\tfrac{N}{2} \tr_{R}(\epsilon)	
		}
		\times
		\\&\hspace{.4cm}
		\times
		\frac{(-1)^{(1+M+N)\sum_{a=1}\eps_a}c_0^{-1}}{\prod_{a<b}^Mq^{\tfrac 1 2 \eps_{ab}^2}(q^{-\tfrac 12 \eps_{ab}}-q^{\tfrac 1 2 \eps_{ab}})}
		\frac{\prod_{a<b}^M(q^{\eps_{ab}};q)_\infty (q^{-\eps_{ab}};q)_\infty}{\prod_{a=1}^M(q^{\epsilon_a};q)_\infty^N}
		\\
		&=
		\big(\log q\, (q;q)_\infty\big)^M q^{-\tfrac{MN}{24}}
		q^{	\big(M-\tfrac{N}{2}\big)\tr_{SU(M)}(\epsilon^2) 
			-\tfrac{N}{2}\tr_{U(1)}(\epsilon^2)
			+\tfrac{N}{2}\tr_{R}(\epsilon)	
			+CS(-\epsilon)
		}
		\times
		\\&\hspace{.4cm}
		\times
		\frac{(-1)^{(1+M+N)\sum_{a=1}\eps_a}}{1-q}
		\frac{\prod_{a<b}^M(q^{\eps_{ab}};q)_\infty (q^{-\eps_{ab}};q)_\infty}{\prod_{a=1}^M(q^{\epsilon_a};q)_\infty^N}\ .
	\end{aligned}
\end{equation}
To express the contour integral as an integral over $X=\text{Gr}(M,N)$, we write terms in the folding factor in terms of characteristic classes of $X$, by identifying $\epsilon_a$'s as rescaled Chern roots of the dual $S^*$ of the tautological bundle on $X$ as in explained in eq.~\eqref{epstoH}. Note that we may write 
\begin{equation}
	\begin{aligned}
		\frac{\prod_{a<b}^M(q^{\eps_{ab}};q)_\infty (q^{-\eps_{ab}};q)_\infty}{\prod_{a=1}^M(q^{\epsilon_a};q)_\infty^N}=
		(q;q)_\infty^{2\binom{M}{2}-MN}
		(1-q^{-1})^{N\sum_{a=1}^M \eps_a}
		\Td_q(X)\cdot   \Gamma_q(X) \cdot \cx E_X(\tx \eps)
	\end{aligned}
\end{equation}
where $\Gamma_{q}(X)$ is the $q$-Gamma class  of $X$ and $\Td_q$ stands for the Todd class with rescaled Chern roots $x_a\mapsto \tx \eps_a=\log \! q\ \eps_a$ (see eq.~\eqref{ToddGammaDef}). Furthermore,
\begin{equation}
	\cx E_X(\tx\eps)= \frac{\prod_{a<b}^M(-\tx \epsilon_{ab}^2)}
	{\prod_{a=1}^M(\tx\epsilon_a)^N}\ ,
\end{equation}
is the integration kernel necessary to map contour integrals to formal integration on $X$ (cf.~\eqref{intform}). Using the Dedekind eta-function we may finally write
\begin{equation}
	\begin{aligned}
		f_{\text{Gr}(M,N)} (q,\epsilon)&=
		\Td_q(X)\cdot   \Gamma_q(X) \cdot \cx E_X(\tx \eps) \cdot \frac{q^{\An(\eps)} N(q,\eps) }{\eta(q)^{M(N-M)}} 
		\ .
	\end{aligned}
\end{equation}
where the total anomaly term is written using the Chern classes \eqref{ChernClassFormal} and Chern character \eqref{ChernCharacter} of $X$
\begin{equation}
	\begin{aligned}
		{\An}(\eps)&=
		\big(M-\tfrac{N}{2}\big)\tr_{SU(M)}(\epsilon^2) 
		-\tfrac{N}{2}\tr_{U(1)}(\epsilon^2)
		-\tfrac{N}{2}\kappa_R \tr_{R}(\epsilon)	
		+CS(-\epsilon)
		\\&
		=-\frac 1 2 c_1(\eps)-\ch_2(\eps)+CS(-\eps)
	\end{aligned}
\end{equation}
and the remaining factor is
\begin{equation}\label{numericalfactor}
	N(q,\eps)=(\log q)^M q^{-\tfrac 1 {24} M^2}
	(-1)^{\tfrac{M+1}{N}c_1(\eps)}
	(1-q)^{c_1(\eps)-1}\ .
\end{equation}

\subsubsection{Half-index for Dirichlet boundary conditions}
\label{app:indexcomputation}
In this appendix we want to derive equation \eqref{indexresult} from the half-index \eqref{halfindexdirichlet}. To start, note that for trivial mass fugacities $y_i=1$ we may write the chiral contribution as
\begin{equation}
\II_\text{chir, N}(q^{ m+ \eps})=\prod_{a=1}^M\frac{1}{(q^{m_a+\eps_a};q)_\infty}
=\prod_{a=1}^M \frac{(q^{\eps_a};q)_{m_a}}{(q^{\eps_a};q)_\infty}
\ .
\end{equation}
where in the last equation we have used \eqref{qpochgeneralindex}. The finite product in the numerator
\begin{equation}
\begin{aligned}
(q^{\eps_a};q)_{m_a}=
\begin{cases}
(1-q^{\eps_a})\prod_{r=1}^{m_a-1}(1-q^{r+\eps_a}) &, \text{ when } m_a>0
\\
\prod_{r=1}^{-m_a}\frac 1 {1-q^{\eps_a-r}} 
&, \text{ when } m_a\leq 0
\end{cases}\ .
\end{aligned}
\end{equation}
shows that the summation over non-negative $m_a$'s contains overall $(1-q^{\eps_a})$ factors. 
Sending $m_a\mapsto -m_a$ for convenience, eq.~\eqref{halfindexdirichlet} becomes to leading order in such factors
\begin{equation}\label{halfindexdirichlet2}
\begin{aligned}
\II_{\cx B=(\cx D,N)}=\frac{1}{(q)_\infty^M }
\bigg[
\sum_{\vec m\in \IZ_{\geq 0}^M} 
\frac
{q^{\tfrac 1 2 \vec m \cdot K\vec m
		-\vec m \cdot K\vec \eps
		+\tk_R \sum_{a=1}^M m_a}
	Q^{\sum_{a=1}^M m_a}}
{\prod_{a\neq b}^M (q^{1- \tx m_{ab}};q)_\infty
	\prod_{a=1}^M(q^{-\tx m_a};q)_\infty^N
}
+\cO\big((1-q^{\eps_a})^N\big)\bigg]\ ,
\end{aligned}
\end{equation}
where $\tx m_a=m_a-\eps_a$ and $\tx m_{ab}=m_{ab}-\eps_{ab}$ and the subscript denotes Dirichlet (Neumann) b.c.~for the gauge (chiral) fields.
The vector contribution may be expressed by the identity
\begin{equation}
\begin{aligned}
\prod_{a<b}^M\frac{1}{(q^{1+\tx m_{ab}};q)_\infty (q^{1- \tx m_{ab}};q)_\infty}
&=
\prod_{a<b}^M
\frac{1}{(q^{1+\eps_{ab}};q)_\infty (q^{1-\eps_{ab}};q)_\infty}
\frac{1}{q^{\tfrac 1 2 \eps_{ab}^2}(q^{-\tfrac 12 \eps_{ab}}-q^{\tfrac 1 2 \eps_{ab}})}
\times 
\\&\hspace{.4cm}
\times 
(-1)^{(M+1)\sum_{a=1}^M m_a}
\prod_{a<b}^M 
q^{\tfrac 1 2 \tx m_{ab}^2}(q^{\tfrac{1}{2}\tx m_{ab}}-q^{-\tfrac{1}{2}\tx m_{ab}})\ .
\end{aligned}
\end{equation}
which follows from equations \eqref{qpochinversionidentity} and \eqref{nonAbelianIdentity}. 
The numerator in the sum \eqref{halfindexdirichlet2} becomes\footnote{With $K$ as in \eqref{CSlevelsdirichlet}, in the notation of footnote \ref{footnotesum}, we have
	\begin{equation*}
	\frac 1 2 \vec m \cdot K \vec m= \frac{\tk_S}{2}\tr_{SU(M)}(m^2)+\frac{\tk_A}{2}\tr_{U(1)}(m^2)
	\end{equation*}
}
\begin{equation}
\begin{aligned}
q^{-\frac{\tk_S}{2}\tr_{SU(M)}(\eps^2) 
	-\frac{\tk_A}{2}\tr_{U(1)}(\eps^2)
	+\tk_R \tr_{R}(\eps)} 
\cdot
q^{
	\frac{\tk_S}{2}\tr_{SU(M)}(\tx m^2) 
	+\frac{\tk_A}{2}\tr_{U(1)}(\tx m^2)
	+\tk_R \tr_{R}(\tx m)
}Q^{\sum_{a=1}^M  m_a}
\end{aligned}
\end{equation}
The chiral contribution is rewritten using eq.~\eqref{chiralIdentity}.
Collecting everything we find
\begin{equation}\label{halfindexdirichletFinal}
\begin{aligned}
\II_{(\cx D,N)}=&
\frac{1}{(q)_\infty^M }
\frac{1}{\prod_{a<b}^M(q^{1+\eps_{ab}};q)_\infty (q^{1-\eps_{ab}};q)_\infty 
	\prod_{a=1}^M(q^{\eps_a};q)_\infty^N}
\\&\times
\frac{1}{q^{\widetilde{CS}(-\eps)}\prod_{a<b}^Mq^{\tfrac 1 2 \eps_{ab}^2}(q^{-\tfrac 12 \eps_{ab}}-q^{\tfrac 1 2 \eps_{ab}})}
\\
&\times
\bigg[\!\!
\sum_{\vec m\in \IZ_{\geq 0}^M} \!\!
q^{
	\widetilde{CS}(\tx m)}
((-1)^{(N+M+1)}Q)^{\sum_{a=1}^M m_a}
\frac
{
	\prod_{a<b}^M 
	q^{\tfrac 1 2 \tx m_{ab}^2}(q^{\tfrac{1}{2}\tx m_{ab}}-q^{-\tfrac{1}{2}\tx m_{ab}})}
{\prod_{a=1}^M\prod_{r=1}^{m_a}(1-q^{r-\eps_a})^N
}
+\cO\big((1-q^{\eps})^N\big)\bigg]
\\=&
\frac{1}{(q)_\infty^M (1-q)}
\frac{((-1)^{(N+M+1)}Q)^{\sum_{a=1}^M \eps_a}}{\prod_{a<b}^M(q^{1+\eps_{ab}};q)_\infty (q^{1-\eps_{ab}};q)_\infty 
	\prod_{a=1}^M(q^{\eps_a};q)_\infty^N}
\\&\times \bigg[
I^{(\tk_S+N,\tk_A+N,\tk_R+\tfrac{N}{2})}_{\text{Gr}(M,N)} \big((-1)^{M+N}Q,q,\eps)
+\cO(\eps^N)\bigg]\ ,
\end{aligned}
\end{equation}
where the CS levels in $I$ are
\begin{equation}
\widetilde{CS}(\tx m)
=\frac{\tk_S+N}{2}\tr_{SU(M)}(\tx m^2) 
+\frac{\tk_A+N}{2}\tr_{U(1)}(\tx m^2)
+(\tk_R+\frac{N}{2}) \tr_{R}(\tx m)\ .
\end{equation}

\subsubsection{Inversion formula for the  Grassmannian $I$-function}
The generalized $q$-hypergeometric function as introduced in \eqref{ISQKGr} and \eqref{IfctAppendix} appears in the literature in different forms. For CS levels as in \eqref{effCS1}, we can write
\begin{equation}
	\begin{aligned}
		I^{(\kh_S,\kh_A,\kh_R)}_{\text{Gr}(M,N)} (Q,q,\epsilon)
		&=
		(1-q)(-Q)^{-\sum_{a=1}^M\eps_a}\times
		\\&\hspace{.3cm}\times
		\!\!
		\sum_{\vec d\in \IZ^M_{\geq 0}}
		\!\!
		\big((-1)^MQ)^{\sum_{a=1}^M d_a}
		q^{CS(d)-\hat \kappa_S \vec d \cdot \vec \eps -\tfrac{\hat \kappa_A-\hat \kappa_S}{M}\sum_{a,b=1}^Md_a\eps_b}
		\frac{\prod_{a\neq b}^M (q^{1-\eps_{ab}};q)_{d_{ab}}}{\prod_{a=1}^M (q^{1-\eps_a};q)_{d_a}^N}\ .
	\end{aligned}
\end{equation}
The sum matches the expressions used in the mathematics literature for the $I$-function of $\Gr(M,N)$ for zero levels \cite{Wen,Giv20}, as well as expressions with level structure \cite{Giv20,dong20,RZ18b} (for specific levels) upon identifying the K-theoretic Chern roots $P_a=q^{-\eps_a}$.

Furthermore, the function  $I_{\Gr(M,N)}^{(\hat \kappa_S,\hat\kappa_A,\hat\kappa_R)}(Q,q,\eps)$ satisfies, by virtue of \eqref{qpochidentity:qzinversion}, the following inversion formula
\begin{equation}\label{inversion}
	I_{\Gr(M,N)}^{(\hat \kappa_S,\hat\kappa_A,\hat\kappa_R)}\!(Q,q^{-1},\eps)
	=
	(-q^{-1})(-1)^{N\sum_{a=1}^M\eps_a} 
	I_{\Gr(M,N)}^{(N-2M-\hat \kappa_S,N-\hat\kappa_A,\tfrac N 2-\hat\kappa_R)}\!\big((-1)^NQ,q,\eps\big)\ .
\end{equation}

\subsection{Further examples of perturbations}

\subsubsection{Perturbed pairing of $\Gr(2,4)$}
\label{app:pertpairingGr24}
The perturbed pairing is given by \eqref{gpairingformalexpansion}, and it also satisfies   $G_{\mu\nu}(t)=\p_{t_\mu}\p_{t_\nu}G_{00}(t)$. Expanding $G_{00}(Q,t)=\frac{1}{1-Q}\sum_{d\geq 0}G_{00}^{(d)}(t)Q^d$, we print out the coefficients $G_{00}^{(d)}(t)$ up to $Q^{d=4}$ and $\mathfrak{t}^5$, where $\mathfrak{t}$ counts homogeneous powers in  $t_\mu$'s:
{\allowdisplaybreaks
	\scriptsize
	\begin{align*}
		G_{00}^{(0)}=&1+\mathfrak{t} (t_{1}+t_{2}+t_{1,1}+t_{2,1}+t_{2,2})
		+\mathfrak{t}^2 (\tfrac{t_{1}^2}{2}+\tfrac{t_{2}^2}{2}+\tfrac{1}{2} t_{1,1}^2+t_{1} (t_{2}+t_{1,1}+t_{2,1}))
		+\mathfrak{t}^3 (\tfrac{t_{1}^3}{6}+t_{1}^2 (\tfrac{t_{2}}{2}+\tfrac{1}{2} t_{1,1}))+\tfrac{1}{12} \mathfrak{t}^4 t_{1}^4
		\\
		G_{00}^{(1)}=& \mathfrak{t}^2 \big(\tfrac{1}{2} t_{2,1}^2+t_{1} t_{2,2}+t_{2,1} t_{2,2}+t_{1,1} (t_{2,1}+t_{2,2})+t_{2} (t_{1,1}+t_{2,1}+t_{2,2})\big)
		\\&+ \mathfrak{t}^3 \big(\tfrac{t_{2}^3}{6}+\tfrac{1}{6} t_{1,1}^3+t_{2}^2 (\tfrac{1}{2} t_{1,1}+\tfrac{1}{2} t_{2,1})
		+t_{1}^2 (\tfrac{1}{2} t_{2,1}+\tfrac{1}{2} t_{2,2})+t_{2} (\tfrac{1}{2} t_{1,1}^2+\tfrac{1}{2} t_{2,1}^2+t_{1,1} (t_{2,1}+t_{2,2}))	
		\\&\hspace{.5cm}
		+\tfrac{1}{2} t_{1,1}^2 t_{2,1}
		+\tfrac{1}{2} t_{1,1} t_{2,1}^2	
		+t_{1} (\tfrac{t_{2}^2}{2}+\tfrac{1}{2} t_{1,1}^2+\tfrac{1}{2} t_{2,1}^2+t_{2,1} t_{2,2}+t_{1,1} (t_{2,1}+t_{2,2})+t_{2} (t_{1,1}+t_{2,1}+t_{2,2}))\big)
		\\&+\mathfrak{t}^4 \big(
		-\tfrac{1}{24} t_{1}^4
		+\tfrac{1}{6} t_{2}^3 t_{1,1}
		+t_{2}^2 (\tfrac{1}{4} t_{1,1}^2+\tfrac{1}{2} t_{1,1} t_{2,1})
		+t_{2} (\tfrac{1}{6} t_{1,1}^3+\tfrac{1}{2} t_{1,1}^2 t_{2,1})
		+t_{1}^3 (\tfrac{t_{2}}{6}+\tfrac{1}{6} t_{1,1}+\tfrac{1}{6} t_{2,1}+\tfrac{1}{6} t_{2,2})
		\\&\hspace{.5cm}
		+t_{1}^2 (\tfrac{t_{2}^2}{4}+\tfrac{1}{4} t_{1,1}^2+\tfrac{1}{4} t_{2,1}^2+t_{1,1} (\tfrac{1}{2} t_{2,1}+\tfrac{1}{2} t_{2,2})+t_{2} (\tfrac{1}{2} t_{1,1}+\tfrac{1}{2} t_{2,1}+\tfrac{1}{2} t_{2,2})+\tfrac{1}{2} t_{2,1} t_{2,2})
		\\&\hspace{.5cm}
		+t_{1} (\tfrac{t_{2}^3}{6}+\tfrac{1}{6} t_{1,1}^3+t_{2}^2 (\tfrac{1}{2} t_{1,1}+\tfrac{1}{2} t_{2,1})+\tfrac{1}{2} t_{1,1}^2 t_{2,1}+\tfrac{1}{2} t_{1,1} t_{2,1}^2+t_{2} (\tfrac{1}{2} t_{1,1}^2+\tfrac{1}{2} t_{2,1}^2+t_{1,1} (t_{2,1}+t_{2,2})))\big)
		\\&+\mathfrak{t}^5 \big(
		\tfrac{t_{1}^5}{120}
		+\tfrac{1}{12} t_{2}^3 t_{1,1}^2
		+t_{1}^4 (\tfrac{t_{2}}{24}+\tfrac{1}{24} t_{1,1}+\tfrac{1}{24} t_{2,1}+\tfrac{1}{24} t_{2,2})
		\\&\hspace{.5cm}
		+\tfrac{1}{12} t_{2}^2 t_{1,1}^3+t_{1} (\tfrac{1}{6} t_{2}^3 t_{1,1}+t_{2}^2 (\tfrac{1}{4} t_{1,1}^2+\tfrac{1}{2} t_{1,1} t_{2,1})+t_{2} (\tfrac{1}{6} t_{1,1}^3+\tfrac{1}{2} t_{1,1}^2 t_{2,1}))
		\\&\hspace{.5cm}
		+t_{1}^2 (\tfrac{t_{2}^3}{12}+\tfrac{1}{12} t_{1,1}^3+t_{2}^2 (\tfrac{1}{4} t_{1,1}+\tfrac{1}{4} t_{2,1})+\tfrac{1}{4} t_{1,1}^2 t_{2,1}+\tfrac{1}{4} t_{1,1} t_{2,1}^2+t_{2} (\tfrac{1}{4} t_{1,1}^2+\tfrac{1}{4} t_{2,1}^2+t_{1,1} (\tfrac{1}{2} t_{2,1}+\tfrac{1}{2} t_{2,2})))
		\\&\hspace{.5cm}
		+t_{1}^3 (\tfrac{t_{2}^2}{12}+\tfrac{1}{12} t_{1,1}^2+\tfrac{1}{12} t_{2,1}^2+t_{1,1} (\tfrac{1}{6} t_{2,1}+\tfrac{1}{6} t_{2,2})+t_{2} (\tfrac{1}{6} t_{1,1}+\tfrac{1}{6} t_{2,1}+\tfrac{1}{6} t_{2,2})+\tfrac{1}{6} t_{2,1} t_{2,2})\big)
		\\
		G_{00}^{(2)}=&
		\tfrac{1}{2}\mathfrak{t}^2 t_{2,2}^2
		+ \mathfrak{t}^3 \big(
		\tfrac{1}{6} t_{2,1}^3
		+\tfrac{1}{2} t_{2}^2 t_{2,2}
		+\tfrac{1}{2} t_{1,1}^2 t_{2,2}
		+\tfrac{1}{2} t_{2,1}^2 t_{2,2}
		+\tfrac{1}{2} t_{1} t_{2,2}^2
		\\&\hspace{.5cm}
		+\tfrac{1}{2} t_{2,1} t_{2,2}^2
		+\tfrac{1}{6} t_{2,2}^3
		+t_{2} (t_{2,1} t_{2,2}+\tfrac{1}{2} t_{2,2}^2)
		+t_{1,1} (t_{2,1} t_{2,2}
		+\tfrac{1}{2} t_{2,2}^2)\big)
		\\&
		+ \mathfrak{t}^4 \big(
		\tfrac{t_{2}^4}{24}
		+\tfrac{1}{24} t_{1,1}^4
		+\tfrac{1}{24} t_{2,1}^4
		+t_{2}^3 (\tfrac{1}{6} t_{2,1}+\tfrac{1}{6} t_{2,2})
		+t_{1,1}^3 (\tfrac{1}{6} t_{2,1}+\tfrac{1}{6} t_{2,2})
		+t_{1,1}^2 (\tfrac{1}{4} t_{2,1}^2+\tfrac{1}{2} t_{2,1} t_{2,2}+\tfrac{1}{4} t_{2,2}^2)
		\\&\hspace{.5cm}
		+\tfrac{1}{6} t_{2,1}^3 t_{2,2}
		+\tfrac{1}{4} t_{1}^2 t_{2,2}^2
		+t_{2}^2 (\tfrac{1}{4} t_{2,1}^2+\tfrac{1}{2} t_{1,1} t_{2,2}+\tfrac{1}{2} t_{2,1} t_{2,2}+\tfrac{1}{4} t_{2,2}^2)
		+t_{1,1} (\tfrac{1}{6} t_{2,1}^3+\tfrac{1}{2} t_{2,1}^2 t_{2,2}+\tfrac{1}{2} t_{2,1} t_{2,2}^2)
		\\&\hspace{.5cm}
		+t_{1} (\tfrac{1}{6} t_{2,1}^3+\tfrac{1}{2} t_{2}^2 t_{2,2}+\tfrac{1}{2} t_{1,1}^2 t_{2,2}+\tfrac{1}{2} t_{2,1}^2 t_{2,2}+\tfrac{1}{2} t_{2,1} t_{2,2}^2+\tfrac{1}{3} t_{2,2}^3+t_{2} (t_{2,1} t_{2,2}+\tfrac{1}{2} t_{2,2}^2)+t_{1,1} (t_{2,1} t_{2,2}+\tfrac{1}{2} t_{2,2}^2))
		\\&\hspace{.5cm}
		+t_{2} (\tfrac{1}{6} t_{2,1}^3+\tfrac{1}{2} t_{1,1}^2 t_{2,2}+\tfrac{1}{2} t_{2,1}^2 t_{2,2}+\tfrac{1}{2} t_{2,1} t_{2,2}^2+t_{1,1} (\tfrac{1}{2} t_{2,1}^2+t_{2,1} t_{2,2}+\tfrac{1}{2} t_{2,2}^2))\big)
		\\&
		+ \mathfrak{t}^5 \big(
		\tfrac{t_{2}^5}{120}
		+\tfrac{1}{120} t_{1,1}^5
		+\tfrac{1}{8} t_{1,1} t_{2,1}^4
		+t_{1,1}^3 (\tfrac{1}{12} t_{2,1}^2+\tfrac{1}{6} t_{2,1} t_{2,2}+\tfrac{1}{12} t_{2,2}^2)
		+t_{1,1}^4 (\tfrac{1}{24} t_{2,1}+\tfrac{1}{24} t_{2,2})
		\\&\hspace{.5cm}
		+t_{2}^4 (\tfrac{1}{24} t_{1,1}+\tfrac{1}{24} t_{2,1}+\tfrac{1}{24} t_{2,2})
		+\tfrac{1}{12} t_{1}^3 t_{2,2}^2
		+t_{2}^3 (\tfrac{1}{12} t_{2,1}^2+t_{1,1} (\tfrac{1}{6} t_{2,1}+\tfrac{1}{6} t_{2,2})+\tfrac{1}{6} t_{2,1} t_{2,2}+\tfrac{1}{12} t_{2,2}^2)
		\\&\hspace{.5cm}
		+t_{1}^2 (\tfrac{1}{4} t_{2}^2 t_{2,2}+\tfrac{1}{4} t_{1,1}^2 t_{2,2}+\tfrac{1}{2} t_{2,1}^2 t_{2,2}+\tfrac{1}{3} t_{2,2}^3+t_{2} (\tfrac{1}{2} t_{2,1} t_{2,2}+\tfrac{1}{4} t_{2,2}^2)+t_{1,1} (\tfrac{1}{2} t_{2,1} t_{2,2}+\tfrac{1}{4} t_{2,2}^2))
		\\&\hspace{.5cm}
		+t_{2}^2 (t_{1,1}^2 (\tfrac{1}{4} t_{2,1}+\tfrac{1}{4} t_{2,2})+\tfrac{1}{2} t_{2,1}^2 t_{2,2}+t_{1,1} (\tfrac{1}{4} t_{2,1}^2+\tfrac{1}{2} t_{2,1} t_{2,2}+\tfrac{1}{4} t_{2,2}^2))
		+\tfrac{1}{2} t_{1,1}^2 t_{2,1}^2 t_{2,2}
		\\&\hspace{.5cm}
		+t_{2} (\tfrac{1}{24} t_{1,1}^4+\tfrac{1}{8} t_{2,1}^4+t_{1,1}^3 (\tfrac{1}{6} t_{2,1}+\tfrac{1}{6} t_{2,2})+t_{1,1} t_{2,1}^2 t_{2,2}+t_{1,1}^2 (\tfrac{1}{4} t_{2,1}^2+\tfrac{1}{2} t_{2,1} t_{2,2}+\tfrac{1}{4} t_{2,2}^2))
		\\&\hspace{.5cm}
		+t_{1} (\tfrac{t_{2}^4}{24}+\tfrac{1}{24} t_{1,1}^4-\tfrac{1}{24} t_{2,1}^4+t_{2}^3 (\tfrac{1}{6} t_{2,1}+\tfrac{1}{6} t_{2,2})+t_{1,1}^3 (\tfrac{1}{6} t_{2,1}+\tfrac{1}{6} t_{2,2})+\tfrac{1}{3} t_{2,1}^3 t_{2,2}
		\\&\hspace{1cm}
		+t_{1,1}^2 (\tfrac{1}{4} t_{2,1}^2+\tfrac{1}{2} t_{2,1} t_{2,2}+\tfrac{1}{4} t_{2,2}^2)+t_{2}^2 (\tfrac{1}{4} t_{2,1}^2+\tfrac{1}{2} t_{1,1} t_{2,2}+\tfrac{1}{2} t_{2,1} t_{2,2}+\tfrac{1}{4} t_{2,2}^2)
		\\&\hspace{1cm}
		+t_{1,1} (\tfrac{1}{3} t_{2,1}^3+t_{2,1} t_{2,2}^2)+t_{2} (\tfrac{1}{3} t_{2,1}^3+\tfrac{1}{2} t_{1,1}^2 t_{2,2}+t_{2,1} t_{2,2}^2+t_{1,1} (\tfrac{1}{2} t_{2,1}^2+t_{2,1} t_{2,2}+\tfrac{1}{2} t_{2,2}^2)))\big)
		\\
		G_{00}^{(3)}=&
		 \mathfrak{t}^4 \big(
		\tfrac{1}{4} t_{2,1}^2 t_{2,2}^2
		-\tfrac{1}{6} t_{1} t_{2,2}^3
		+\tfrac{1}{6} t_{2} t_{2,2}^3
		+\tfrac{1}{6} t_{1,1} t_{2,2}^3
		+\tfrac{1}{6} t_{2,1} t_{2,2}^3
		+\tfrac{1}{12} t_{2,2}^4\big)
		\\&
		+ \mathfrak{t}^5 \big(\tfrac{1}{12} t_{2,1}^4 t_{2,2}+t_{1}^2 (\tfrac{1}{12} t_{2,1}^3-\tfrac{1}{4} t_{2,1}^2 t_{2,2}+\tfrac{1}{4} t_{2,1} t_{2,2}^2-\tfrac{1}{4} t_{2,2}^3)+t_{2}^2 (\tfrac{1}{12} t_{2,1}^3-\tfrac{1}{4} t_{2,1}^2 t_{2,2}+\tfrac{1}{4} t_{2,1} t_{2,2}^2+\tfrac{1}{12} t_{2,2}^3)
		\\&\hspace{.5cm}
		+t_{1,1}^2 (\tfrac{1}{12} t_{2,1}^3-\tfrac{1}{4} t_{2,1}^2 t_{2,2}+\tfrac{1}{4} t_{2,1} t_{2,2}^2+\tfrac{1}{12} t_{2,2}^3)+t_{1,1} (-\tfrac{1}{12} t_{2,1}^4+\tfrac{1}{6} t_{2,1}^3 t_{2,2}+\tfrac{1}{4} t_{2,1}^2 t_{2,2}^2+\tfrac{1}{6} t_{2,1} t_{2,2}^3+\tfrac{1}{12} t_{2,2}^4)
		\\&\hspace{.5cm}
		+t_{2} (-\tfrac{1}{12} t_{2,1}^4+\tfrac{1}{6} t_{2,1}^3 t_{2,2}+\tfrac{1}{4} t_{2,1}^2 t_{2,2}^2+\tfrac{1}{6} t_{2,1} t_{2,2}^3+\tfrac{1}{12} t_{2,2}^4+t_{1,1} (\tfrac{1}{6} t_{2,1}^3-\tfrac{1}{2} t_{2,1}^2 t_{2,2}+\tfrac{1}{2} t_{2,1} t_{2,2}^2+\tfrac{1}{6} t_{2,2}^3))
		\\&\hspace{.5cm}
		+\tfrac{1}{6} t_{2,1}^2 t_{2,2}^3
		+t_{1} (\tfrac{1}{12} t_{2,1}^4-\tfrac{1}{6} t_{2,1}^3 t_{2,2}+\tfrac{1}{4} t_{2,1}^2 t_{2,2}^2
		+t_{2} (-\tfrac{1}{6} t_{2,1}^3+\tfrac{1}{2} t_{2,1}^2 t_{2,2}-\tfrac{1}{2} t_{2,1} t_{2,2}^2+\tfrac{1}{6} t_{2,2}^3)
		\\&\hspace{1cm}
		+\tfrac{1}{6} t_{2,1} t_{2,2}^3+\tfrac{1}{12} t_{2,2}^4
		+t_{1,1} (-\tfrac{1}{6} t_{2,1}^3+\tfrac{1}{2} t_{2,1}^2 t_{2,2}-\tfrac{1}{2} t_{2,1} t_{2,2}^2+\tfrac{1}{6} t_{2,2}^3))\big)
		\\
		G_{00}^{(4)}=&
		-\tfrac{1}{24}  \mathfrak{t}^4 t_{2,2}^4+\mathfrak{t}^5 \big(\tfrac{1}{120} t_{2,1}^5-\tfrac{1}{24} t_{2,1}^4 t_{2,2}+\tfrac{1}{12} t_{2,1}^3 t_{2,2}^2-\tfrac{1}{12} t_{2,1}^2 t_{2,2}^3
		\\&\hspace{1cm}
		-\tfrac{1}{24} t_{1} t_{2,2}^4-\tfrac{1}{24} t_{2} t_{2,2}^4-\tfrac{1}{24} t_{1,1} t_{2,2}^4+\tfrac{1}{24} t_{2,1} t_{2,2}^4+\tfrac{1}{120} t_{2,2}^5\big)
	\end{align*}	
}

\subsubsection{Perturbations of $\Gr(2,5)\simeq \Gr(3,5)$}
\label{app:Gr25Pert}
We state some of the results on the perturbations of $\Gr(2,5)$ with canonical Chern-Simons levels. The computation of the perturbations follows the steps of the example of $\Gr(2,4)$ in the main text (see \eqref{J24} onwards). The results agree with those of $\Gr(3,5)$ under the expected geometric duality that sends  $\mu\mapsto \mu^T$ for partitions labeling (perturbations along) structure sheaves ($t_\mu$) $\cx O_\mu$ on Schubert cycles $\sigma_\mu$, up to a redefinition $Q\mapsto (-1)^M Q$. \\ 
Our starting point is the $I$-function \eqref{ISQKGr}
\begin{equation}
  J_{\Gr(2,5)}(t=0)=(-Q)^{\eps_1+\eps_2}I^{(0,0,0)}_{\text{Gr}(2,5)} (Q,q,\epsilon)=\!c_0\!\! \sum_{d_1,d_2 \geq 0} \!\! (-Q)^{d_1+d_2} \, \frac{ q^{\tfrac{1}{2}\tilde{d}_{12}^2} (q^{\tfrac{1}{2}\tilde{d}_{12}} - q^{-\tfrac{1}{2}\tilde{d}_{21}})}{\prod\limits_{a=1}^2\prod\limits_{r=1}^{d_a} (1-q^{r-\epsilon_a})^5}.
\end{equation}
The input $t\in \cx K_+$ is
\begin{equation}
	t= t_{1} \cx{O}_{1}
	+ t_{2} \cx{O}_{2}
	+ t_{1,1} \cx{O}_{1,1}
	+ t_{3} \cx{O}_{3}
	+ t_{2,1} \cx{O}_{2,1} 
	+ t_{3,1} \cx{O}_{3,1}
	+ t_{2,2} \cx{O}_{2,2}
	+ t_{3,2} \cx{O}_{3,2}
	+ t_{3,3} \cx{O}_{3,3}.
\end{equation}
The $J$-function with this input is computed analogously to \eqref{reconstructedGr24}. The deformation parameter $\tau(Q,t)=\sum_{\mu\in\cx B_{2,5}} \tau_\mu(Q,t)\O_\mu$ can be computed recursively. Up to print-friendly orders $Q, \mathfrak t^2$, where $\mathfrak{t}$ tracks homogeneous powers of $t_\mu$'s, we have:
\begin{equation}
	\begin{aligned}
		\tau_0(Q,t)&=Q \mathfrak{t}^2 \big(t_{3} ((q-1) t_{3,1}-t_{2,1})+t_{2} (-t_{3,1}-t_{3})+\tfrac{1}{2} (2 q-1) t_{3}^2\big),
		\\
		\tau_1(Q,t)&=\mathfrak{t} t_{1}+Q \mathfrak{t}^2 \big(t_{3} (-q t_{3,1}+(q-1) t_{3,2}+t_{2,1}-t_{2,2})-t_{2,1} t_{3,1}
		\\&
		\hspace{1.8cm}
		+\tfrac{1}{2} (q-1) t_{3,1}^2+t_{2} (t_{3,1}-t_{3,2}+t_{3})-\tfrac{3}{2} q t_{3}^2\big),
		\\
		\tau_2(Q,t)&=\mathfrak{t} t_{2}+Q \mathfrak{t}^2 \big(t_{3} (-q t_{3,2}+(q-1) t_{3,3}+t_{3,1})	-\tfrac{1}{2} q t_{3,1}^2+t_{2,1} (t_{3,1}-t_{3,2})
		\\&
		\hspace{1.8cm}
		+t_{3,1} ((q-1) t_{3,2}-t_{2,2})+t_{2} (t_{3,2}-t_{3,3})+\tfrac{1}{2} (q+1) t_{3}^2\big),
		\\
		\tau_{1,1}(Q,t)&=\mathfrak{t} t_{1,1} + Q \mathfrak{t}^2 \big(-q t_{3,1}^2+t_{3} (-q t_{3,1}-2 q t_{3,2}+t_{2,2})+t_{2,1} t_{3,1}+t_{2} t_{3,2}\big),
		\\
		\tau_3(Q,t)&= \mathfrak{t} t_{3}+ Q \mathfrak{t}^2 \big(t_{3,1} ((q-1) t_{3,3}-q t_{3,2})+\tfrac{1}{2} (q-1) t_{3,2}^2
		\\&
		\hspace{1.8cm}
		+\tfrac{1}{2} t_{3,1}^2-t_{2,2} t_{3,2}+t_{2,1} (t_{3,2}-t_{3,3})\big),
		\\
		\tau_{2,1}(Q,t)&=\mathfrak{t} t_{2,1} + Q \mathfrak{t}^2 \big(\tfrac{1}{2} (2 q+1) t_{3,1}^2+t_{3,1} (t_{2,2}-2 q t_{3,2})+t_{2,1} (t_{3,2}-t_{3,1})
		\\&
		\hspace{1.8cm}
		+t_{3} (q t_{3,1}+(2 q+1) t_{3,2}-2 q t_{3,3})+t_{2} (t_{3,3}-t_{3,2})\big),
		\\
		\tau_{3,1}(Q,t)&=\mathfrak{t} t_{3,1} + Q \mathfrak{t}^2 \big(t_{3,1} ((2 q+1) t_{3,2}-2 q t_{3,3})-q t_{3,2}^2
		\\&
		\hspace{1.8cm}
		-\tfrac{1}{2} t_{3,1}^2+t_{2,2} t_{3,2}+t_{2,1} (t_{3,3}-t_{3,2})\big),
		\\
		\tau_{2,2}(Q,t)&=\mathfrak{t} t_{2,2} + Q \mathfrak{t}^2 \big((q+1) t_{3,1} t_{3,2}+(q+1) t_{3} t_{3,3}\big),
		\\
		\tau_{3,2}(Q,t)&=\mathfrak{t} t_{3,2} + Q \mathfrak{t}^2 \big(\tfrac{1}{2} (q+1) t_{3,2}^2+t_{3,1} ((-q-1) t_{3,2}+(q+1) t_{3,3})\big),
		\\
		\tau_{3,3}(Q,t)&=\mathfrak{t} t_{3,3}.
		\\
	\end{aligned}
\end{equation}
The resulting correlators \eqref{Jcorr} are again up to $Q ,\mathfrak{t}^2$:
{\allowdisplaybreaks
	\scriptsize
	\begin{align*}
		J^0&= Q \Big(
		\tfrac{2 q+1}{(q-1)^4}
		+\mathfrak{t} \big(-\tfrac{t_{1,1}}{(q-1)^3}+\tfrac{t_{2,1}}{(q-1)^2}+\tfrac{t_{3,1}}{1-q}+\tfrac{(2 q+1) t_{1}}{(q-1)^4}+\tfrac{(-q-1) t_{2}}{(q-1)^3}+\tfrac{t_{3}}{(q-1)^2})
		\\&\hspace{0.5cm}
		+\mathfrak{t}^2 \big(t_{1} (-\tfrac{t_{1,1}}{(q-1)^3}+\tfrac{t_{2,1}}{(q-1)^2}+\tfrac{t_{3,1}}{1-q}+\tfrac{(-q-1) t_{2}}{(q-1)^3}+\tfrac{t_{3}}{(q-1)^2})
		\\&\hspace{1cm}
		+\tfrac{t_{3} t_{1,1}}{1-q}+t_{2} (\tfrac{t_{1,1}}{(q-1)^2}+\tfrac{t_{2,1}}{1-q})+\tfrac{(2 q+1) t_{1}^2}{2 (q-1)^4}+\tfrac{t_{2}^2}{2 (q-1)^2}\big)
		\Big),
		\\
		J^1&= Q \Big(
		\tfrac{5 (q^2+q)}{(q-1)^5}
		+\mathfrak{t} \big(-\tfrac{q t_{1,1}}{(q-1)^4}+\tfrac{t_{2,2}}{(q-1)^2}+\tfrac{t_{3,2}}{1-q}+\tfrac{(3 q^2+4 q) t_{1}}{(q-1)^5}+\tfrac{(-q^2-2 q) t_{2}}{(q-1)^4}+\tfrac{q t_{3}}{(q-1)^3}\big)
		\\&\hspace{0.5cm}
		+\mathfrak{t}^2 \big(t_{1} (-\tfrac{q t_{2,1}}{(q-1)^3}+\tfrac{t_{2,2}}{(q-1)^2}+\tfrac{q t_{3,1}}{(q-1)^2}+\tfrac{t_{3,2}}{1-q}-\tfrac{q t_{2}}{(q-1)^4})
		\\&\hspace{1cm}
		-\tfrac{t_{1,1}^2}{2 (q-1)^3}-\tfrac{t_{2,1}^2}{2 (q-1)}+\tfrac{t_{3} t_{2,1}}{1-q}+t_{1,1} (\tfrac{t_{2,1}}{(q-1)^2}+\tfrac{t_{3,1}}{1-q})
		\\&\hspace{1cm}
		+t_{2} (\tfrac{q t_{2,1}}{(q-1)^2}+\tfrac{t_{2,2}}{1-q}+\tfrac{t_{3,1}}{1-q}+\tfrac{t_{3}}{(q-1)^2})+\tfrac{(q^2+3 q) t_{1}^2}{2 (q-1)^5}-\tfrac{q t_{2}^2}{2 (q-1)^3}\big)
		\Big),
		\\
		J^{2}&= 
		-\tfrac{\mathfrak{t}^2 t_{1}^2}{2 (q-1)}
		+Q \Big(
		\tfrac{9 q^3+12 q^2-q}{(q-1)^6}
		\\&\hspace{0.5cm}
		+\mathfrak{t} \big(-\tfrac{q t_{1,1}}{(q-1)^4}-\tfrac{q t_{2,1}}{(q-1)^4}+\tfrac{q t_{2,2}}{(q-1)^3}
		+\tfrac{t_{3,3}}{1-q}+\tfrac{q^2 t_{3}}{(q-1)^4}+\tfrac{(4 q^3+7 q^2-q) t_{1}}{(q-1)^6}+\tfrac{(-q^3-3 q^2+q) t_{2}}{(q-1)^5}\big)
		\\&\hspace{0.5cm}
		+\mathfrak{t}^2 \big(t_{1} (\tfrac{q t_{1,1}}{(q-1)^5}-\tfrac{q t_{2,1}}{(q-1)^4}+\tfrac{q t_{3,2}}{(q-1)^2}+\tfrac{t_{3,3}}{1-q}-\tfrac{q t_{2}}{(q-1)^4})-\tfrac{q t_{1,1}^2}{2 (q-1)^4}
		\\&\hspace{1cm}
		+\tfrac{q t_{2,1}^2}{2 (q-1)^2}+\tfrac{t_{3} t_{3,1}}{1-q}+t_{2,1} (\tfrac{t_{2,2}}{1-q}+\tfrac{t_{3,1}}{1-q})	+t_{1,1} (\tfrac{t_{2,2}}{(q-1)^2}+\tfrac{t_{3,2}}{1-q})
		\\&\hspace{1cm}
		+t_{2} (\tfrac{q t_{3,1}}{(q-1)^2}+\tfrac{t_{3,2}}{1-q})+\tfrac{(q^3+3 q^2-q) t_{1}^2}{2 (q-1)^6}-\tfrac{q t_{2}^2}{2 (q-1)^4}+\tfrac{t_{3}^2}{2 (q-1)^2}\big)\Big),
		\\
		J^{1,1}&=-\tfrac{\mathfrak{t}^2 t_{1}^2}{2 (q-1)}
		+ Q \Big(
		\tfrac{8 q^3+9 q^2-2 q}{(q-1)^6}
		\\&\hspace{0.5cm}
		+\mathfrak{t} \big(\tfrac{(-q^2-2 q) t_{2,1}}{(q-1)^4}+\tfrac{2 q t_{1,1}}{(q-1)^5}+\tfrac{2 q t_{2,2}}{(q-1)^3}+\tfrac{q t_{3,1}}{(q-1)^3}
		\\&\hspace{1cm}
		-\tfrac{q t_{3,2}}{(q-1)^2}+\tfrac{(3 q^3+4 q^2-2 q) t_{1}}{(q-1)^6}+\tfrac{(-q^2-2 q) t_{2}}{(q-1)^4}+\tfrac{q t_{3}}{(q-1)^3}\big)
		\\&\hspace{0.5cm}
		+\mathfrak{t}^2 \big(t_{1} (\tfrac{(q^2+2 q) t_{1,1}}{(q-1)^5}-\tfrac{q (q+2) t_{2,1}}{(q-1)^4}+\tfrac{q t_{2,2}}{(q-1)^3}+\tfrac{q t_{3,1}}{(q-1)^3}+\tfrac{(q^2+2 q) t_{2}}{(q-1)^5}-\tfrac{q t_{3}}{(q-1)^4})
		\\&\hspace{1cm}
		-\tfrac{q t_{1,1}^2}{(q-1)^4}+\tfrac{q t_{1,1} t_{2,1}}{(q-1)^3}+\tfrac{t_{3} t_{2,2}}{1-q}+\tfrac{t_{2,1} t_{3,1}}{1-q}-\tfrac{q t_{1}^2}{(q-1)^6}-\tfrac{q (q+2) t_{2}^2}{2 (q-1)^4}
		\\&\hspace{1cm}
		+t_{2} (-\tfrac{q t_{1,1}}{(q-1)^4}+\tfrac{q t_{2,1}}{(q-1)^3}+\tfrac{t_{3,2}}{1-q}+\tfrac{q t_{3}}{(q-1)^3})\big)\Big),
		\\
		J^{3}&= \tfrac{\mathfrak{t}^2 t_{1} t_{2}}{1-q}
		+Q \Big(\tfrac{14 q^4+23 q^3-3 q^2+q}{(q-1)^7}
		\\&\hspace{0.5cm}
		+\mathfrak{t} \big(-\tfrac{q t_{1,1}}{(q-1)^4}-\tfrac{q t_{2,1}}{(q-1)^4}+\tfrac{q t_{2,2}}{(q-1)^3}-\tfrac{q t_{3,1}}{(q-1)^4}+\tfrac{q t_{3,2}}{(q-1)^3}
		\\&\hspace{1cm}
		-\tfrac{q t_{3,3}}{(q-1)^2}+\tfrac{(5 q^4+11 q^3-2 q^2+q) t_{1}}{(q-1)^7}+\tfrac{(-q^4-4 q^3+q^2-q) t_{2}}{(q-1)^6}+\tfrac{(q^3+q) t_{3}}{(q-1)^5}\big)
		\\&\hspace{0.5cm}
		+\mathfrak{t}^2 \big(t_{1} (\tfrac{q t_{1,1}}{(q-1)^5}+\tfrac{q t_{2,1}}{(q-1)^5}-\tfrac{q t_{2,2}}{(q-1)^4}-\tfrac{q t_{3,1}}{(q-1)^4}+\tfrac{q t_{3,2}}{(q-1)^3}+\tfrac{(-q^3-q) t_{2}}{(q-1)^6}+\tfrac{q t_{3}}{(q-1)^5})
		\\&\hspace{1cm}
		-\tfrac{q t_{1,1}^2}{2 (q-1)^4}+\tfrac{q t_{2,1}^2}{2 (q-1)^3}-\tfrac{t_{2,2}^2}{2 (q-1)}-\tfrac{t_{3,1}^2}{2 (q-1)}
		+t_{2} (\tfrac{q t_{1,1}}{(q-1)^5}-\tfrac{q t_{2,1}}{(q-1)^4}+\tfrac{q t_{3,1}}{(q-1)^3}-\tfrac{q t_{3}}{(q-1)^4})
		\\&\hspace{1cm}
		+\tfrac{t_{2,1} t_{3,2}}{1-q}+t_{1,1} (-\tfrac{q t_{2,1}}{(q-1)^4}+\tfrac{q t_{2,2}}{(q-1)^3}+\tfrac{t_{3,3}}{1-q})
		+\tfrac{(q^4+4 q^3-q^2+q) t_{1}^2}{2 (q-1)^7}+\tfrac{q t_{2}^2}{2 (q-1)^5}+\tfrac{q t_{3}^2}{2 (q-1)^3}\big)
		\Big),
		\\
		J^{2,1}&= \mathfrak{t}^2 \big(t_{1} (\tfrac{t_{1,1}}{1-q}+\tfrac{t_{2}}{1-q})+\tfrac{t_{1}^2}{2 (q-1)}\big)
		+Q \Big(
		+\tfrac{16 q^4+27 q^3-7 q^2-q}{(q-1)^7}
		\\&\hspace{0.5cm}
		+\mathfrak{t} \big(\tfrac{q^2 t_{3,1}}{(q-1)^4}+\tfrac{(q^3+8 q^2+q) t_{1,1}}{(q-1)^6}+\tfrac{(-q^3-6 q^2-q) t_{2,1}}{(q-1)^5}+\tfrac{(2 q^2+q) t_{2,2}}{(q-1)^4}
		\\&\hspace{1cm}
		-\tfrac{q t_{3,3}}{(q-1)^2}+\tfrac{q^2 t_{3}}{(q-1)^4}+\tfrac{(4 q^3+10 q^2+q) t_{1}}{(q-1)^6}+\tfrac{(-q^4-2 q^3+7 q^2+q) t_{2}}{(q-1)^6}\big)
		\\&\hspace{0.5cm}
		+\mathfrak{t}^2 \big(t_{1} (\tfrac{(q^3+6 q^2+q) t_{1,1}}{(q-1)^6}+\tfrac{(-3 q^2-q) t_{2,1}}{(q-1)^5}+\tfrac{q t_{3,2}}{(q-1)^3}-\tfrac{q^2 t_{3}}{(q-1)^5}+\tfrac{(q^3+6 q^2+q) t_{2}}{(q-1)^6}) +\tfrac{(-2 q^2-q) t_{1,1}^2}{2 (q-1)^5}
		\\&\hspace{1cm}
		+t_{2} (-\tfrac{q^2 t_{1,1}}{(q-1)^5}+\tfrac{q t_{3,1}}{(q-1)^3}+\tfrac{t_{3,2}}{q-1}+\tfrac{t_{3,3}}{1-q})+\tfrac{q t_{2,1}^2}{2 (q-1)^3}-\tfrac{t_{3,1}^2}{2 (q-1)}+\tfrac{q t_{1,1} t_{2,2}}{(q-1)^3}+\tfrac{t_{2,2} t_{3,1}}{1-q}
		\\&\hspace{1cm}
		+\tfrac{t_{3} t_{3,2}}{1-q}+t_{2,1} (\tfrac{t_{3,1}}{q-1}+\tfrac{t_{3,2}}{1-q})+\tfrac{(-4 q^3-10 q^2-q) t_{1}^2}{2 (q-1)^7}+\tfrac{(-3 q^2-q) t_{2}^2}{2 (q-1)^5}+\tfrac{q t_{3}^2}{2 (q-1)^3}\big)
		\Big),
		\\
		J^{3,1}&=  \mathfrak{t}^2\big(-\tfrac{t_{2}^2}{2 (q-1)}+\tfrac{t_{1,1} t_{2}}{1-q}+t_{1} (\tfrac{t_{2}}{q-1}+\tfrac{t_{3}}{1-q}+\tfrac{t_{2,1}}{1-q})\big)
		+Q \Big(
		+\tfrac{26 q^5+56 q^4-14 q^3+q^2+q}{(q-1)^8}
		\\&\hspace{0.5cm}
		+\mathfrak{t}\big(\tfrac{(5 q^4+14 q^3-3 q^2-q) t_{1}}{(q-1)^7}+\tfrac{(-q^3-5 q^2-q) t_{2}}{(q-1)^5}+\tfrac{(q^4-q^3+4 q^2+q) t_{3}}{(q-1)^6}+\tfrac{(2 q^4+15 q^3-q^2-q) t_{1,1}}{(q-1)^7}
		\\&\hspace{1cm}
		+\tfrac{(-q^4-7 q^3+2 q^2+q) t_{2,1}}{(q-1)^6}+\tfrac{(2 q^2+q) t_{2,2}}{(q-1)^4}+\tfrac{(q^3-3 q^2-q) t_{3,1}}{(q-1)^5}+\tfrac{(2 q^2+q) t_{3,2}}{(q-1)^4}-\tfrac{q (q+1) t_{3,3}}{(q-1)^3}\big) 
		\\&\hspace{0.5cm}
		+ \mathfrak{t}^2\big(
		t_{1}(\tfrac{(q^3+5 q^2+q) t_{2}}{(q-1)^6}+\tfrac{(-q^3+3 q^2+q) t_{3}}{(q-1)^6}+\tfrac{(q^4+7 q^3-2 q^2-q) t_{1,1}}{(q-1)^7}+t_{1,1} (\tfrac{(-2 q^2-q) t_{2,1}}{(q-1)^5}+\tfrac{q (q+1) t_{2,2}}{(q-1)^4})
		\\&\hspace{1cm}
		+\tfrac{(-2 q^2-q) t_{2,2}}{(q-1)^5}+\tfrac{t_{3,1}^2}{2 (q-1)}
		+\tfrac{(-2 q^2-q) t_{3,1}}{(q-1)^5}+\tfrac{q (q+1) t_{3,2}}{(q-1)^4}) 
		+\tfrac{(-q^3+3 q^2+q) t_{2,1}}{(q-1)^6}
		\\&\hspace{1cm}
		+t_{2} (\tfrac{(-2 q^2-q) t_{3}}{(q-1)^5}+\tfrac{(-q^3+3 q^2+q) t_{1,1}}{(q-1)^6}+\tfrac{(-2 q^2-q) t_{2,1}}{(q-1)^5}+\tfrac{q (q+1) t_{3,1}}{(q-1)^4})\big)
		\\&\hspace{1cm}
		+\tfrac{t_{3,1} t_{3,2}}{1-q}+t_{2,1} (\tfrac{t_{3,2}}{q-1}+\tfrac{t_{3,3}}{1-q})+\tfrac{(-5 q^4-14 q^3+3 q^2+q) t_{1}^2}{2 (q-1)^8}+\tfrac{t_{2,2} t_{3,2}}{1-q}
		\\&\hspace{1cm}
		+\tfrac{(-q^3+3 q^2+q) t_{2}^2}{2 (q-1)^6}+\tfrac{q (q+1) t_{3}^2}{2 (q-1)^4}+\tfrac{(-2 q^2-q) t_{1,1}^2}{2 (q-1)^5}+\tfrac{q (q+1) t_{2,1}^2}{2 (q-1)^4}
		\Big),
		\\
		J^{2,2}&= \mathfrak{t}^2 \big(-\tfrac{t_{1,1}^2}{2 (q-1)}+\tfrac{t_{1} t_{2,1}}{1-q}-\tfrac{t_{2}^2}{2 (q-1)}\big)
		+Q \Big(
		\tfrac{5 (4 q^4+9 q^3+q^2)}{(q-1)^7}
		\\&\hspace{0.5cm}
		+\mathfrak{t} \big(\tfrac{2 q^3 t_{2,2}}{(q-1)^5}+\tfrac{q^2 t_{3,1}}{(q-1)^4}-\tfrac{q^2 t_{3,3}}{(q-1)^3}+\tfrac{2 (q^4+8 q^3+q^2) t_{1,1}}{(q-1)^7}+\tfrac{(-q^5-q^4+15 q^3+2 q^2) t_{2}}{(q-1)^7}
		\\&\hspace{1cm}
		+\tfrac{q^2 t_{3,2}}{(q-1)^4}+\tfrac{(-q^4-8 q^3-q^2) t_{2,1}}{(q-1)^6}+\tfrac{q^2 t_{3}}{(q-1)^4}+\tfrac{(4 q^5-2 q^4-33 q^3-4 q^2) t_{1}}{(q-1)^8}\big)
		\\&\hspace{0.5cm}
		+\mathfrak{t}^2 \big(
		t_{1} (-\tfrac{2 q^3 t_{2,1}}{(q-1)^6}-\tfrac{q^2 t_{2,2}}{(q-1)^5}-\tfrac{q^2 t_{3,1}}{(q-1)^5}+\tfrac{q^2 t_{3,2}}{(q-1)^4}+\tfrac{(q^4+8 q^3+q^2) t_{1,1}}{(q-1)^7}-\tfrac{q^2 t_{3}}{(q-1)^5}+\tfrac{(q^4+8 q^3+q^2) t_{2}}{(q-1)^7})
		\\&\hspace{1cm}
		-\tfrac{q^3 t_{1,1}^2}{(q-1)^6}+\tfrac{q^2 t_{2,1}^2}{2 (q-1)^4}+t_{1,1} (\tfrac{q^2 t_{2,2}}{(q-1)^4}-\tfrac{q^2 t_{2,1}}{(q-1)^5})+\tfrac{(-8 q^4-24 q^3-3 q^2) t_{1}^2}{2 (q-1)^8}-\tfrac{q^3 t_{2}^2}{(q-1)^6}
		\\&\hspace{1cm}
		+t_{2} (-\tfrac{q^2 t_{1,1}}{(q-1)^5}-\tfrac{q^2 t_{2,1}}{(q-1)^5}+\tfrac{q^2 t_{3,1}}{(q-1)^4}-\tfrac{q^2 t_{3}}{(q-1)^5})+\tfrac{t_{3,1} t_{3,2}}{1-q}+\tfrac{t_{3} t_{3,3}}{1-q}+\tfrac{q^2 t_{3}^2}{2 (q-1)^4}\big)\Big),
		\\
		J^{3,2}&=  \mathfrak{t}^2\big(\tfrac{t_{2}^2}{2 (q-1)}+(\tfrac{t_{3}}{1-q}+\tfrac{t_{2,1}}{1-q}) t_{2}+\tfrac{t_{1,1} t_{2,1}}{1-q}+t_{1} (\tfrac{t_{2,1}}{q-1}+\tfrac{t_{2,2}}{1-q}+\tfrac{t_{3,1}}{1-q})\big)
		+Q \Big(\tfrac{5 (7 q^5+20 q^4+2 q^3-q^2)}{(q-1)^8}
		\\&\hspace{0.5cm}
		+ \mathfrak{t}(\tfrac{5 (q^6-q^5-14 q^4-q^3+q^2) t_{1}}{(q-1)^9}+\tfrac{(-q^5-3 q^4+14 q^3+5 q^2) t_{2}}{(q-1)^7}+\tfrac{(q^5-2 q^4+11 q^3+5 q^2) t_{3}}{(q-1)^7}+\tfrac{(4 q^5+33 q^4+2 q^3-4 q^2) t_{1,1}}{(q-1)^8}
		\\&\hspace{1cm}
		+\tfrac{(-q^4-10 q^3-4 q^2) t_{2,1}}{(q-1)^6}+\tfrac{(2 q^4-4 q^3-3 q^2) t_{2,2}}{(q-1)^6}+\tfrac{(q^4-7 q^3-4 q^2) t_{3,1}}{(q-1)^6}+\tfrac{(4 q^3+3 q^2) t_{3,2}}{(q-1)^5}+\tfrac{(-q^3-2 q^2) t_{3,3}}{(q-1)^4})
		\\&\hspace{0.5cm}
		+\mathfrak{t}^2(-\tfrac{5 (2 q^5+7 q^4-q^3-q^2) t_{1}^2}{2 (q-1)^9}+t_{3,1} (\tfrac{t_{3,2}}{q-1}+\tfrac{t_{3,3}}{1-q}))
		-\tfrac{t_{3,2}^2}{2 (q-1)}+t_{1,1} (\tfrac{(-4 q^3-3 q^2) t_{2,1}}{(q-1)^6}+\tfrac{(q^3+2 q^2) t_{2,2}}{(q-1)^5})
		\\&\hspace{1cm}
		+ t_{1}(\tfrac{(q^5+3 q^4-14 q^3-5 q^2) t_{2}}{(q-1)^8}+\tfrac{(-q^4+7 q^3+4 q^2) t_{3}}{(q-1)^7}+\tfrac{(q^4+10 q^3+4 q^2) t_{1,1}}{(q-1)^7}+\tfrac{(q^4+10 q^3+4 q^2) t_{2,1}}{(q-1)^7}
		\\&\hspace{1cm}
		+\tfrac{(-4 q^3-3 q^2) t_{2,2}}{(q-1)^6}+\tfrac{(-4 q^3-3 q^2) t_{3,1}}{(q-1)^6}+\tfrac{(q^3+2 q^2) t_{3,2}}{(q-1)^5})
		+\tfrac{(q^4+10 q^3+4 q^2) t_{2}^2}{2 (q-1)^7}+\tfrac{(q^3+2 q^2) t_{3}^2}{2 (q-1)^5}
		\\&\hspace{1cm}+t_{2} (\tfrac{(-4 q^3-3 q^2) t_{3}}{(q-1)^6}+\tfrac{(-q^4+7 q^3+4 q^2) t_{1,1}}{(q-1)^7}+\tfrac{(-4 q^3-3 q^2) t_{2,1}}{(q-1)^6}+\tfrac{(q^3+2 q^2) t_{3,1}}{(q-1)^5})
		\\&\hspace{1cm}+\tfrac{(-2 q^4+4 q^3+3 q^2) t_{1,1}^2}{2 (q-1)^7}+\tfrac{(q^3+2 q^2) t_{2,1}^2}{2 (q-1)^5} \Big),
		\\
		J^{3,3}&=\mathfrak{t}^2\big(-\tfrac{t_{3}^2}{2 (q-1)}-\tfrac{t_{2,1}^2}{2 (q-1)}+\tfrac{t_{1,1} t_{2,2}}{1-q}+\tfrac{t_{2} t_{3,1}}{1-q}+\tfrac{t_{1} t_{3,2}}{1-q}\big) 
		+Q \Big(\tfrac{10 (4 q^5+13 q^4+4 q^3)}{(q-1)^8}
		\\&\hspace{0.5cm}
		+\mathfrak{t}\big(-\tfrac{(q+2) t_{3,3} q^3}{(q-1)^5}+\tfrac{5 (q^6-2 q^5-20 q^4-7 q^3) t_{1}}{(q-1)^9}+\tfrac{(-q^7-q^6+14 q^5-56 q^4-26 q^3) t_{2}}{(q-1)^9}
		\\&\hspace{1cm}
		+\tfrac{(q^6-3 q^5+23 q^4+14 q^3) t_{3}}{(q-1)^8}+\tfrac{5 (q^5+9 q^4+4 q^3) t_{1,1}}{(q-1)^8}
		+\tfrac{(-q^6-7 q^5+27 q^4+16 q^3) t_{2,1}}{(q-1)^8}
		\\&\hspace{1cm}
		+\tfrac{(2 q^5-9 q^4-8 q^3) t_{2,2}}{(q-1)^7}+\tfrac{(q^5-12 q^4-9 q^3) t_{3,1}}{(q-1)^7}+\tfrac{5 (q^4+q^3) t_{3,2}}{(q-1)^6}\big)
		\\&\hspace{0.5cm}
		+\mathfrak{t}^2\big(-\tfrac{5 (q^6+2 q^5-7 q^4-3 q^3) t_{1}^2}{(q-1)^{10}}
		+\tfrac{(2 q^5+21 q^4+12 q^3) t_{2}^2}{2 (q-1)^8}+\tfrac{(q^4+2 q^3) t_{3}^2}{2 (q-1)^6}+t_{1,1} (\tfrac{q^3 (q+2) t_{2,2}}{(q-1)^6}-\tfrac{5 (q^4+q^3) t_{2,1}}{(q-1)^7})
		\\&\hspace{1cm}
		+ t_{1}(\tfrac{(q+2) t_{3,2} q^3}{(q-1)^6}+\tfrac{(q^6-3 q^5-47 q^4-21 q^3) t_{2}}{(q-1)^9}+\tfrac{(-q^5+12 q^4+9 q^3) t_{3}}{(q-1)^8}+\tfrac{(q^6+7 q^5-27 q^4-16 q^3) t_{1,1}}{(q-1)^9}
		\\&\hspace{1cm}
		+\tfrac{(2 q^5+21 q^4+12 q^3) t_{2,1}}{(q-1)^8}-\tfrac{5 (q^4+q^3) t_{2,2}}{(q-1)^7}-\tfrac{5 (q^4+q^3) t_{3,1}}{(q-1)^7})
		+\tfrac{(-2 q^5+9 q^4+8 q^3) t_{1,1}^2}{2 (q-1)^8}+\tfrac{(q^4+2 q^3) t_{2,1}^2}{2 (q-1)^6}
		\\&\hspace{1cm}
		+t_{2} (\tfrac{(q+2) t_{3,1} q^3}{(q-1)^6}-\tfrac{5 (q^4+q^3) t_{3}}{(q-1)^7}+\tfrac{(-q^5+12 q^4+9 q^3) t_{1,1}}{(q-1)^8}-\tfrac{5 (q^4+q^3) t_{2,1}}{(q-1)^7})\big) \Big) .
	\end{align*}
}
The non-constant pairing $G_{\mu\nu}$ \eqref{GTid} at zero deformation is
\begin{equation}
	G_{\mu\nu}(t=0)=
	\frac 1 {1-Q}
	\begin{pmatrix}
		1 & 1 & 1 & 1 & 1 & 1 & 1 & 1 & 1 & 1 \\
		1 & 1 & 1 & 1 & 1 & 1 & 1 & 1 & 1 & Q \\
		1 & 1 & 1 & 1 & 1 & 1 & 1 & Q & Q & Q \\
		1 & 1 & 1 & 1 & Q & 1 & Q & 1 & Q & Q \\
		1 & 1 & 1 & Q & 1 & Q & Q & Q & Q & Q \\
		1 & 1 & 1 & 1 & Q & 1 & Q & Q & Q & Q \\
		1 & 1 & 1 & Q & Q & Q & Q & Q & Q & Q \\
		1 & 1 & Q & 1 & Q & Q & Q & Q & Q & Q^2 \\
		1 & 1 & Q & Q & Q & Q & Q & Q & Q & Q^2 \\
		1 & Q & Q & Q & Q & Q & Q & Q^2 & Q^2 & Q^2
	\end{pmatrix}.
\end{equation}
At non-zero deformation we use that $G_{\mu\nu}=\partial_\mu\partial_\nu G_{00}$, with $G_{00}=\frac 1{1-Q}\sum_{d,n\geq 0}G^{(d)}_{00}(t)Q^d $.  The coefficients $G^{(d)}_{00}(t)$ read up to order $Q^{d=3}$ and $\mathfrak{t}^3$:

{\allowdisplaybreaks
	\scriptsize
	
\begin{align*}
	G_{00}^{(0)}(t)=&
	1+\mathfrak{t} \big(t_{1}+t_{2}+t_{3}+t_{1,1}+t_{2,1}+t_{2,2}+t_{3,1}+t_{3,2}+t_{3,3}\big)
	\\&
	+\tfrac{1}{2} \mathfrak{t}^2 \big(t_{1}^2+2 t_{1} t_{2}+t_{2}^2+2 t_{1} t_{3}+2 t_{2} t_{3}+t_{3}^2+2 t_{1} t_{1,1}+2 t_{2} t_{1,1}+t_{1,1}^2+2 t_{1} t_{2,1}
	\\&\hspace{.5cm}
	+2 t_{2} t_{2,1}+2 t_{1,1} t_{2,1}+t_{2,1}^2+2 t_{1} t_{2,2}+2 t_{1,1} t_{2,2}+2 t_{1} t_{3,1}+2 t_{2} t_{3,1}+2 t_{1} t_{3,2}\big)
	\\&
	+\tfrac{1}{6} \mathfrak{t}^3 \big(t_{1}^3+3 t_{1}^2 t_{2}+3 t_{1} t_{2}^2+t_{2}^3+3 t_{1}^2 t_{3}+6 t_{1} t_{2} t_{3}+3 t_{1}^2 t_{1,1}+6 t_{1} t_{2} t_{1,1}+3 t_{2}^2 t_{1,1}
	\\&\hspace{.5cm}
	+3 t_{1} t_{1,1}^2+t_{1,1}^3+3 t_{1}^2 t_{2,1}+6 t_{1} t_{2} t_{2,1}+6 t_{1} t_{1,1} t_{2,1}+3 t_{1}^2 t_{2,2}+3 t_{1}^2 t_{3,1}\big)
	\\
	G_{00}^{(1)}(t)=&
	\tfrac{1}{2}  \mathfrak{t}^2 \big(2 t_{3} t_{1,1}+2 t_{3} t_{2,1}+2 t_{2} t_{2,2}+2 t_{3} t_{2,2}+2 t_{2,1} t_{2,2}+t_{2,2}^2+2 t_{3} t_{3,1}+2 t_{1,1} t_{3,1}+t_{3,2}^2
	\\&\hspace{.5cm}
	+2 t_{2,1} t_{3,1}+2 t_{2,2} t_{3,1}+t_{3,1}^2+2 t_{2} t_{3,2}+2 t_{3} t_{3,2}+2 t_{1,1} t_{3,2}+2 t_{2,1} t_{3,2}+2 t_{2,2} t_{3,2}
	\\&\hspace{.5cm}
	+2 t_{3,1} t_{3,2}+2 t_{1} t_{3,3}+2 t_{2} t_{3,3}+2 t_{3} t_{3,3}+2 t_{1,1} t_{3,3}+2 t_{2,1} t_{3,3}+2 t_{3,1} t_{3,3}\big)
	\\&
	+\tfrac{1}{6}  \mathfrak{t}^3 \big(
	3 t_{2}^2 t_{3}+3 t_{1} t_{3}^2+3 t_{2} t_{3}^2+t_{3}^3+6 t_{1} t_{3} t_{1,1}+6 t_{2} t_{3} t_{1,1}+3 t_{3}^2 t_{1,1}+3 t_{2} t_{1,1}^2
	\\&\hspace{.5cm}
	+3 t_{3} t_{1,1}^2+3 t_{2}^2 t_{2,1}+6 t_{1} t_{3} t_{2,1}+6 t_{2} t_{3} t_{2,1}+3 t_{3}^2 t_{2,1}+6 t_{2} t_{1,1} t_{2,1}+6 t_{3} t_{1,1} t_{2,1}
	\\&\hspace{.5cm}
	+3 t_{1,1}^2 t_{2,1}+3 t_{1} t_{2,1}^2+3 t_{2} t_{2,1}^2+3 t_{3} t_{2,1}^2+3 t_{1,1} t_{2,1}^2+t_{2,1}^3+6 t_{1} t_{2} t_{2,2}+3 t_{2}^2 t_{2,2}
	\\&\hspace{.5cm}
	+6 t_{1} t_{3} t_{2,2}+6 t_{2} t_{3} t_{2,2}+6 t_{1} t_{1,1} t_{2,2}+6 t_{2} t_{1,1} t_{2,2}+6 t_{3} t_{1,1} t_{2,2}+3 t_{1,1}^2 t_{2,2}+6 t_{1} t_{2,1} t_{2,2}
	\\&\hspace{.5cm}
	+6 t_{2} t_{2,1} t_{2,2}+6 t_{3} t_{2,1} t_{2,2}+6 t_{1,1} t_{2,1} t_{2,2}+3 t_{2,1}^2 t_{2,2}+3 t_{1} t_{2,2}^2+3 t_{2} t_{2,2}^2+3 t_{3} t_{2,2}^2
	\\&\hspace{.5cm}
	+6 t_{1} t_{2} t_{3,1}+3 t_{2}^2 t_{3,1}+6 t_{1} t_{3} t_{3,1}+6 t_{2} t_{3} t_{3,1}+3 t_{3}^2 t_{3,1}+6 t_{1} t_{1,1} t_{3,1}+6 t_{2} t_{1,1} t_{3,1}
	\\&\hspace{.5cm}
	+6 t_{3} t_{1,1} t_{3,1}+3 t_{1,1}^2 t_{3,1}+6 t_{1} t_{2,1} t_{3,1}+6 t_{2} t_{2,1} t_{3,1}+6 t_{3} t_{2,1} t_{3,1}+6 t_{1,1} t_{2,1} t_{3,1}
	\\&\hspace{.5cm}
	+3 t_{2,1}^2 t_{3,1}+6 t_{1} t_{2,2} t_{3,1}+6 t_{2} t_{2,2} t_{3,1}+6 t_{1,1} t_{2,2} t_{3,1}+6 t_{2,1} t_{2,2} t_{3,1}+3 t_{1} t_{3,1}^2+3 t_{2} t_{3,1}^2
	\\&\hspace{.5cm}
	+3 t_{3} t_{3,1}^2+3 t_{1,1} t_{3,1}^2+3 t_{2,1} t_{3,1}^2+3 t_{1}^2 t_{3,2}+6 t_{1} t_{2} t_{3,2}+3 t_{2}^2 t_{3,2}+6 t_{1} t_{3} t_{3,2}
	\\&\hspace{.5cm}
	+6 t_{2} t_{3} t_{3,2}+6 t_{1} t_{1,1} t_{3,2}+6 t_{2} t_{1,1} t_{3,2}+6 t_{3} t_{1,1} t_{3,2}+3 t_{1,1}^2 t_{3,2}+6 t_{1} t_{2,1} t_{3,2}+6 t_{2} t_{2,1} t_{3,2}
	\\&\hspace{.5cm}
	+6 t_{3} t_{2,1} t_{3,2}+6 t_{1,1} t_{2,1} t_{3,2}+3 t_{2,1}^2 t_{3,2}+6 t_{1} t_{2,2} t_{3,2}+6 t_{2} t_{2,2} t_{3,2}+6 t_{1} t_{3,1} t_{3,2}+6 t_{2} t_{3,1} t_{3,2}
	\\&\hspace{.5cm}
	+6 t_{1,1} t_{3,1} t_{3,2}+3 t_{1} t_{3,2}^2+3 t_{1}^2 t_{3,3}+6 t_{1} t_{2} t_{3,3}+3 t_{2}^2 t_{3,3}+6 t_{1} t_{3} t_{3,3}+6 t_{1} t_{1,1} t_{3,3}
	\\&\hspace{.5cm}
	+6 t_{2} t_{1,1} t_{3,3}+6 t_{3} t_{1,1} t_{3,3}+6 t_{1} t_{2,1} t_{3,3}+6 t_{2} t_{2,1} t_{3,3}+6 t_{1} t_{3,1} t_{3,3}\big)
	\\
	G_{00}^{(2)}(t)
	=&\tfrac{1}{2}  \mathfrak{t}^2 \big(2 t_{2,2} t_{3,3}+2 t_{3,2} t_{3,3}+t_{3,3}^2\big)
	\\&
	+\tfrac{1}{6}  \mathfrak{t}^3 \big(
	3 t_{3}^2 t_{2,2}+3 t_{1,1} t_{2,2}^2+3 t_{2,1} t_{2,2}^2+t_{2,2}^3+6 t_{3} t_{2,2} t_{3,1}+3 t_{2,2}^2 t_{3,1}+3 t_{2,2} t_{3,1}^2+t_{3,1}^3
	\\&\hspace{.5cm}
	+3 t_{3}^2 t_{3,2}+6 t_{3} t_{2,2} t_{3,2}+6 t_{1,1} t_{2,2} t_{3,2}+6 t_{2,1} t_{2,2} t_{3,2}+3 t_{2,2}^2 t_{3,2}+6 t_{3} t_{3,1} t_{3,2}+6 t_{2,1} t_{3,1} t_{3,2}
	\\&\hspace{.5cm}
	+6 t_{2,2} t_{3,1} t_{3,2}+3 t_{3,1}^2 t_{3,2}+3 t_{2} t_{3,2}^2+3 t_{3} t_{3,2}^2+3 t_{1,1} t_{3,2}^2+3 t_{2,1} t_{3,2}^2+3 t_{2,2} t_{3,2}^2
	\\&\hspace{.5cm}
	+3 t_{3,1} t_{3,2}^2+t_{3,2}^3+6 t_{2} t_{3} t_{3,3}+3 t_{3}^2 t_{3,3}+3 t_{1,1}^2 t_{3,3}+6 t_{3} t_{2,1} t_{3,3}+6 t_{1,1} t_{2,1} t_{3,3}
	\\&\hspace{.5cm}
	+3 t_{2,1}^2 t_{3,3}+6 t_{1} t_{2,2} t_{3,3}+6 t_{2} t_{2,2} t_{3,3}+6 t_{3} t_{2,2} t_{3,3}+6 t_{1,1} t_{2,2} t_{3,3}+6 t_{2,1} t_{2,2} t_{3,3}
	\\&\hspace{.5cm}
	+3 t_{2,2}^2 t_{3,3}+6 t_{2} t_{3,1} t_{3,3}+6 t_{3} t_{3,1} t_{3,3}+6 t_{1,1} t_{3,1} t_{3,3}+6 t_{2,1} t_{3,1} t_{3,3}+6 t_{2,2} t_{3,1} t_{3,3}
	\\&\hspace{.5cm}
	+3 t_{3,1}^2 t_{3,3}+6 t_{1} t_{3,2} t_{3,3}+6 t_{2} t_{3,2} t_{3,3}+6 t_{3} t_{3,2} t_{3,3}+6 t_{1,1} t_{3,2} t_{3,3}+6 t_{2,1} t_{3,2} t_{3,3}
	\\&\hspace{.5cm}
	+6 t_{2,2} t_{3,2} t_{3,3}+6 t_{3,1} t_{3,2} t_{3,3}+3 t_{3,2}^2 t_{3,3}+3 t_{1} t_{3,3}^2+3 t_{2} t_{3,3}^2+3 t_{1,1} t_{3,3}^2+3 t_{2,1} t_{3,3}^2+3 t_{2,2} t_{3,3}^2\big)
	\\
	G_{00}^{(3)}(t)
	=&\tfrac{1}{6} \mathfrak{t}^3 \big(3 t_{3} t_{3,3}^2+3 t_{3,1} t_{3,3}^2+3 t_{3,2} t_{3,3}^2+t_{3,3}^3\big)
\end{align*}
}

\newpage

\bibliographystyle{JHEP}
\bibliography{QK}
\end{document}